\documentclass[preprint]{elsarticle}

\usepackage[colorlinks,citecolor=blue,linktoc=all,linkcolor=cyan]{hyperref}
\usepackage{graphicx}

\usepackage[T1]{fontenc}
\usepackage{dsfont}               
\usepackage{mathrsfs}             
\usepackage{slashed}              
\usepackage{amsmath}
\usepackage{amssymb}
\usepackage{amsbsy}
\usepackage{amsfonts}

\numberwithin{equation}{section}
\numberwithin{table}{section}
\numberwithin{figure}{section}

\journal{Progress in Particle and Nuclear Physics}

\usepackage{graphicx}
\usepackage{dcolumn}
\usepackage{bm}
\usepackage[normalem]{ulem}
\usepackage{tikz}
\usetikzlibrary{positioning}

\usepackage{braket}
\usepackage{multirow}
\usepackage{dsfont}
\usepackage{mathtools}
\usepackage{xcolor}
\usepackage[normalem]{ulem}

\usepackage{cancel}
\usepackage{epsfig,dsfont}
\usepackage{subfigure}
\usepackage{bm}
\usepackage{amssymb}
\usepackage{mathrsfs}
\usepackage{amsmath}
\usepackage{enumitem}
\usepackage{dcolumn}

\newcommand{\bp}{\boldsymbol{p}}

\newcommand{\bA}{\boldsymbol{A}}

\renewcommand{\vec}[1]{\mbox{\boldmath$#1$\unboldmath}}

\newcommand{\I}{\mathrm{i}}
\newcommand{\bT}{\boldsymbol{T}}
\def\lsi{\raise0.3ex\hbox{$<$\kern-0.75em\raise-1.1ex\hbox{$\sim$}}}
\def\gsi{\raise0.3ex\hbox{$>$\kern-0.75em\raise-1.1ex\hbox{$\sim$}}}

\newcommand{\bea}{\begin{eqnarray}}
\newcommand{\nn}{\nonumber}
\newcommand{\eea}{\end{eqnarray}}
\newcommand{\beq}{\begin{equation}}
\newcommand{\eeq}{\end{equation}}

\usepackage{titlesec}
\usepackage{sectsty}
\titleformat{\section}{\normalfont\Large\bfseries}{\thesection}{1em}{}
\titleformat{\subsection}{\normalfont\large\bfseries}{\thesubsection}{1em}{}
\titleformat{\subsubsection}{\normalfont\normalsize\bfseries}{\thesubsubsection}{1em}{}

\begin{document}
	
	\begin{frontmatter}

	\title{Chiral spin symmetry and hot/dense QCD.}	
		
		\author[mymainaddress,mysecondaryaddress]{L. Ya. Glozman}

		\ead{leonid.glozman@uni-graz.at}
		
		\address[mymainaddress]{Institute of Physics, University of Graz, 8010 Graz, Austria}

		\begin{abstract}
Above the chiral symmetry restoration crossover around $T_{ch} ~\sim ~155$ MeV
a new regime arises in QCD, a stringy fluid, which is characterized by an approximate chiral spin symmetry of the thermal partition function. This symmetry is not a symmetry of the Dirac Lagrangian and is a symmetry of the electric part of
the QCD Lagrangian. In this regime the medium consists of the chirally symmetric
and approximately chiral spin symmetric  hadrons that are made of the chirally symmetric quarks connected into the color singlet compounds
by  a confining chromoelectric
field. This regime is evidenced by the approximate chiral spin symmetry of the spatial and temporal correlators and by the breakdown of the thermal perturbation theory at the crossover between the partonic (the quark-gluon plasma) and the stringy fluid regimes at $\sim 3 T_{ch}$. The chiral spin symmetry smoothly disappears above  $\sim 3 T_{ch}$ which means that the chromoelectric
confining  interaction gets screened.
A direct evidence  that  the stringy
fluid medium consists of densely packed hadrons is the  pion spectral function that shows a distinct pion state and its first radial excitation above $T_{ch}$. Another direct evidence of the hadron degrees of freedom in the stringy fluid is the bottomonium spectrum with the 1S,2S,3S and 1P,2P radial and orbital excitations that become broad with temperature. The hadrons between $T_{ch}$ and $\sim 3 T_{ch}$ in the stringy fluid  interact
strongly which makes the stringy fluid more a liquid rather than a gas. We  discuss how this chiral spin symmetric regime extends into the 
 finite chemical potentials domain and present a
qualitative sketch of the QCD phase diagram.			
		\end{abstract}
		
		\begin{keyword}
			chiral spin symmetry\sep hot and dense QCD\sep stringy fluid \sep QCD phase diagram 
			
		\end{keyword}
		
	\end{frontmatter}
	
	\newpage
	
	\thispagestyle{empty}
	\tableofcontents
	

	\newpage
\section{A little bit of history. Introduction.}

Before the RHIC era there was a general belief
that at some critical temperature $T_c$ there should happen
a deconfinement phase transition from a hadron resonance gas (HRG) phase to
a quark gluon plasma phase (QGP) \cite{Hag,Cab}. A crucial difference between
two phases are degrees of freedom. While in the HRG phase the degrees of freedom 
are hadrons that practically do not interact, in the QGP phase these are
partons - quarks and gluons. There is no spontaneous breaking of
chiral symmetry in
a system of free partons. Consequently it was expected that at the
same temperature $T_c$ a chiral restoration happens, so the critical
temperature $T_c$ should be a common temperature of both deconfinement
and of chiral restoration phase transitions.

A lot of experimental efforts for the last 30 years were invested into a search of  the quark-gluon plasma
in heavy ion collisions at AGS (BNL),  SPS (CERN), RHIC (BNL) 
and LHC (CERN). Experimental findings first at RHIC and then at LHC indicate that  assuming local thermal equilibrium within the fireball  
the hot QCD matter is different from a dilute hadron resonance gas at low temperatures
\cite{BNL1,BNL2,BNL3,BNL4,BNL5}, indeed. The most prominent result is the observation of
the elliptic flow. The fit of hadron spectra and elliptic flow at RHIC and LHC
by means of viscous hydrodynamic, see \cite{H1,H2} and references therein,
suggests  rather small values of eta/s within the fireball
at the RHIC and LHC temperatures, of the order 0.2, that is only slightly above the limiting value $1/4\pi$. This result tells that the system is strongly coupled with a small mean free path of the constituents, i.e., highly collective.  It implies that the fireball cannot
be a dilute gas of mesons like at low temperatures and zero net baryon density. Consequently, we can  say that a new form of matter is seen experimentally.
To answer the question about the origin and structure 
of this new form of matter one needs information about nature of the constituents. It was a priori
assumed that these constituents should be strongly
interacting quark and gluon quasiparticles as no other
degrees of freedom would exist above the Hagedorn
temperature. 
Another prominent indication of the
strongly interacting matter within the fireball is a modification of
a jet that propagates through the fireball. In a dilute meson gas
one should not expect a significant modification of
a jet as compared to the vacuum.

In parallel with the experimental activity the same
field was developing on the lattice. On the lattice the only
a priori ingredient is the QCD Lagrangian. In Euclidean space-time
one can quantize the theory at vanishing baryon chemical potential
in the equilibrium at a given temperature  using existing 
Monte-Carlo algorithms  and calculate observables.
It was concluded 
that in reality there is no a phase transition and instead a smooth analytic
crossover takes place  \cite{A1}. The quark condensate 
decreases from its zero temperature value to practically zero at temperatures 
from 120 MeV to 180 MeV with a pseudocritical temperature of chiral
symmetry restoration $T_{ch} \sim 155$ MeV \cite{F} and approximately
at the same temperature (or slightly above) the Polyakov loop, which is
an order parameter for center symmetry ("deconfinement") in a pure glue
theory, showed an inflection point. So the community took the point that in QCD there
is a fast common deconfinement - chiral restoration crossover from
hadron gas to QGP around the pseudocritical temperature $T_{pc} \sim 155$ MeV.
This result was confirmed by a few lattice groups.

\begin{figure}
\centering
\includegraphics[angle=0,width=0.45\linewidth]{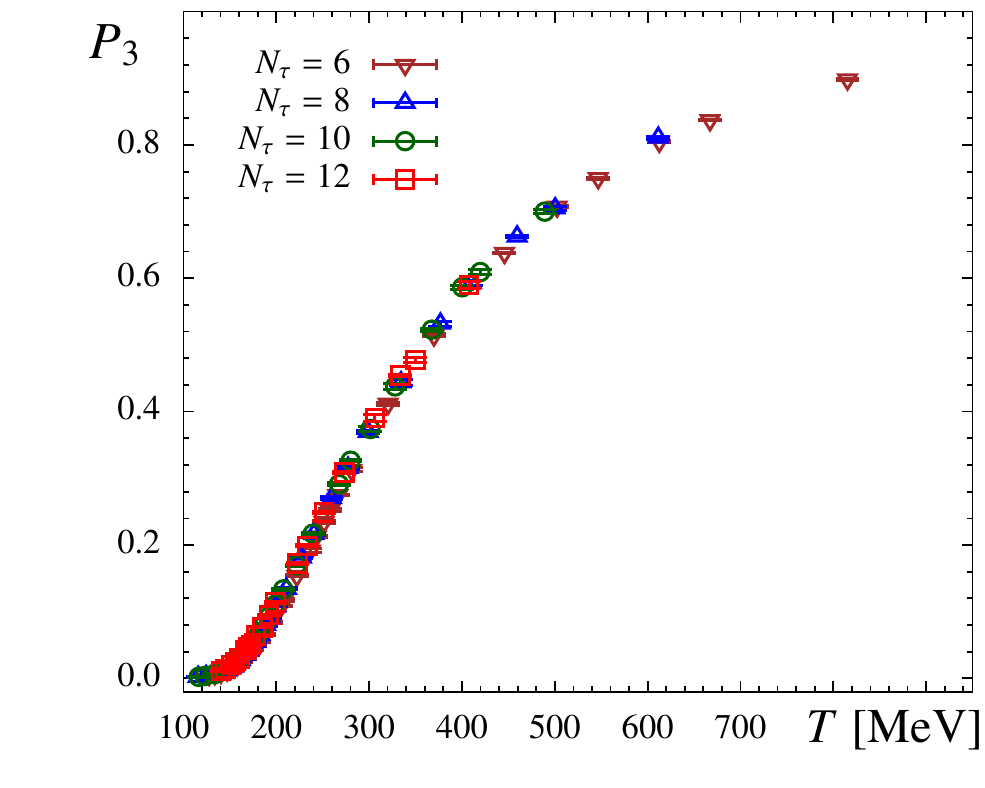}
\caption{ Renormalized Polyakov loop in $N_f = 2 + 1$ QCD at  physical quark masses.
From Ref. \cite{S}.}
\label{St}
\end{figure}

However, the inflection point of the non renormalized Polyakov loop
was used to see a pseudocritical temperature for "deconfinement".
Some time ago the evolution of the renormalized Polyakov was
obtained, e.g., in Ref. \cite{S},  see Fig ~\ref{St}.  
One clearly observes that there is no hint of a deconfinement crossover
around $T_{ch} \sim 155$ MeV since the deconfinement transition
should be accompanied by the  Polyakov loop evolution from 0 to 1.
The renormalized Polyakov loop evolves from 0 to 1 in a broad temperature
interval up to temperatures of $T \sim 1$ GeV and its inflection point
is around 300 MeV, as can be seen from  Fig ~\ref{St}. 
This suggests that above the chiral symmetry
restoration crossover between 120 and 180 MeV QCD is still in the
confining regime and there cannot be any (quasi)parton degrees freedom.
Unfortunately this circumstance was largely ignored by the authors and the community.

\begin{figure}
\centering
\includegraphics[angle=0,width=0.45\linewidth]{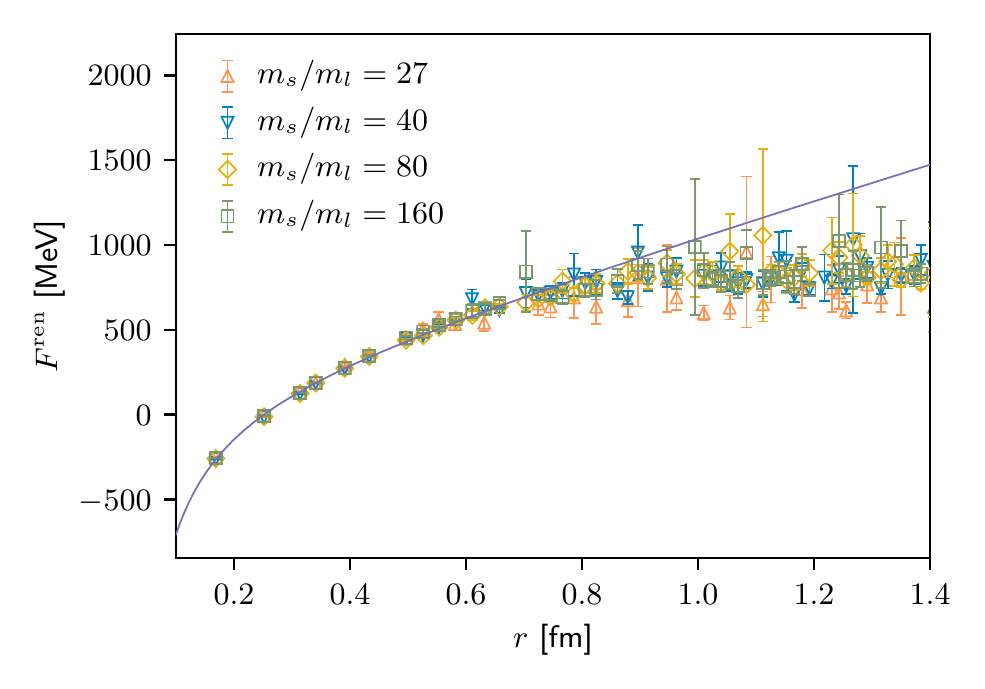}
\includegraphics[angle=0,width=0.45\linewidth]{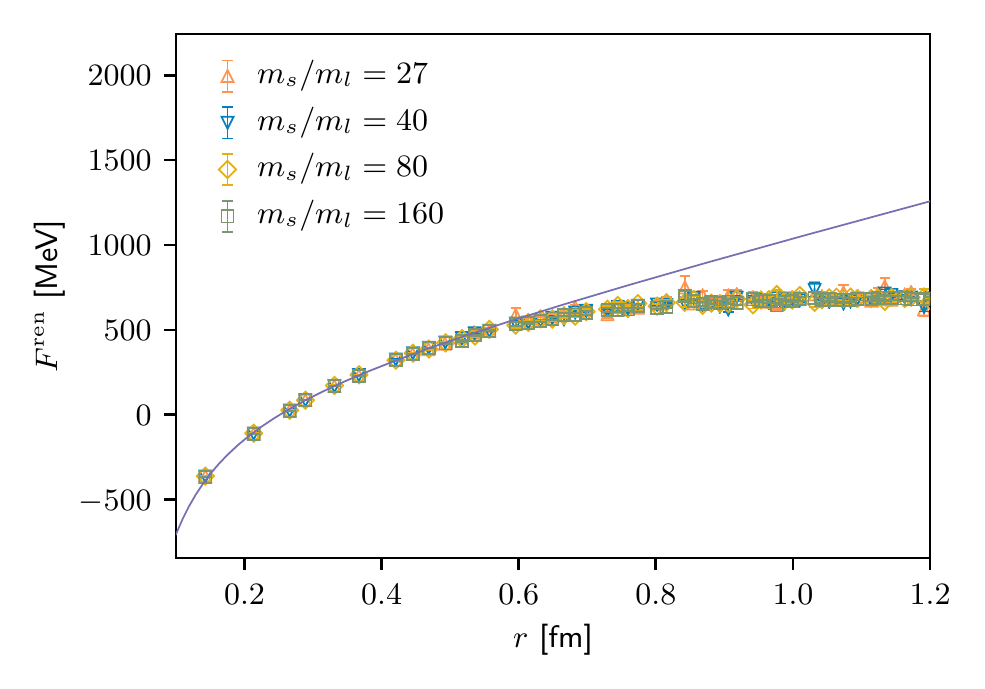}
\caption{ The color-singlet part of the Polyakov loop correlator in 
$N_f = 2 + 1$ QCD at physical and lower than physical quark masses.
The left and right panels are for $T = 141$ MeV and $T = 166$ MeV,
respectively.
From Ref. \cite{K}.}
\label{Karsch}
\end{figure}

Another potential evidence for a deconfinement phase transition was suggested
by Matsui and Satz in Ref. \cite{MS}. They argued that at a
critical temperature the familiar linear + Coulomb confining potential
between the static charges (infinitely heavy quarks) should be
Debye screened and becomes weaker than a negative Coulomb potential. Such a potential,
$\sim -1/r \exp( - m_D r)$, does not support any bound state
between heavy quarks and consequently signals a deconfining
transition to QGP. 
A potential obtained from the correlators of the Polyakov loops at  temperatures significantly
above $T_{ch}$ demonstrates a flattening
of the linear part of the potential which is, however, not yet
 a Debye screening. In particular, no Debye
screening is seen at the chiral restoration temperature, see 
Fig. ~\ref{Karsch}, where such a potential
 is shown for $N_f =2+1$ at physical quark masses \cite{K}. The extracted potential 
 with obvious flattening is exactly the same for temperatures below and above $T_{ch}$. I.e., it cannot be related to "deconfinement" and simply
 signals that a process of production of two heavy-light mesons
 takes place. Actually the concept of an effective potential between
 static sources is a model dependent construction. If one assumes
 that an optical potential should take place instead of a pure real
 potential as in Fig. ~\ref{Karsch}, then the real part of the potential
 turns out to be a linear confining potential up to  large temperatures
 with rising with temperature imaginary part \cite{optical}.
 Recently it was concluded by the Bielefeld lattice group that no evidence
 of deconfinement from the Polyakov loop exists on the lattice in the vicinity of
 the chiral pseudocritical temperature \cite{La}.

 There is no
 reliable and accepted definition and order parameter for deconfinement
 in QCD with light quarks. The only sensible question that should
 be answered is about degrees of freedom that drive the system. If it
 turns out that these degrees of freedom are (quasi)quarks and (quasi)gluons, then 
 this would mean that it is a QGP. This situation is expected at a
 very large temperature where the asymptotic freedom forces the
 strong coupling constant to vanish \cite{Col}. However, there is no evidence
 either theoretical or experimental that above  $T_{ch} \sim 155$ MeV
 in QCD the degrees of freedom are (quasi)quarks and (quasi)gluons.
 So it is a key problem to establish effective degrees of freedom
 above the chiral crossover.

\begin{figure}
\centering
\includegraphics[scale=0.3]{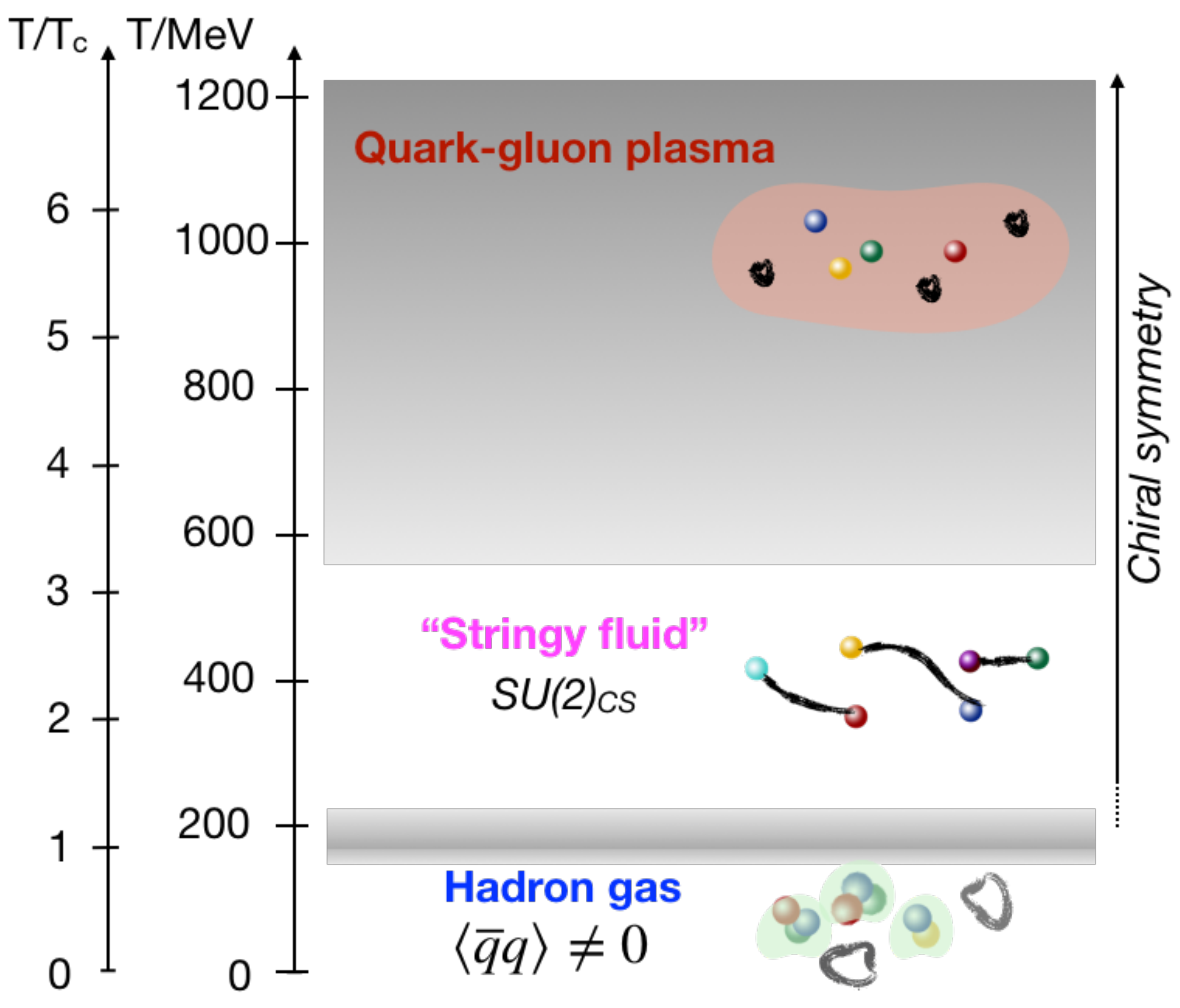} 
\caption{
Sketch for the temperature evolution of the QCD effective degrees of freedom.
From Ref. \cite{R2}.}
\label{fig:sketch}
\end{figure}

Some time ago it was predicted that at finite temperatures
above the chiral
symmetry restoration crossover 
QCD should be still in the confining regime with
hadron-like degrees of freedom \cite{G3}. Such a regime
should be evidenced by a chiral spin symmetry \cite{G1,G2}
of the QCD correlators above $T_{ch}$. A year later first
results on approximate chiral spin symmetry of spatial correlators above the chiral
crossover  were presented by a collaboration of theorists from Graz, Ljubljana  and  JLQCD \cite{R1}. Those results were limited by a
temperature $T \sim 400$ MeV. In a subsequent study \cite{R2} the temperatures
were extended up to $T \sim 1$ GeV and it was established that the approximate
chiral spin symmetry smoothly disappears above $ T \sim 3 T_{ch}$. Three regimes
of QCD were identified with clearly distinguishable symmetries,
with spontaneously broken chiral symmetry below $T_{ch}$, with chiral
symmetries and approximate chiral spin symmetry between $T_{ch}$ and
$ \sim 3 T_{ch}$ and with chiral symmetry at higher temperatures, see
Fig. \ref{fig:sketch}. 
These regimes are different by symmetries and degrees of
freedom. The stringy fluid regime is characterized by the
approximate chiral spin symmetry of the thermal QCD partition
function. The degrees of freedom are the color-singlet hadron-like
states where chirally symmetric quarks are bound by the confining
electric field.
The chiral spin symmetry of the thermal partition function was verified
in temporal correlators above $T_{ch}$ \cite{R3}. These results have
been summarized in  a talk  "Three regimes of QCD" \cite{three}.

Since then an important development in the 
field happened. Namely, an evidence for existence of such intermediate
regime independent of symmetry arguments was obtained from  published screening mass mass spectra
which demonstrate the breakdown of partonic description of the system
below $ 3 T_{ch}$ \cite{GPP}. Very recently  another direct evidence 
for hadron-like degrees of freedom in the stringy fluid regime was presented \cite{LP}: The pseudoscalar spectral function
extracted from the spatial lattice correlators   
demonstrates a distinct pion state and its first radial excitation.
Further evidence for hadron-like degrees of freedom above $T_{ch}$
was obtained for heavy quarks: The bottomonium spectral function 
above $T_{ch}$ is not flat and contains  radial and orbital excitations 1S,2S,3S
and 1P,2P
that become broader with temperature \cite{Lar}.
The stringy fluid medium consists of densely packed hadrons
that interact strongly, in contrast to the dilute meson gas below $T_{ch}$ with a large
mean free path of mesons.
Hence the stringy fluid is more a liquid rather than a gas.

We also discuss a simple physical picture for chirally symmetric
and approximately chiral spin symmetric mesons above $T_{ch}$. They can be presented as color-electric strings with massless chiral quarks at the ends. Such a 
view automatically explains the observed chiral spin symmetry.

Since the quark chemical potential in the QCD action is manifestly
chiral spin symmetric \cite{G4} one should expect that the chiral spin symmetric regime extends at finite baryon
density as a chiral spin symmetric band downwards across the QCD phase diagram \cite{GPP}. In the cold and dense region the baryon parity doublet matter is proposed as a possible candidate for chiral spin symmetric
matter. 

Finally we  discuss  available experimental data on dileptons
both at zero baryon density as well as at large baryon chemical potential
and show that they are consistent with existence of the chiral spin symmetric band.

 \section{Chiral spin symmetry} 
 
 The history of the chiral spin symmetry begins with the observation
 of an unexpected degeneracy of isovector $J=1$ mesons seen on the
 lattice upon artificial truncation of the near-zero modes of
 the Dirac operator \cite{D1}.
 The quark condensate of the vacuum is connected with
the density of the near-zero modes of the Euclidean Dirac
operator via the Banks-Casher relation \cite{BC}

\begin{equation}
 \langle \bar{\psi}\psi\rangle=-\pi\rho(0)\;,\\ 
 \label{banks}
 \end{equation}

\begin{equation}
\rho(0) =
  \lim_{m \rightarrow 0}  
\frac{1}{\pi } \int_{-\infty}^{+\infty}
\frac{d \lambda \rho(\lambda)}{m - i\lambda}\;.\nn 
 \end{equation}

\noindent
 The hermitian Euclidean Dirac operator, $i \gamma_\mu D_\mu$, 
 has in a finite volume $V$ a discrete spectrum with real eigenvalues $\lambda_n$:

\begin{equation}
i \gamma_\mu D_\mu  \psi_n(x) = \lambda_n \psi_n(x).
\label{ev}
\end{equation}
Consequently, removing by hands $k$ lowest lying modes
of the Dirac operator from the quark propagators,
 
 \begin{equation}
 S =S_{Full}-
  \sum_{i=1}^{k}\,\frac{1}{\lambda_i}\,|\lambda_i\rangle \langle \lambda_i|,
 \end{equation} 
one a priori expects restoration of chiral 
$SU(2)_L \times SU(2)_R$ and possibly of $U(1)_A$
symmetries, if hadrons survive. This should be signalled by a degeneracy
of hadrons connected by the $SU(2)_L \times SU(2)_R$ and $U(1)_A$
transformations. However, it turned out that all isovector $J=1$ mesons
get degenerate, not only those that are connected by the chiral
transformations.
The symmetry groups that are responsible for this degeneracy, the
transformation laws as well as their physical meaning  were obtained in Refs. \cite{G1,G2}.

 In Ref. \cite{G1} the  $SU(2)_{CS}$ chiral spin transformation 
 was defined as a transformation that mixes  the
 right- and left-handed Weyl quark spinors

\begin{equation}
\left(\begin{array}{c}
R\\
L
\end{array}\right)\; \rightarrow
\left(\begin{array}{c}
R'\\
L'
\end{array}\right)=
\exp \left(i  \frac{\varepsilon^n \sigma^n}{2}\right) \left(\begin{array}{c}
R\\
L
\end{array}\right)\; .
\label{eq:RL}
\end{equation}

\noindent 
  So the fundamental irreducible representation of $SU(2)_{CS}$
 is {\it two-dimensional}. In terms of the Dirac spinors $\psi$ the same
 transformation can be written via four-dimensional $\gamma$-matrices \cite{G2}

\begin{equation}
\label{V-defsp}
  \psi \rightarrow  \psi^\prime = \exp \left(i  \frac{\varepsilon^n \Sigma^n}{2}\right) \psi\; ,
\end{equation}

\noindent
where the generators $\Sigma^n$ of the four-dimensional reducible
representation are

\begin{equation}
 \Sigma^n = \{\gamma_0,-i \gamma_5\gamma_0,\gamma_5\}, ~~~[\Sigma^a,\Sigma^b]=2i\epsilon^{abc}\Sigma^c.
\label{SIGCS}
\end{equation}

\noindent
The $U(1)_A$ group is a subgroup of $SU(2)_{CS}$.

In Euclidean space with the $O(4)$ symmetry all four directions are 
equivalent and the $SU(2)_{CS}$ transformations can be generated
by any Euclidean  hermitian $\gamma$-matrix $\gamma_k$, $k=1,2,3,4$ instead
of Minkowskian $\gamma_0$:

\begin{equation}
 \Sigma^n = \{\gamma_k,-i \gamma_5\gamma_k,\gamma_5\},
\label{SIGCS}
\end{equation} 

\begin{equation}
\gamma_i\gamma_j + \gamma_j \gamma_i =
2\delta^{ij}; \qquad \gamma_5 = \gamma_1\gamma_2\gamma_3\gamma_4.
\label{gamma}
\end{equation}
The $su(2)$ algebra 
is satisfied with any $k=1,2,3,4$. A choice of $k$ is limited
by the spatial $O(3)$ invariance: only those $k$ can be used that
do not mix operators with different spatial $O(3)$ spins $J$.

Note that the chiral spin transformations mix quarks with different
chiralities (i.e., they mix different irreducible representations
of the Lorentz group) and consequently the $SU(2)_{CS}$ symmetry is not a
symmetry of the Dirac Lagrangian. 

The direct product of the $SU(2)_{CS}$ group with the flavor group
$SU(N_F)$ can be embedded into a $SU(2N_F)$ group.
 This group contains the chiral
symmetry  $SU(N_F)_L \times SU(N_F)_R \times U(1)_A$ as a subgroup.
 The set of $(2N_F)^2-1$ generators of $SU(2N_F)$ is

\begin{align}
\{
(\tau^a \otimes \mathds{1}_D),
(\mathds{1}_F \otimes \Sigma^n),
(\tau^a \otimes \Sigma^n)
\}
\end{align}
with $\tau$  being the flavor generators (with the flavor index $a$) and $n=1,2,3$ is the $SU(2)_{CS}$ index.
The fundamental
vector of $SU(2N_F)$ at $N_F=2$ is

\begin{equation}
\Psi =\begin{pmatrix} u_{\textsc{R}} \\ u_{\textsc{L}}  \\ d_{\textsc{R}}  \\ d_{\textsc{L}} \end{pmatrix}. 
\label{fourcomp}
\end{equation}
\noindent

The chiral spin and $SU(2N_F)$ symmetries above should not be
confused with the  Pauli-G\"ursey $SU(2N_F)$ symmetry \cite{PG1,PG2},
which is a symmetry of the free Dirac Lagrangian and mixes the
right quark with the left antiquark (and vice versa). It should also
not be mixed up with the non-relativistic $SU(2N_F)$ symmetry with
heavy quarks. The multiplets of the latter group contain states
of only a given spatial parity.

While the $SU(2)_{CS}$ and $SU(2N_F)$ symmetries are not
symmetries of the Dirac Lagrangian, they are symmetries
of the Lorentz-invariant color  charge
\begin{equation}
Q^a =  \int d^3x  
\psi^\dagger(x) T^a \psi(x),
\label{Q}
\end{equation}
with $T^a$ the $SU(3)$ color generators. The color charge remains
invariant under the unitary $SU(2)_{CS}$ and $SU(2N_F)$ transformations.

The latter important feature allows us to use the 
$SU(2)_{CS}$ and $SU(2N_F)$ symmetries to distinguish the
chromoelectric and chromomagnetic fields in a given reference frame
because  the chromoelectric field is defined through its interaction
with the color charge while the chromomagnetic field is defined via its
action on 
the spatial current. The latter current is not 
$SU(2)_{CS}$ and $SU(2N_F)$ symmetric. This can be made
explicit as follows.

 In Minkowski space in a given reference frame the electric and
 magnetic fields are
 different fields.
Interaction of  fermions with the gauge field in Minkowski space-time
can be split in a given reference frame into temporal and spatial parts:

\begin{equation}
\overline{\psi}   \gamma^{\mu} D_{\mu} \psi = \overline{\psi}   \gamma^0 D_0  \psi 
  + \overline{\psi}   \gamma^i D_i  \psi ,
\label{cl}
\end{equation}
\noindent
where the covariant derivative $D_{\mu}$  includes
interaction of the matter field $\psi$ with the  gauge field $\bA_\mu$,

\begin{equation}
D_{\mu}\psi =( \partial_\mu - ig \bT \cdot \bA_\mu)\psi.
\end{equation}
The temporal term contains  interaction of the color-octet
 charge density 

\begin{equation}
\bar \psi (x)  \gamma^0  \bT \psi(x) = \psi (x)^\dagger  \bT \psi(x)
\label{density}
\end{equation}
with the electric  
part of the gluonic field. 
It is invariant  under 
 $SU(2)_{CS}$  and  $SU(2N_F)$. Note that the $SU(2)_{CS}$ transformations
defined  via the Euclidean
Dirac matrices can be identically applied to Minkowski Dirac spinors without
any modification of the generators.
 The spatial part contains the quark kinetic term
and   interaction with the chromomagnetic field.  It breaks 
 $SU(2)_{CS}$ and $SU(2N_F)$.   We conclude that $SU(2)_{CS}$ and $SU(2N_F)$
 symmetries are symmetries of the electric part of the QCD Lagrangian 
  in a given reference frame and can be used to distinguish
  the electric and magnetic interactions: A symmetry of the electric part of the QCD
 Lagrangian is larger than the chiral symmetry of the QCD
 Lagrangian as a whole.\footnote{Notice that it is a gauge-invariant
 statement since it is based on the gauge-invariant definition of
 the electric field, $\vec F = Q^a \vec E^a$.} 
Of course, in order to discuss the electric and magnetic
components of the gauge field
one needs to fix a reference frame. The invariant mass of the hadron
is  the rest frame energy. Consequently, to discuss physics
of hadron mass  it is natural to use the hadron rest frame.
At high temperatures the Lorentz invariance is broken and 
 the preferred frame  is the medium rest frame. 

This analysis suggests  the necessary and sufficient conditions for emergence
of approximate $SU(2)_{CS}$ and $SU(2N_F)$ symmetries: (i) both chiral symmetries must be at least approximately restored and (ii) the color-electric
quark-gluon interaction must strongly dominate over the color-magnetic one and over kinetic
terms. The latter condition implies that the color-electric field
of an effective action must strongly dominate over the color-magnetic one. Within  perturbative description this cannot happen since the symmetry
of the perturbation theory is the symmetry of the Dirac Lagrangian, i.e.
only chiral symmetry. In addition, the perturbative gluons, like photons, contain both electric and magnetic
parts with equal magnitude.

\section{Representations of chiral and chiral spin groups for mesons}

Consider first  spin $J=0$ mesons  within $N_F=2$,
which are the $ \pi (1,0^{-+}), f_0 (0,0^{++}), a_0(1,0^{++})$ and $\eta(0,0^{-+})$ mesons with the $u,d$
quark content only.
Their local interpolating fields
are given as

\begin{equation}
 O_\pi(x)  = \bar q(x) \vec \tau \imath \gamma_5 q(x),
\label{ppi}
\end{equation}

\begin{equation}
 O_{f_0}(x)  = \bar q(x)  q(x),
\label{ff0}
\end{equation}

\begin{equation}
 O_{\eta}(x)  = \bar q(x)  \imath \gamma_5 q(x),
\label{eeta}
\end{equation}

\begin{equation}
 O_{a_0}(x)  = \bar q(x) \vec \tau  q(x).
\label{aa0}
\end{equation}

\noindent
These four operators belong to an
irreducible representation
of the group
$U(2)_L\times U(2)_R  \supset SU(2)_L\times SU(2)_R \times U(1)_A$. 
It is instructive to see how these interpolating fields transform
under different subgroups of the group above.

 The $SU(2)_L\times SU(2)_R$ transformations consist of vectorial
 and axial transformations in the isospin space. The axial
transformation $$q \rightarrow exp(i \gamma_5 \frac{\theta_A^a \tau^a}{2})  q$$ mixes fields of opposite parity. For instance,

\begin{equation}
  \bar q(x)  q(x) \longrightarrow  \bar q(x) e^{ i \gamma_5 \theta^a_A \tau^a} q(x)
 = \cos {|\vec \theta_A|} \bar q(x)  q(x) + \sin {| \vec \theta_A|}  
 \frac{\vec \theta_A}{|\vec \theta_A|} \cdot 
\bar q(x) \vec \tau \imath \gamma_5 q(x).  
\label{f0mix}
\end{equation}

\noindent 
 Hence, under the axial part of the $SU(2)_L\times SU(2)_R$ transformation
the following fields get mixed

\begin{equation}
 O_\pi(x)  \leftrightarrow O_{f_0}(x). 
\label{pif0}
\end{equation} 

\noindent
Similarly one obtains

\begin{equation}
 O_{a_0}(x)  \leftrightarrow O_{ \eta}(x).
\label{a0eta}
\end{equation} 

\noindent
The fields (\ref{pif0}) form the basis functions of the $(1/2,1/2)_a$
irreducible representation of the $SU(2)_L \times SU(2)_R$  group, while the fields
(\ref{a0eta}) transform as $(1/2,1/2)_b$. \footnote{The irreducible representations of the 
$SU(2)_L \times SU(2)_R$  group are described by the total isospins of the right
and left-handed quarks, $(I_R,I_L)$. The total usual isospin of quarks can take values
according to the standard angular momentum addition rules, $ |I_R - I_L| \leq I \leq I_R + I_L$.
The indices $a$ and $b$ distinguish two different representations $(1/2,1/2)$.}

The $U(1)_A$ transformation  $$q \rightarrow exp(i \theta_A \gamma_5) q$$ mixes  fields
of the same isospin but opposite parity:

\begin{equation}
 O_\pi(x)  \leftrightarrow O_{a_0}(x) 
\label{pia0}
\end{equation} 

\noindent
as well as

\begin{equation}
 O_{f_0}(x)  \leftrightarrow O_{ \eta}(x).
\label{f0eta}
\end{equation}

\noindent
All four interpolators together belong to the  representation
$(1/2,1/2)_a \oplus (1/2,1/2)_b$ which
is an irreducible representation of the groups $U(2)_L\times U(2)_R $  and
$SU(2)_L\times SU(2)_R \times U(1)_A $.

With the  spin $J=0$ local fields it is impossible to construct
irreducible representations of the chiral spin group (\ref{eq:RL}).
Indeed, applying the $SU(2)_{CS}$ transformations (\ref{V-defsp}) to the fields
(\ref{ppi}),(\ref{ff0}),(\ref{eeta}) and (\ref{aa0}) one obtains
that these fields get mixed with  
$\bar q(x) \gamma_0 \gamma_5 q(x)$   (and similar for the isovector
operators), that represents the axial charge density. It does not create
a physical state with $J=0$ and consequently
is not a proper $J=0$ operator. This means that the $O(3)$ spatial invariance
is not consistent with the $SU(2)_{CS}$ symmetry for $J=0$ mesons
(see also chapter 10 below).
If the spatial rotational invariance is preserved, like it is in an
isotropic medium, an approximate $SU(2)_{CS}$ symmetry of an effective action
and of the thermal partition function cannot be observed with the $J=0$
mesons. One needs the higher spin mesons to see this symmetry of an effective action.

\begin{figure}
\centering
\includegraphics[angle=0,width=0.65\linewidth]{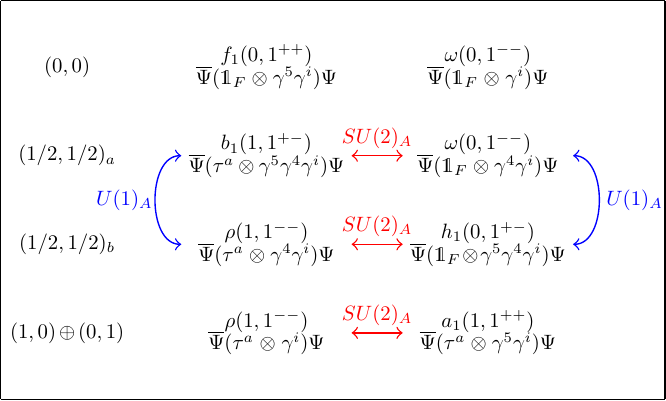}
\includegraphics[angle=0,width=0.65\linewidth]{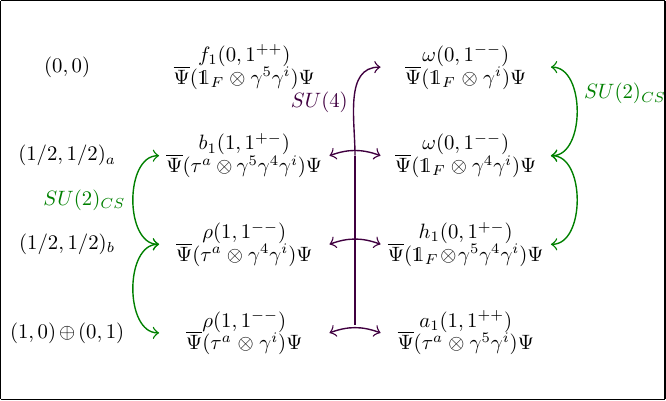}
\caption{Transformations between $J=1$ operators, $i=1,2,3$.
The left columns indicate the $SU(2)_L \times SU(2)_R$ 
representation for every
operator. Red and blue arrows connect operators which transform into 
each other under $SU(2)_L \times SU(2)_R$ and $U(1)_A$, respectively.
Green arrows connect operators that belong to
$SU(2)_{CS}$ irreducible triplets. Purple arrow shows the $SU(4)$
15-plet. The $f_1$ operator  is a singlet of $SU(4)$.
From Ref. \cite{G2}.}
\label{algebra}
\end{figure}

The chiral  as well as 
the chiral spin and $SU(4)$ multiplets for $J=1$ are given
in Fig. \ref{algebra} \cite{G2}. The
local fields  presented in this figure   are characterized
by usual quantum numbers $I,J^{PC}$ and by a representation
of the $SU(2)_L \times SU(2)_R$ group. All these fields
are orthogonal since each of them has a unique set of quantum
numbers. The $U(1)_A$ and $SU(2)_L \times SU(2)_R$ transformations
of these fields obtained like for $J=0$ fields are depicted in
the upper part of the figure.
The chiral spin transformations (\ref{V-defsp})
connect  operators
for different $J=1$ fields and the CS transformations are
consistent with the $O(3)$ invariance. Note that these 
representations (\ref{V-defsp})\footnote{The Minkowskian $\gamma_0$
matrix coincides with the Euclidean $\gamma_4$.} are suited only for study of symmetries of
the Hamiltonian,
i.e. symmetries of the temporal Euclidean correlators.
For the spatial correlators one needs to use multiplets
discussed in detail in Ref. \cite{R2} and in Sec. 6.2 below.

A few important comments are in order. One observes from  Fig.
\ref{algebra} that there are, e.g., two different $\rho$ operators. They both 
have the same spin, isospin as well as spatial and charge parities.
They differ by the gamma-structure as well as by  chiral representations.
In  vacuum with broken chiral symmetry both these operators create from the vacuum
 one and the same $\rho$-meson, though with different couplings. These couplings are
 determined by the chiral symmetry breaking in the physical  $\rho$-meson wave function,
 i.e. by a mixture of $(1,0)+ (0,1)$ and $(1/2,1/2)_b$ components in the meson
 wave function.
 This issue is well understood on the lattice \cite{GLL1,GLL2}. However, in the chirally
 symmetric world above $T_{ch}$ the index of the chiral representation becomes an
 exact and conserved quantum number of the physical state. This means that in the chirally symmetric world
 there are two different mesons with $(1,1^{--})$ usual quantum numbers that differ by the
 chiral quantum number: one of them has the chiral quantum number $(1,0)+ (0,1)$ while
 another one is described by $(1/2,1/2)_b$. {\it These are different orthogonal states}.
 The $SU(2)_{CS}$ symmetry requires that the orthogonal states within an irreducible representation
 of $SU(2)_{CS}$ must be degenerate. Consequently a prediction of the chiral spin symmetry
 is existence of three degenerate states, two of them carry $(1,1^{--})$ usual quantum numbers
 but differ by the chiral index and  $b_1$ meson. The same situation takes place for the isoscalar
 mesons.

Similar transformation properties as well as representations of the chiral spin
and $SU(4)$ groups can be obtained for $J=2$  and higher spin meson operators.
Note that the latter operators are necessarily nonlocal \cite{D2}.

\section{Observation of the chiral spin symmetry in truncation studies
in mesons and its implications in vacuum and for hot QCD}

\begin{figure}
\centering 
\includegraphics[width=0.55\linewidth]{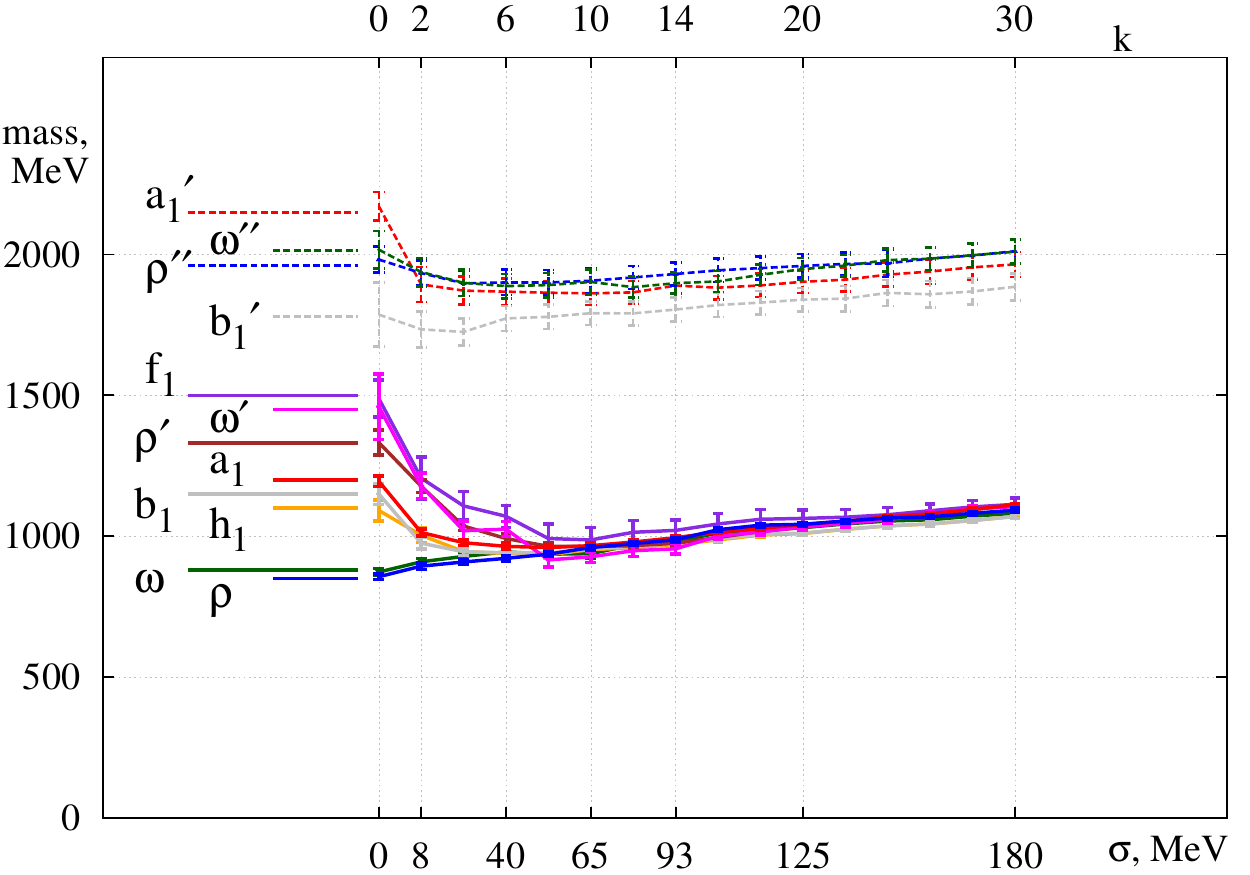}
\caption{$J=1$ meson masses  as a function of the
truncation number $k$ where $k$ represents the amount of removed
lowest modes of the Dirac operator. $\sigma$ shows the energy gap in the Dirac spectrum. From Ref. \cite{D2}.} 
\label{den}
\end{figure}

The symmetry predictions from the $SU(2)_{CS}$ and $SU(2N_F)$
groups for J=1,2 mesons
have been tested in $N_F=2$ QCD in Refs. \cite{D2,D3}, 
see as an example a degeneracy pattern of all $J=1$
mesons \cite{D2} in Fig. \ref{den}. 
This large degeneracy, presumably only approximate, represents the $SU(2)_{CS}$  and the $SU(4)$ symmetries since it contains irreducible representations of both groups, see Fig. \ref{algebra}.
These  results imply, given the symmetry classification of
the QCD Lagrangian, that while the confining chromoelectric
interaction is distributed among
 all modes of the Dirac operator, the
chromomagnetic interaction, which breaks both  symmetries, is located 
at least predominantly in the near-zero
modes. Consequently  an artificial removal of the near-zero modes leads to
the emergence of $SU(2)_{CS}$  and $SU(4)$ in hadron spectrum. {\it Chiral symmetry
breaking and confinement in QCD are not directly related phenomena.}
The highly degenerate level seen in Fig. \ref{den} represents a $SU(4)$-
symmetric level of the pure electric confining interaction.
The hadron
spectra  could be viewed as a  splitting of
the  level of the  QCD string by means of dynamics
contained in the near-zero modes of the Dirac operator, i.e., dynamics
of $SU(2)_L \times SU(2)_R$ and $U(1)_A$ chiral symmetry breaking 
that also includes  magnetic effects
in QCD \cite{G1}. A possible candidate for latter dynamics could
be local instanton or other topological fluctuations of the gluonic
field \cite{hooft,ss}. Note also that a confining interaction
can also lead to the accumulation of the near-zero modes of the Dirac operator, i.e. to some
contribution to the quark condensate.  The present
results seen in Fig. \ref{den} imply, however,  that chiral symmetry
might be restored due to some specific reasons, e.g. in the QCD
medium at some temperature or baryon density,
but confinement would be still there.

Analytical studies \cite{lang,cgl} conclude the following.
Some specific gluonic dynamics leads to the accumulation
of the near-zero modes of the Dirac operator and consequently 
to the breaking of $SU(2)_L \times SU(2)_R$ and $U(1)_A$ chiral symmetries.
A gap in the Dirac spectrum provided by the artificial truncation of
the near-zero modes in the Dirac operator necessarily implies
restoration of both symmetries. The root of this statement is
precisely the same as of the Banks-Casher relation. It is a general
statement. We do not need to know which dynamics and why it leads
to the accumulation of the near-zero modes.
Emergence of larger approximate  $SU(2)_{CS}$  and $SU(4)$ symmetries,
seen in Fig.  \ref{den}, requires that the electric confining interaction
should be the most important for the higher-lying modes.

In
reality the degeneracy of  Fig. \ref{den} represents a larger
symmetry since both the 15-plet and the singlet of $SU(4)$ are
also degenerate. What symmetry is it \cite{G2,Cohen}? The latter
question was answered in Ref. \cite{G5}. It is a $SU(4) \times SU(4)$.
Indeed  the irreducible 16-plet of $SU(4) \times SU(4)$
is a direct sum of the 15-plet and of the singlet of $SU(4)$.
A transparent physical reason for emergence of the larger $SU(4) \times SU(4)$ symmetry
is a simple one. A confining electric flux tube binds a quark and an antiquark
and has two independent quark-gluon vertices. Each vertex has its own
$SU(4)$ symmetry.\footnote{
 Consider the Minkowski QCD Hamiltonian in Coulomb gauge in the hadron rest frame \cite{Lee}:
\begin{equation}
H_{QCD} = H_E + H_B
 + \int d^3 x \Psi^\dag({\boldsymbol{x}}) 
[-i \boldsymbol{ \alpha} \cdot \boldsymbol{\nabla} ]  \Psi(\boldsymbol{x})
+ H_T + H_C,
\label{ham}
\end{equation}
\noindent
with the transverse  and instantaneous "Coulombic" interactions to be:
\begin{equation}
H_T = -g \int d^3 x \, \Psi^\dag({\boldsymbol{x}}) \boldsymbol{\alpha} 
\cdot t^a \boldsymbol{A}^a(\boldsymbol{x}) \, \Psi(\boldsymbol{x}) \; , 
\end{equation}
\begin{equation} 
H_C = \frac{g^2}{2} \int  d^3 x \, d^3 y\, J^{-1} \ \rho^a(\boldsymbol{x})  F^{ab}(\boldsymbol{x},\boldsymbol{y}) \, J \, \rho^b(\bf y) \; .
\label{coul}
\end{equation}
\noindent
Here $J$ is the  Faddeev-Popov determinant, $\rho^a(\boldsymbol{x})$ and $\rho^a(\boldsymbol{y})$
are color-charge densities of quarks (\ref{density}) and gluons at the space points  $\boldsymbol{x}$
and $\boldsymbol{y}$
and $F^{ab}(\boldsymbol{x},\boldsymbol{y})$ is a 
"Coulombic" kernel. 

The kinetic and transverse  parts of the Hamiltonian are
 chirally symmetric. The confining "Coulombic" part (\ref{coul})
carries the $SU(2N_F)$ symmetry, because the quark color charge density
operator is $SU(2N_F)$ symmetric. The gluonic part of the color charge density is trivially
 $SU(2N_F)$ invariant. 
However, both $\rho^a(\boldsymbol{x})$ and $\rho^b(\boldsymbol{y})$
 are independently $SU(2N_F)$ symmetric because the $SU(2N_F)$ transformations
 at  spatial points $\boldsymbol{x}$
and $\boldsymbol{y}$ can be completely independent, with different
rotations angles. A contribution with $\boldsymbol{x} = \boldsymbol{y}$ is absent because
of Grassmannian nature of quarks.
This means that the confining "Coulombic" interaction
 is actually $SU(2N_F) \times SU(2N_F)$-symmetric.}
 
The results of truncation studies, discussed above, have direct implications for
QCD at temperatures above the pseudocritical temperature of chiral symmetry restoration
 around $T_{ch} \sim 155$ MeV. Here the quark condensate vanishes
and consequently the near-zero modes of the Dirac operator
are suppressed by temperature. There are strong indications from the lattice
that the $U(1)_A$ symmetry is also at least approximately effectively
restored \cite{JLQCD1,JLQCD3}. Given these observations and given results
on emerging symmetries obtained at $T=0$ upon artificial truncation
of the near-zero modes of the Dirac operator it was
predicted that above the chiral restoration crossover the 
$SU(2)_{CS}$  and $SU(4)$ symmetries should naturally emerge,
without any truncation, and QCD should 
still be in a confining mode with the hadron-like degrees of freedom \cite{G3}.

\section{Representations of  chiral spin group for nucleons}

Emergence of the chiral spin and $SU(4)$ symmetries in baryons
upon truncation of the near zero modes was observed and studied
in Ref. \cite{D4}, see degeneracy patterns of the correlators
presented in this paper\footnote{This Section is technically more involved and can be omitted at the first reading.}. In the cited paper a complete classification
of the chiral spin representations for nucleons was absent. Hence for
future possible applications we present here such classification obtained in Ref.
\cite{CG}.

Lorentz and Fierz-invariance of the local color-singlet
three-quark operators restricts the number of such linear independent
operators to be equal two \cite{Dm}.
However, the chiral spin symmetry is not
a symmetry of the Dirac equation and the chiral spin transformations mix 
irreducible representations of  the Lorentz group. Consequently
if one discusses properties of  operators under the chiral spin
transformations we need a complete set of such operators with respect
to $SU(2)_{CS}$. A single-quark field transforms under a two-dimensional
irreducible representation  of $SU(2)_{CS}$. Consequently  a 
complete set of the local three-quark nucleon operators with respect to $SU(2)_{CS}$
should contain eight independent interpolators of positive and negative
parity because 
$\bm{2}\otimes\bm{2}\otimes\bm{2}= \bm{2}_1 \oplus \bm{2}_2 \oplus \bm{4}$.

A complete set of local nucleon operators ($J=1/2,I=1/2$) with positive and negative spatial parity
with spin-zero  and isospin-zero diquark   has the following structure:
\begin{equation}
N_{\pm}^{(i)} = \epsilon_{abc}\mathcal{P}_{\pm}\Gamma_1^{(i)}u_a\{d_b^T \Gamma_2^{(i)}u_c - u_b^T \Gamma_2^{(i)}d_c\},
\label{eq:nucl_int1}
\end{equation}
with $a,b,c$ being the color index and the parity projector $\mathcal{P}_{\pm} = \frac{1}{2}\left(\mathds{1} \pm \gamma_4 \right)$. 
The matrices $\Gamma_1^{(i)}$ and $\Gamma_2^{(i)}$  ($i=1,2,3,4$)
for these four operators have the following explicit form: 
 $\mathds{1}$ and $C\gamma_5$ for $i=1$;  $\gamma_5$ and $C$  for $i=2$;
  $\I\mathds{1}$ and $C\gamma_5\gamma_4$  for $i=3$ as well as
  $\I\gamma_5$ and $C\gamma_4$  for $i=4$. 

\begin{figure}
\centering
\includegraphics[scale=0.8]{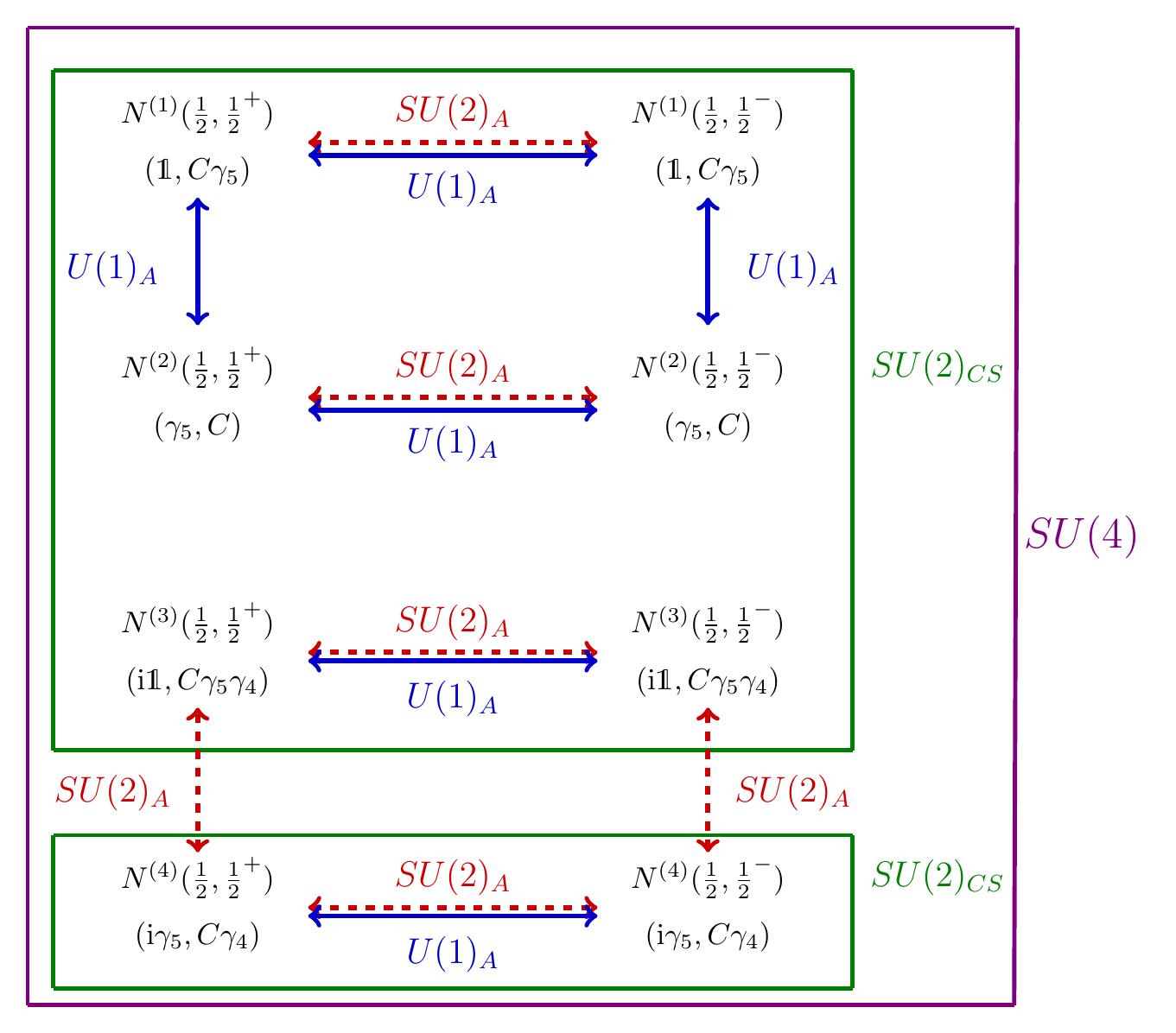}
\caption{ The nucleon fields linked by dashed red arrows are connected by $SU(2)_A$, 
by blue arrows are connected by $U(1)_A$. 
The interpolators inside  green boxes are all connected via $SU(2)_{CS}$  transformations
of the Dirac spinors (\ref{V-defsp})
and inside the violet box are connected via $SU(4)$. From Ref. \cite{cgl}.
}
\label{fig:nuclink}
\end{figure}

Notice that these three-quark fields are {\it not orthogonal}, in contrast
to the $J=0,1$ meson interpolators, discussed earlier. This feature
makes their classification with respect to $U(1)_A$,
$SU(2)_L \times SU(2)_R$, $SU(2)_{CS}$ and $SU(4)$ more complicated.
Applying the $U(1)_A$ transformation on the given operator
one obtains a linear combination of operators that are connected
by blue arrows in Fig. \ref{fig:nuclink}. The irreducible representations of
$U(1)_A$ are one-dimensional, hence the operators that are connected
by blue arrows form reducible representations of $U(1)_A$. The irreducible representations
can be obtained as linear combinations of operators linked by blue arrows.

The axial part of  $SU(2)_L \times SU(2)_R$
transforms the given operator into a superposition
of operators connected by dashed red lines. This is true for both positive and negative parity  operators $N_{\pm}^{(1)}$ and  $N_{\pm}^{(2)}$.
For the operators $N_{\pm}^{(3)}$ and  $N_{\pm}^{(4)}$ the situation is more complicated.
In this case applying the axial part  $SU(2)_L \times SU(2)_R$ one obtains linear combinations
of these operators and of $I=3/2$ $\Delta$-operators
with spin $J=1/2$. This is because both nucleon
and delta operators of the same spin form irreducible representations $(1,1/2) + (1/2,1)$
of the parity-chiral group. The latter $\Delta$-
operators are not depicted in Fig. \ref{fig:nuclink}.
The $SU(2)_{CS}$ transformations of the quark spinors  (\ref{V-defsp}) connect operators inside the green boxes.  
The $SU(4)$ transformations connect all eight operators of Fig. \ref{fig:nuclink}
along with the respective $\Delta$-partners.

A set of nucleon operators that
transform under irreducible representations of $SU(2)_{CS}$ 
consists of linear combinations of nonorthogonal operators $N_{\pm}^{(i)}$ \cite{CG,cgl}:

\begin{equation}
\begin{split}
&B_{2_1}(-1/2) =  \frac{1}{4\sqrt{2}} \gamma_-\left[-(N^{(1)}_{+}-N^{(1)}_{-}) +(N^{(2)}_{+}-N^{(2)}_{-})- \I (N^{(3)}_{+}+N^{(3)}_{-}) +\I (N^{(4)}_{+}+N^{(4)}_{-})\right]\\
&B_{2_1}(1/2) =\frac{1}{4\sqrt{2}} \gamma_-\left[(N^{(1)}_{+}+N^{(1)}_{-}) -(N^{(2)}_{+}+N^{(2)}_{-})+ \I (N^{(3)}_{+}-N^{(3)}_{-}) -\I (N^{(4)}_{+}-N^{(4)}_{-})\right]\\
&B_{2_2}(-1/2) =\frac{1}{8}\sqrt{\frac{2}{3}}\gamma_-\left[-(N^{(1)}_{+}-N^{(1)}_{-}) + (N^{(2)}_{+}-N^{(2)}_{-}) - \I (N^{(3)}_{+}+N^{(3)}_{-}) -3\I (N^{(4)}_{+}+N^{(4)}_{-})\right]\\
&B_{2_2}(1/2) =\frac{1}{8}\sqrt{\frac{2}{3}}\gamma_-\left[ (N^{(1)}_{+}+N^{(1)}_{-}) - (N^{(2)}_{+}+N^{(2)}_{-}) + \I (N^{(3)}_{+}-N^{(3)}_{-}) +3\I (N^{(4)}_{+}-N^{(4)}_{-})\right]\\
&B_4 (-3/2) = \frac{1}{4}\gamma_-\left[(N^{(1)}_{+}+N^{(1)}_{-}) + (N^{(2)}_{+}+N^{(2)}_{-})\right]\\
&B_4 (-1/2) = \frac{1}{4}\sqrt{\frac{1}{3}}\gamma_- \left[ (N^{(1)}_{+}-N^{(1)}_{-}) - (N^{(2)}_{+}-N^{(2)}_{-}) -2\I (N^{(3)}_{+}+N^{(3)}_{-})\right]\\
&B_4 (1/2) = \frac{1}{4}\sqrt{\frac{1}{3}}\gamma_- \left[ (N^{(1)}_{+}+N^{(1)}_{-}) - (N^{(2)}_{+}+N^{(2)}_{-}) -2\I (N^{(3)}_{+}-N^{(3)}_{-})\right]\\
&B_4 (3/2) = \frac{1}{4}\gamma_-\left[(N^{(1)}_{+}-N^{(1)}_{-}) + (N^{(2)}_{+}-N^{(2)}_{-})\right]\\
\end{split}
\end{equation}
\noindent
Here $\gamma_\pm=\frac{1}{2}({\mathds{1}\pm\gamma_5})$ and $B_r (\chi_z)$ is the nucleon interpolator in the irreducible
representation  of dimension $r = 2\chi + 1$ of $SU(2)_{CS}$  with  $\chi_z$ being the $z$-projection of the chiral spin $\chi$. 
Upon
the chiral spin transformation  (\ref{eq:RL})
 only those nucleon operators $B$ are connected that belong to 
 the same irreducible representation of $SU(2)_{CS}$.

 The cross-correlation matrix  is

\begin{equation}
C (t)_{\chi,\chi\prime,\chi_z, {\chi\prime}_z} = \langle 0 \vert B_r(\chi_z;t)B_{r\prime}( {\chi\prime}_z;0)^{\dagger}\vert 0 \rangle.
\end{equation}

\noindent
The $SU(2)_{CS}$ restoration requires  the cross-correlators of operators
from different
representations of $SU(2)_{CS}$ to vanish. The cross-correlators of operators  within 
a given representation of $SU(2)_{CS}$ that are diagonal in indices
 $\chi_z$ and  ${\chi\prime}_z$ must coincide while the off-diagonal must vanish.
 In this case the diagonal correlators $C(t)_{N_{\pm}^{(i)}}$ of nucleon interpolators $N_{\pm}^{(i)}$
 are given as \cite{CG}
\begin{equation}
\begin{array}{lll}
C(t)_{N_{\pm}^{(i)}} &= \frac{2}{3}C(t)_{3/2} + \frac{1}{12}C(t)_{1/2_2} +& \frac{1}{4}C(t)_{1/2_1}\\
 & & \mbox{for}\;i=1,2,3\\
\\
C(t)_{N_{\pm}^{(4)}} &= \frac{1}{4}C(t)_{1/2_2} + \frac{3}{4}C(t)_{1/2_1}.
\end{array}
\label{eq:corr_nucl}
\end{equation}
\noindent
Here $C(t)_{3/2}$ is a correlator $C (t)_{\chi,\chi\prime,\chi_z, {\chi\prime}_z}$ with $\chi=\chi\prime = 3/2$ and with any $\chi_z = {\chi\prime}_z$, and similar for $C(t)_{1/2_1}$ and $C(t)_{1/2_2}$.
We conclude that in the $SU(2)_{CS}$-symmetric regime
 all correlators $C(t)_{N_{\pm}^{(i)}}$ with $i=1,2,3$, inside the large green box in Fig. \ref{fig:nuclink} should be degenerate. Such a degeneracy
was indeed observed at zero temperature upon truncation of the near-zero modes of the Dirac
operator in Ref. \cite{D4}.

Note that the nucleon operators
 described in this section are appropriate only for study of temporal correlators.

\section{Emergence of approximate chiral spin and $SU(4)$
symmetries above chiral restoration crossover}

\subsection{Correlators and spectral function}

Symmetry properties of QCD can be studied via symmetries of
correlators calculated at a given temperature. For  meson operators $O_\Gamma(t,x,y,z)=\bar{\psi}(t,x,y,z)\Gamma \frac{\boldsymbol{\tau}}{2}\psi(t,x,y,z)$ with 
$\Gamma\in\{1,\gamma_5,\gamma_\mu,\gamma_5\gamma_\mu,\sigma_{\mu\nu},\gamma_5\sigma_{\mu\nu}\}$, the Euclidean correlation functions,
\begin{equation}
C_\Gamma(t,x,y,z)=\langle O_\Gamma(t,x,y,z)\,O_\Gamma(0,\mathbf{0})^\dagger\rangle\;,
\end{equation}
carry the full spectral information of all isovector excitations with $J=0,1$ in their
associated spectral functions $\rho_\Gamma(\omega,\bp)$\footnote{Note 
that at a finite temperature the
correlation functions   are automatically
calculated in the medium rest frame which is the preferred reference frame.}:

\begin{equation}
C_\Gamma(t,\bp) =\int_0^\infty \frac{d\omega}{2\pi}\;K(t,\omega)\rho_\Gamma(\omega,\bp),  \\
\label{eq:corr}
\end{equation}

\begin{equation}
K(t,\omega)=\frac{\cosh(\omega(t-1/2T))}{\sinh(\omega/2T)}\;.
\end{equation}

\noindent
The spatial and temporal correlators are defined as

\begin{equation}
C_\Gamma^s(z)=\sum_{x,y,t} C_\Gamma(t,x,y,z)\;,\label{eq:c_z}\\
\end{equation}

\begin{equation}
C_\Gamma^t(t)=\sum_{x,y,z}C_\Gamma(t,x,y,z)\;.
\label{eq:c_t}
\end{equation}

\noindent
They collect the spectral information projected on the $(p_x=p_y=\omega=0)$ and $(p_x=p_y=p_z=0)$  axes, respectively. In thermal equilibrium the system is isotropic and momentum distributions are the same in all spatial
directions. Consequently it is sufficient to study a propagation of the
excitation only along one direction, e.g. $z$. 

The temporal correlators reflect dynamics of the QCD Hamiltonian
since
$H$ translates states  in Euclidean time
\begin{equation}
|\psi(t+1;x,y,z)\rangle =\exp(-aH)|\psi(t;x,y,z)\rangle\;.\\
\label{Hz}
\end{equation}
The spatial correlators are connected to the dynamics of the analogous operator $H_z$ translating states in 
$z$-direction
\begin{equation}
|\psi(t; x,y,z+1)\rangle = \exp(-aH_z)|\psi(t; x,y,z)\rangle\;.
\label{Hz}
\end{equation}

While the temporal correlator is completely
determined by the spectral function of a hadron at rest 
$C_\Gamma^t(t) = C_\Gamma(t,\bp =0)$, the spatial z-direction correlator
requires integration of the spectral function over all possible spatial
momenta

\begin{equation}
C_\Gamma^s(z)=\int_{-\infty}^{+\infty} \frac{dp_z}{2\pi}e^{ip_zz}
\int_0^\infty\frac{d\omega}{\pi \omega} \rho_\Gamma(\omega,p_x =0,p_y =0,p_z)
\;.
\label{eq:c_zpr}\\
\end{equation}
Observing approximate
chiral spin symmetry both in spatial and temporal correlators is
sufficient to conclude that it is also a symmetry of the spectral function
$\rho_\Gamma(\omega,\bp)$.

Different quantum number channels at a given temperature are evaluated
with the same  effective action for QCD in  the medium. Hence  symmetries
of the effective action that describes the medium in the rest frame
at the temperature $T$ \footnote{This effective action is not known and should be eventually reconstructed; this is similar to classical electrodynamics,
where the Maxwell equations in vacuum and in  medium are different.}
can be obtained from symmetries of the correlators. Observed degeneracy patterns reflect symmetries of the non-perturbative effective action, and
hence of the thermal partition function of QCD. 

Complete information about degrees of freedom in
the thermal medium is contained in the experimentally measurable
spectral functions. Given the continuous spectral function in Minkowski space
one can directly calculate both temporal and spatial
Euclidean correlators. However, our goal is just opposite: to extract
the spectral function
from the correlators that we can calculate on the lattice.  Naively
the  spectral density of a hadron at rest $\rho_\Gamma(\omega,\bp=0)$
could be obtained from the temporal correlators (\ref{eq:c_t})
via the inverse transform to the Eq. (\ref{eq:corr}).
 However, on the lattice one calculates Euclidean correlators only
on a finite number of discrete points. Then the inverse transform
is ill-posed and extraction of spectral functions requires some
additional assumptions, e.g. input from phenomenological modelling
and the perturbation theory 
at large $\omega$ combined with statistical methods like the
maximum entropy method, etc., for reviews see e.g. \cite{hat,me,rot}.

\subsection{Spatial correlators and their symmetries}

\begin{table}
\center
\begin{tabular}{cccll}
\hline\hline
 Name        &
 Dirac structure &
 Abbreviation    &
 \multicolumn{2}{l}{
 } \\\hline
\textit{Pseudoscalar}        & $\gamma_5$                 & $PS$         & \multirow{2}{*}{$\left.\begin{aligned}\\ \end{aligned}\right] U(1)_A$} &\\
\textit{Scalar}              & $\mathds{1}$               & $S$          & &\\\hline
\textit{Axial-vector}        & $\gamma_k\gamma_5$         & $\mathbf{A}$ & \multirow{2}{*}{$\left.\begin{aligned}\\ \end{aligned}\right] SU(2)_A$}&\\
\textit{Vector}              & $\gamma_k$                 & $\mathbf{V}$ & & \\
\textit{Tensor-vector}       & $\gamma_k\gamma_3$         & $\mathbf{T}$ & \multirow{2}{*}{$\left.\begin{aligned}\\ \end{aligned}\right] U(1)_A$} &\\
\textit{Axial-tensor-vector} & $\gamma_k\gamma_3\gamma_5$ & $\mathbf{X}$ & &\\
\hline\hline
\end{tabular}
\caption{
A complete set of isovector $J=0,1$ operators in spatial correlators
and their
$U(1)_A$ and $SU(2)_L \times SU(2)_R$ transformation properties.  The
open vector index $k$ denotes the components $1,2,4$, \textit{i.e.} $x,y,t$.}
\label{t1}
\end{table}

\begin{figure}
  \centering
  \includegraphics[scale=0.5]{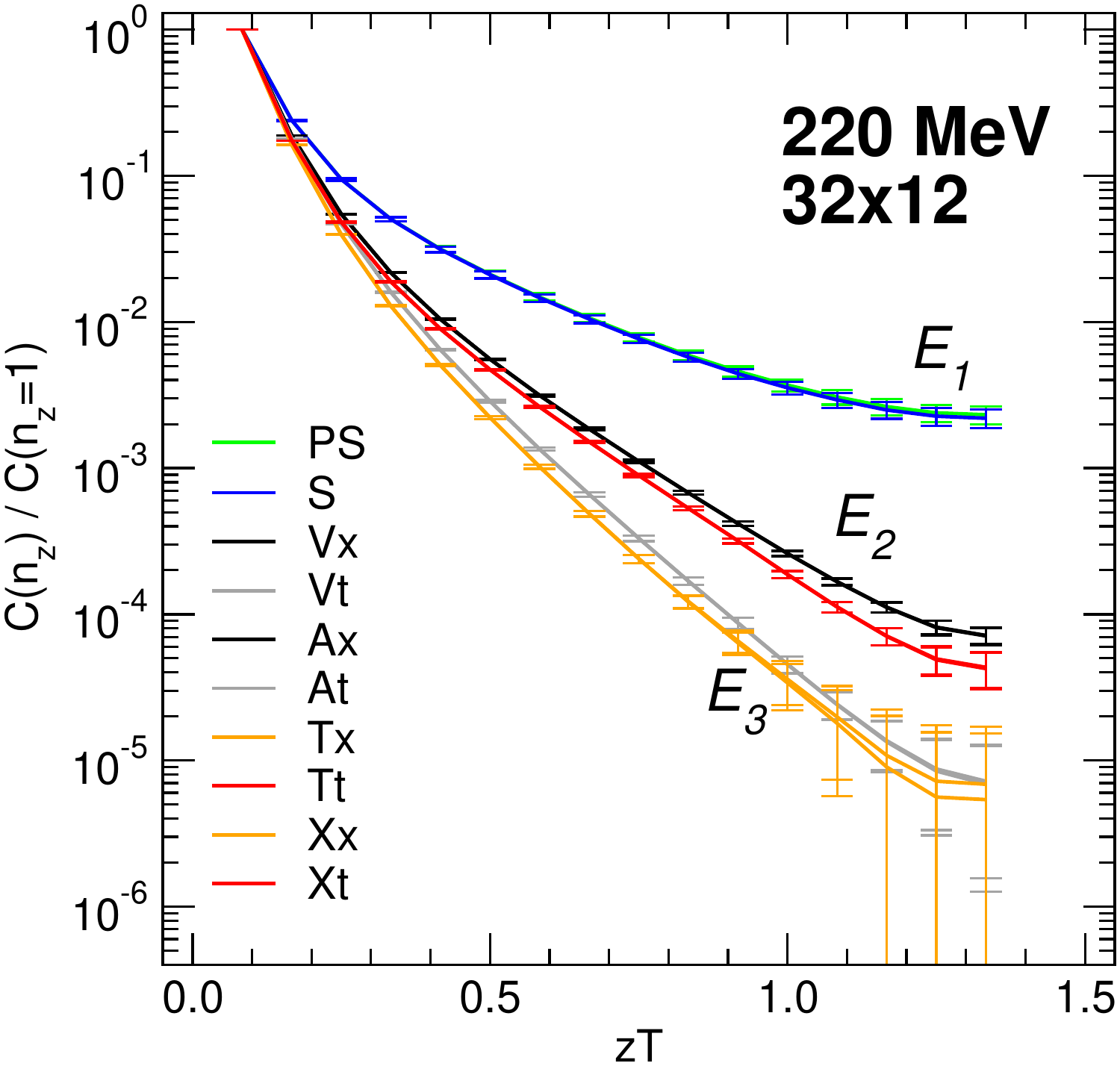} 
  \includegraphics[scale=0.5]{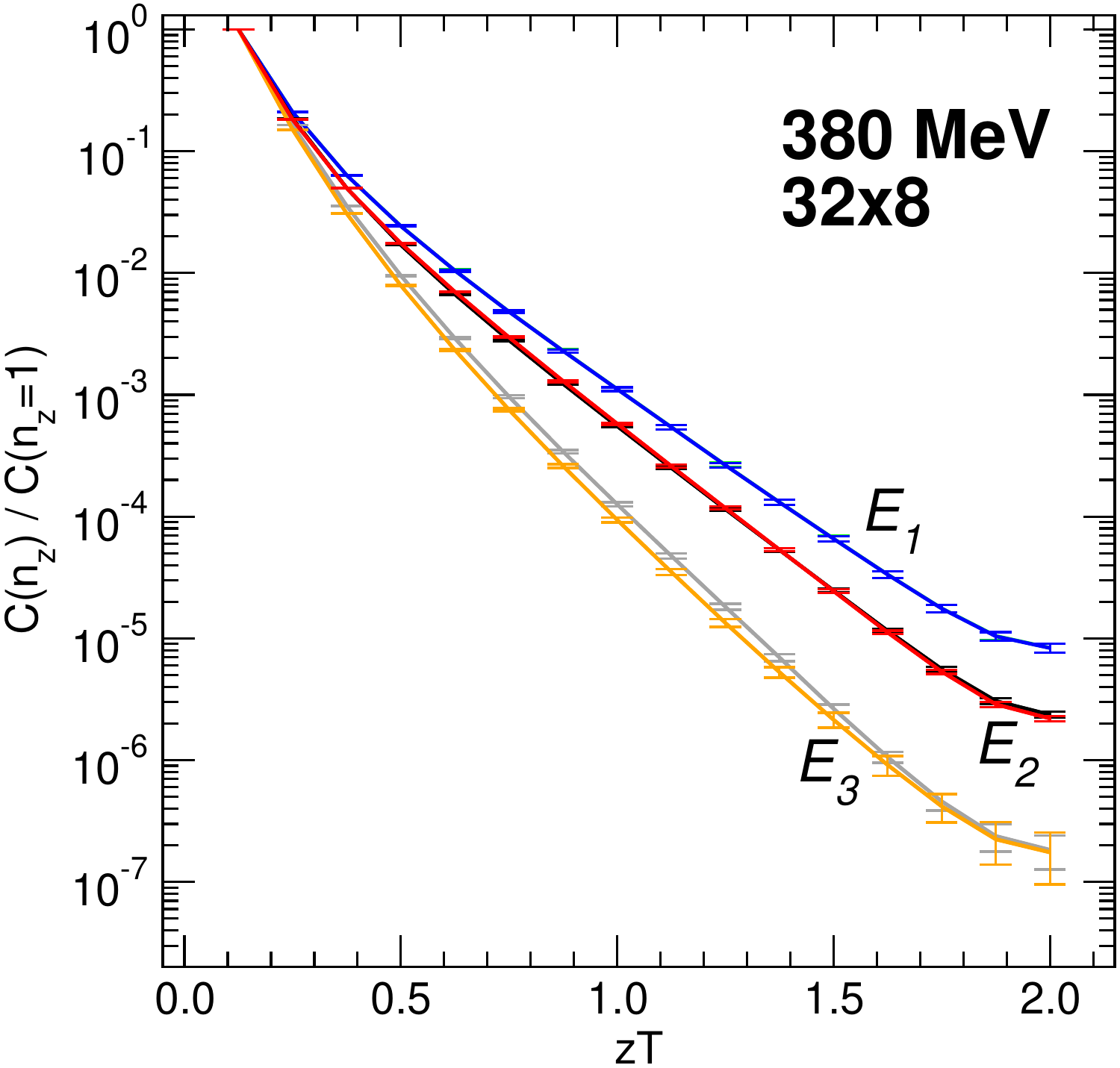} 
  \includegraphics[scale=0.5]{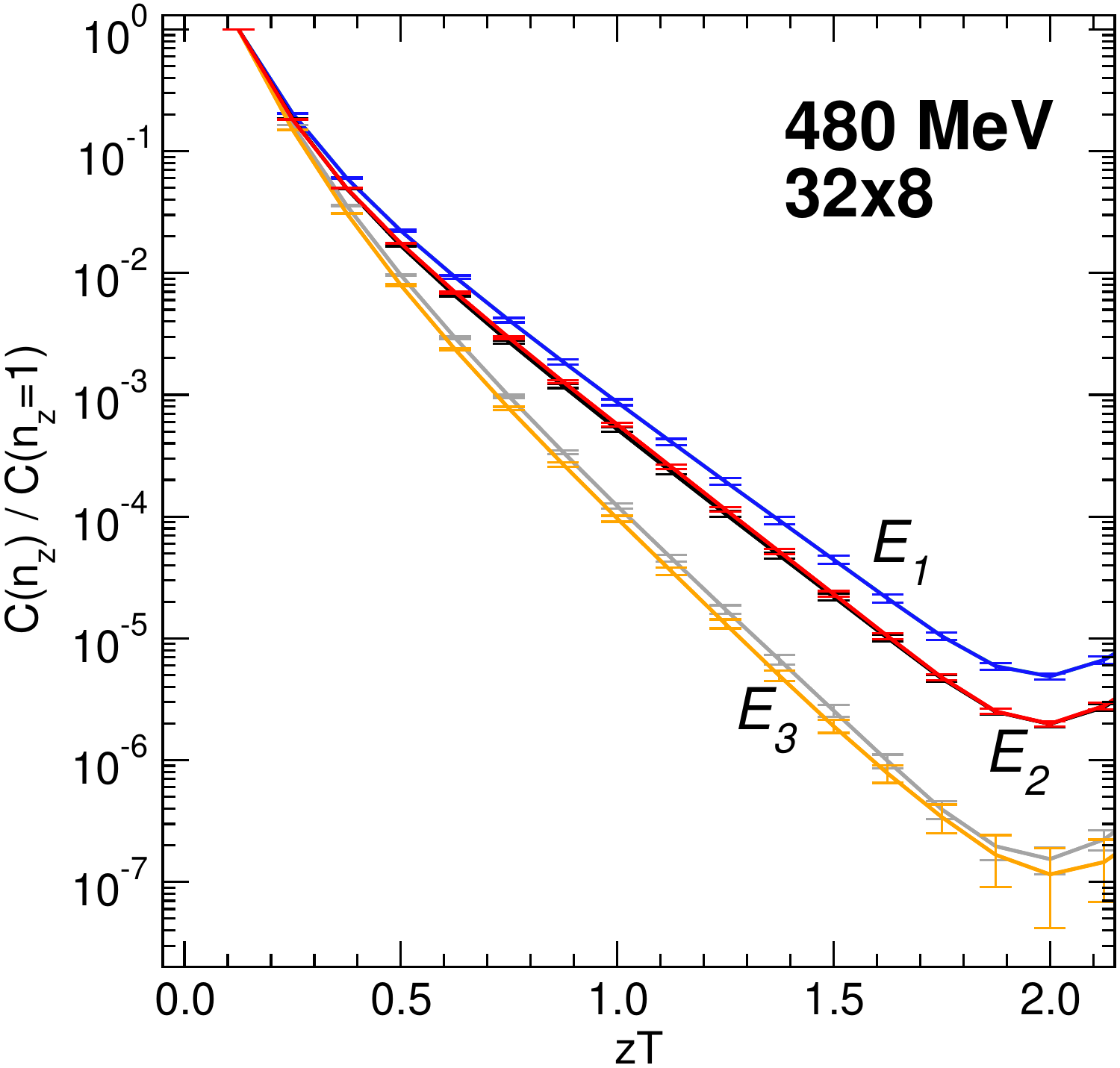}
  \includegraphics[scale=0.5]{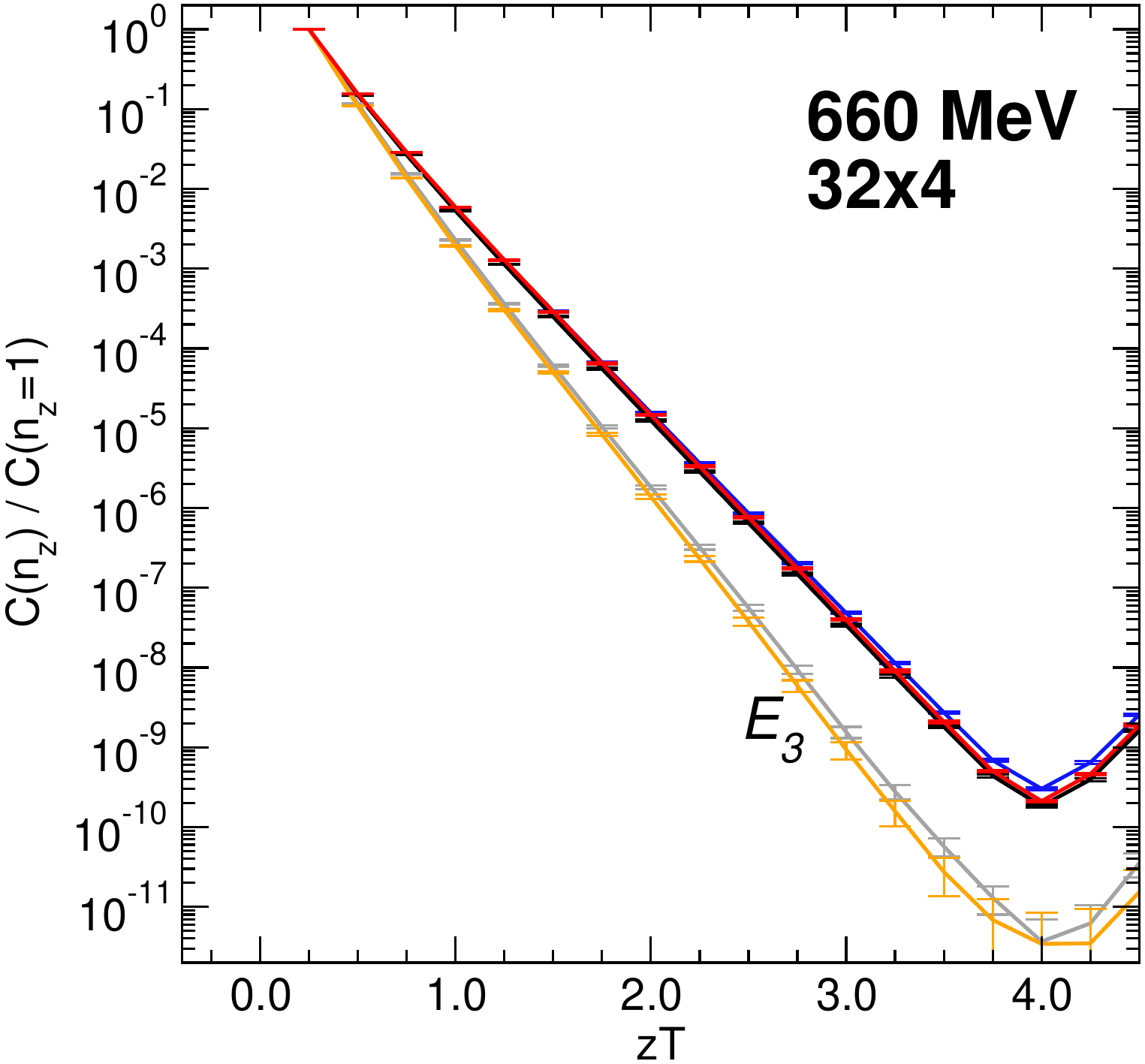}
\caption{Spatial
correlation functions of all possible isovector $J=0,1$ bilinears.
For notations of the operators and their content see Table  \ref{t1}.
  From
Ref. \cite{R2}.}
\label{spatial}
\end{figure}

A complete set of all possible isovector local $J=0,1$ operators relevant
for spatial correlators is given in Table \ref{t1}. This Table makes it also
clear how these operators transform under  $U(1)_A$ and axial part of
$SU(2)_L \times SU(2)_R$. Restoration of these symmetries requires
correlators of the corresponding operators to be degenerate.
The $SU(2)_{CS}$ and $SU(4)$
transformation properties relevant to the spatial propagators
were discussed Refs. \cite{R1,R2} and are shortly summarized below.

The chiral spin transformations (\ref{SIGCS}) with $k=1,2$ together with the
$x \leftrightarrow y$ symmetry generate the following multiplets:
\begin{align}
(V_x,V_y); \; (A_x, A_y, T_t, X_t) \; ,  \label{equ:s2a} \\
(V_t); \;   (A_t, T_x, T_y, X_x, X_y) \; . \label{equ:s2b} 
\end{align}

\noindent
Considering $SU(4)$ one obtains larger 
multiplets of the isovector operators:
\begin{align}
(V_x, V_y, A_x, A_y, T_t, X_t) \; , \label{equ:ss2a} \\
(V_t, A_t, T_x, T_y, X_x, X_y) \; . \label{equ:ss2b} 
\end{align}
Complete  $SU(4)$ multiplets contain also the isoscalar partners
of the operators $A_x, A_y, T_t, X_t$ in Eq. (\ref{equ:ss2a})
and isoscalar partners of the $A_t, T_x, T_y, X_x, X_y$ operators
in Eq. (\ref{equ:ss2b}).

In Fig. \ref{spatial} we show spatial correlators  (\ref{eq:c_z})
evaluated
at different temperatures with chirally symmetric domain wall
Dirac operator at physical quark masses using the $N_F=2$ JLQCD ensembles
\cite{R2}. Here a complete set of all possible isovector
local $J=0,1$ operators has been used.  We see a distinct multiplet structure of the correlators.
This multiplet structure  reflects symmetry properties of the
effective action at the given temperature.

The multiplet $E_1$ consists of isovector scalar (S) and pseudoscalar (PS)
correlators. The degeneracy of S and PS correlators evidences restored
$U(1)_A$ symmetry. If there is still a tiny breaking of $U(1)_A$ it should be
too small to be seen in the present data.

The multiplet $E_2$ contains four approximately
degenerate correlators obtained with $V_x, A_x, T_t, X_t$ 
 $J=1$ isovector operators. The $V_x$  and $A_x$  operators 
 are connected by the
 axial part of the $SU(2)_L \times SU(2)_R$ transformation and 
their degeneracy is a signal of restored
$SU(2)_L \times SU(2)_R$ symmetry.
The $T_t$  and
 $X_t$  operators are connected by the
$U(1)_A$ transformation and a degeneracy of the corresponding correlators is
required by the restored  $U(1)_A$ symmetry. The operators $(A_x, T_t, X_t)$
form a triplet of the $SU(2)_{CS}$ group. An approximate degeneracy
of the correlators indicates emerged approximate $SU(2)_{CS}$ symmetry.
All four operators 
$(V_x, A_x, T_t, X_t)$ are connected by the $SU(4)$ transformation 
and a degeneracy of the corresponding correlators shows
emergent  approximate $SU(4)$ symmetry.

The $E_3$ multiplet consists of four approximately
degenerate correlators obtained with $V_t, A_t, T_x, X_x$ operators.
Notice that the $V_t, A_t$ operators represent the charge and axial
charge densities, respectively. These operators do not create physical
states. The current conservation connects the $T_x, X_x$ to $V_t, A_t$
so the former operators are not independent from the latter.
If the $V_t, A_t, T_x, X_x$ correlators are normalized, as it is in 
Fig. \ref{spatial}, then in the case of noninteracting quarks they
must be identical \cite{R2}. Consequently a degeneracy of the normalized
$V_t, A_t, T_x, X_x$ correlators
is consistent 
with both the $SU(2)_L \times SU(2)_R \times U(1)_A$ symmetry alone
and with the $SU(4)$ symmetry. This
is precisely the reason why the $E_3$ multiplet persists at all temperatures.
So it cannot be used as an indicator
of emerged chiral spin symmetry and of its $SU(4)$ extension. 

We observe approximate emerged $SU(2)_{CS}$ and $SU(4)$ symmetry up
to temperatures of about $\sim 500$ MeV. At higher temperatures
two different distinct multiplets $E_1$ and $E_2$ disappear. This
happens because the full QCD correlators approach at high temperatures
correlators of the free quark gas, as will become evident below.

\begin{figure}
  \centering
  \includegraphics[scale=0.5]{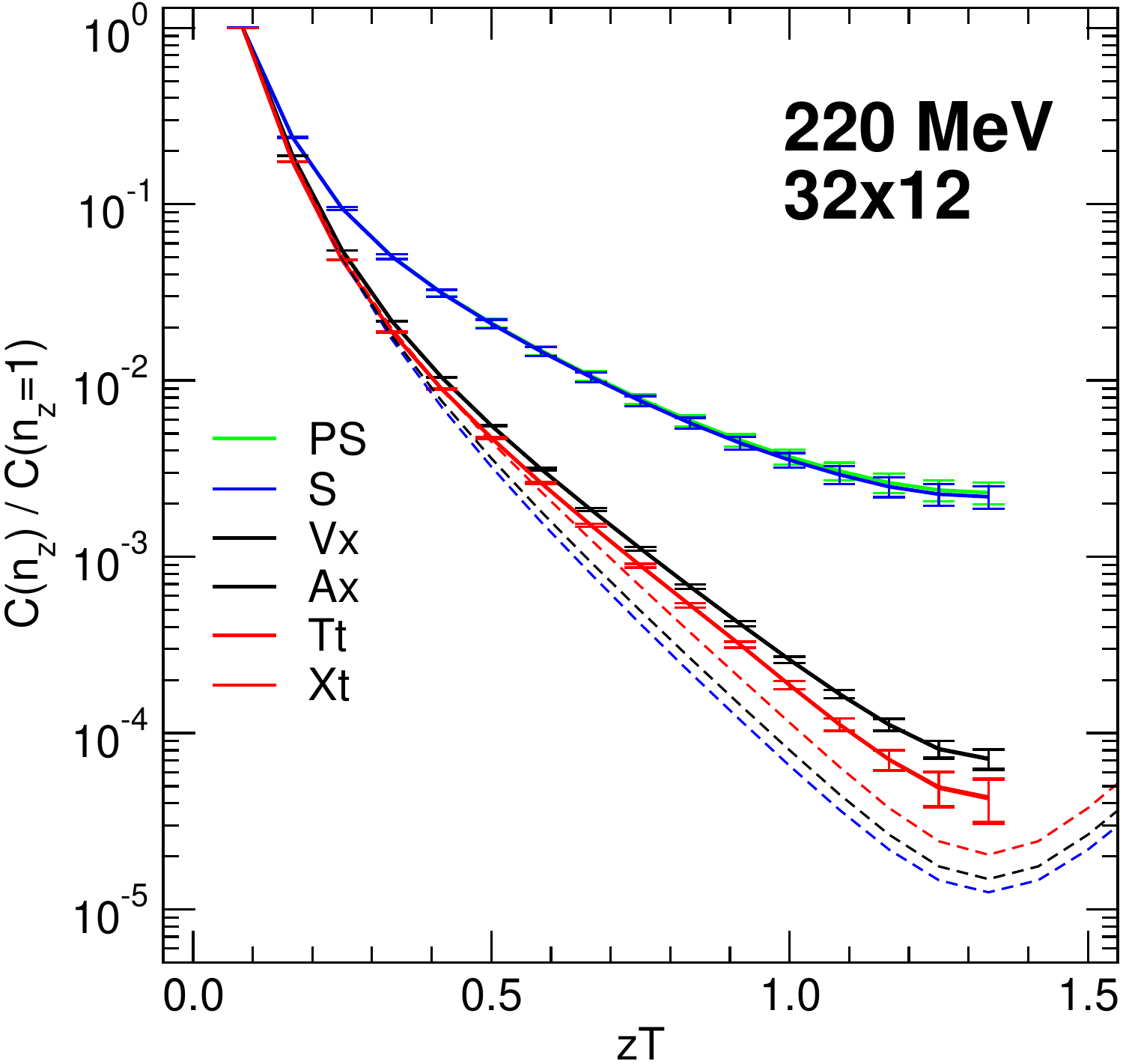} 
  \includegraphics[scale=0.5]{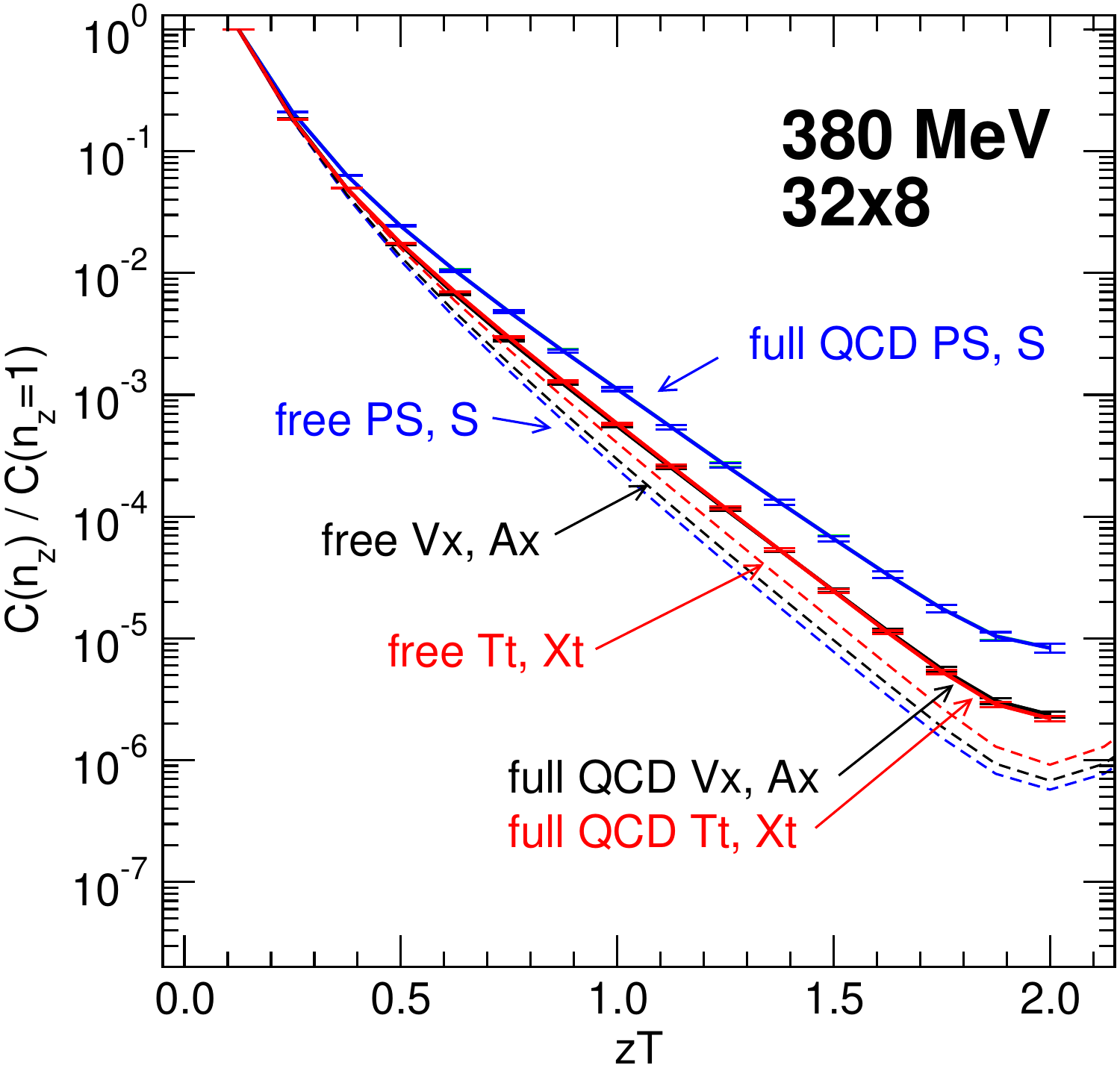} 
  \includegraphics[scale=0.5]{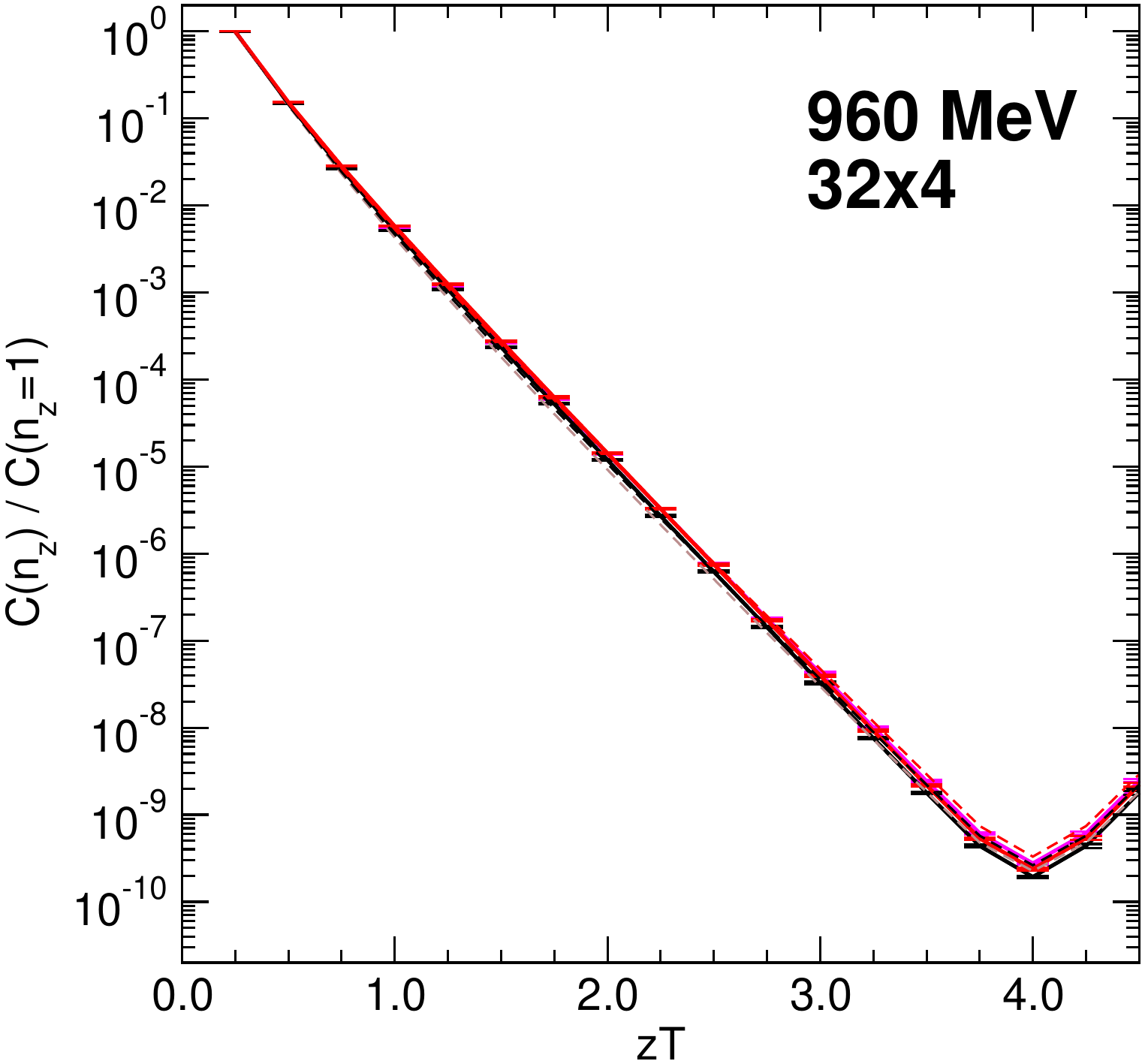}
\caption{
Correlation functions of the bilinears $PS, S, V_x, A_x, T_t, X_t$. The
solid curves represent full QCD calculation and the dashed lines
are correlators calculated with free noninteracting quarks. From
Ref. \cite{R2}.}
\label{fig:e2_withfreedata}
\end{figure}

In Fig. \ref{fig:e2_withfreedata}
we compare correlators from the
$E_1$ and $E_2$ multiplets evaluated in full QCD
with the corresponding correlators obtained with
a free quark gas.
 The full
QCD correlators are given by the solid lines while the
correlators calculated with noninteracting quarks are described
by the dashed curves.
Note that the correlators calculated with noninteracting
quarks reflect physics at a very high temperature and only chiral
$SU(2)_L \times SU(2)_R$ and $U(1)_A$ symmetries are present in this case.
No $SU(2)_{CS}$ and $SU(4)$ symmetries exist for free quarks.

The quark gluon plasma, which is a system of (quasi) free partons
is characterized by chiral symmetries, i.e. symmetries of the Dirac
equation. The presence of approximate $SU(2)_{CS}$ and $SU(4)$ symmetries 
below 500 MeV tells that the degrees of freedom should be
the quark-antiqiark systems with
chirally symmetric quarks bound into color singlet objects by a
confining electric field.

A dramatic difference between the S and PS correlators in full
QCD and in free quark gas is obvious. This immediately tells
us that there must be some color singlet resonances with pion
and sigma quantum numbers below 500 MeV. This implies that
the medium below 500 MeV is by far not a system
 are of quasi-free partons. This issue will be discussed below in
 chapter 8. 
 
At the highest temperature of this study, $T \sim 960$ MeV, the situation has changed significantly:
All full QCD correlators are very close to the corresponding free
correlators.
Hence at $T \sim 960$ MeV we have reached the region where only chiral
$U(1)_A$ and $SU(2)_L \times SU(2)_R$ symmetries exist and the near coincidence
with the free correlators suggests a gas of quasi-free quarks. Notice that
this near coincidence is a consequence of the log scale used in Fig. \ref{fig:e2_withfreedata}.
The QCD correlators are not identical to the free quark gas correlators, which
is  well seen in more detailed plots in Ref. \cite{R2}.

\subsection{Temporal correlators and their symmetries}

\begin{figure}
  \centering
  \includegraphics[scale=0.6]{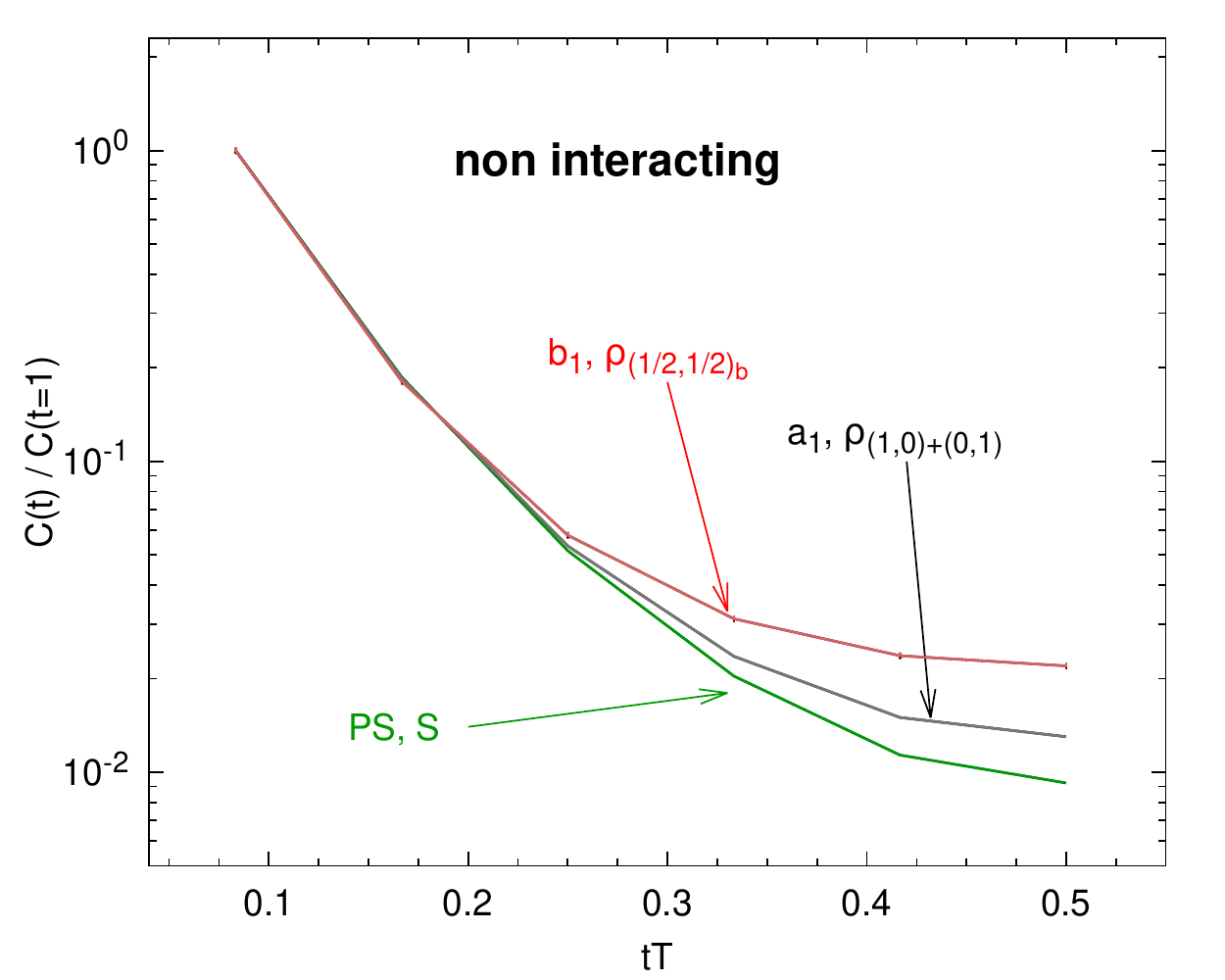} 
  \includegraphics[scale=0.6]{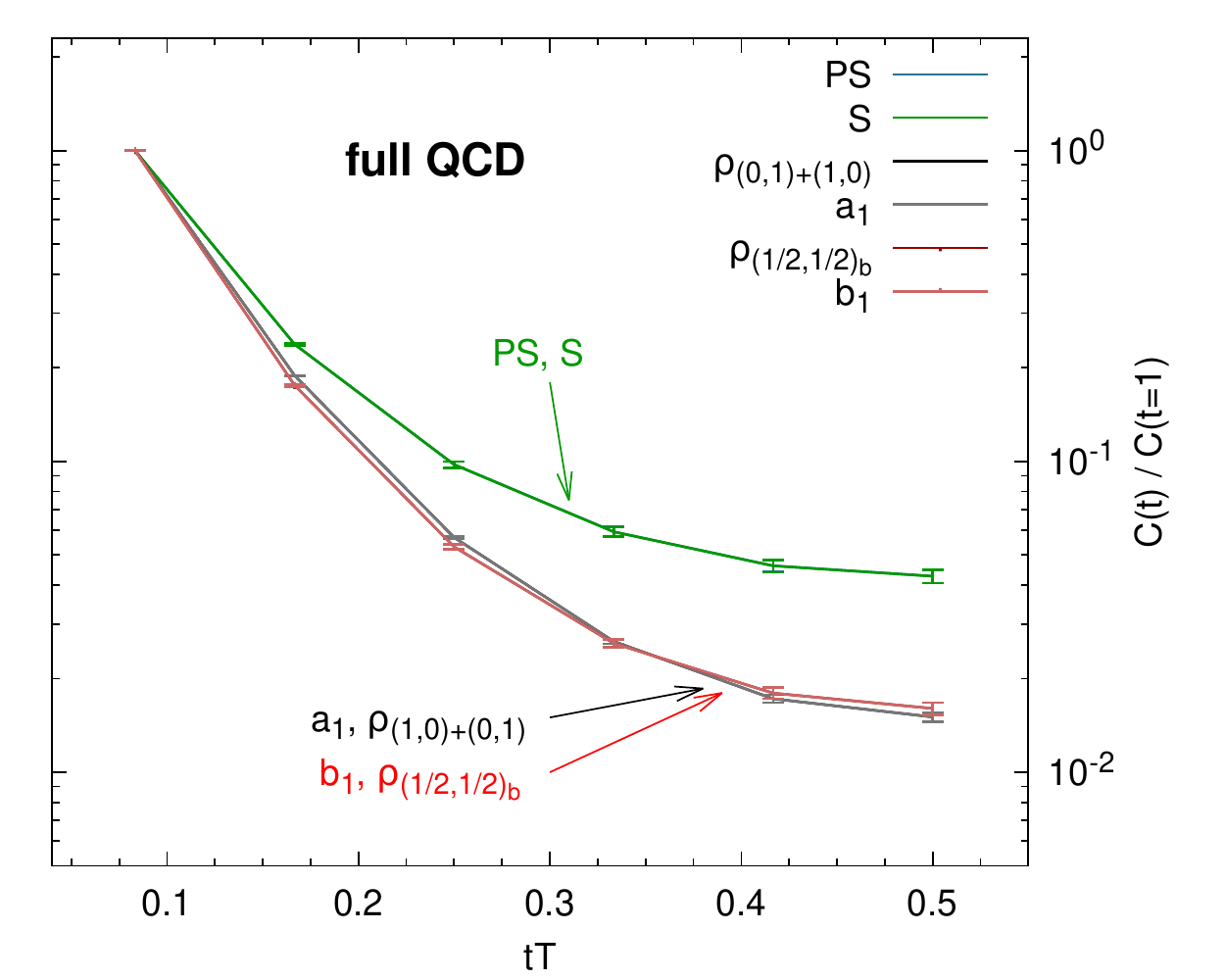} 
\caption{ Temporal correlation functions for $12 \times 48^3$
lattices. The l.h.s. shows correlators calculated with free
noninteracting quarks with manifest $U(1)_A$  and $SU(2)_L \times SU(2)_R$
symmetries. The r.h.s. presents full QCD results at a temperature 220 MeV,
which shows multiplets of all  $U(1)_A$, $SU(2)_L \times SU(2)_R$, $SU(2)_{CS}$  and $SU(4)$ groups. From Ref. \cite{R3}.
}
\label{tcorr}
\end{figure}

On the right side of Fig.~\ref{tcorr} we show temporal correlators
(\ref{eq:c_t}) at $T=220$ MeV calculated with the domain wall Dirac operator
at physical quark masses with $N_F=2$ JLQCD ensembles
\cite{R3}.
Transformation properties of the local  $J=1$  quark-antiquark bilinears 
$\mathcal{O}_\Gamma(x,y,z,t)$
 with respect to $U(1)_A$, $SU(2)_L \times SU(2)_R$, $SU(2)_{CS}$  and  
 $SU(4)$, relevant for temporal correlators,  are given in  
Fig.~\ref{algebra}.
Emergence of the respective symmetries is signalled by  degeneracy of
the correlators (\ref{eq:c_t}) calculated with operators
that are connected by the corresponding transformations.

On the l.h.s of Fig.~\ref{tcorr} we demonstrate correlators
calculated with noninteracting quarks on the same lattice.
They represent a QGP at a very high temperature where due to asymptotic freedom
the quark-gluon interaction can be neglected.
Dynamics of free quarks are governed by the Dirac
equation and only $U(1)_A$ and $SU(2)_L \times SU(2)_R$ chiral symmetries
exist. A qualitative difference between the pattern on the l.h.s.   and the pattern on the r.h.s of Fig.~\ref{tcorr}
is obvious. In the latter case we clearly see approximate
$SU(2)_{CS}$ and $SU(4)$ symmetries. 
$SU(2)_{CS}$ and $SU(4)$ symmetries of the spatial and temporal
correlators imply the same 
symmetries of spectral densities and of the thermal partition function.

\subsection{Conclusions to symmetry studies}

There are a few most important conclusions from the symmetry
studies of the meson correlators.

The QCD effective action and thermal partition function above 
the chiral crossover have not only chiral symmetries
but are approximately symmetric with respect to
$SU(2)_{CS}$ chiral spin group and its flavor extension $SU(4)$
for $N_F=2$. This is true in the medium rest frame
which is  the preferred frame. (At a nonzero temperature there is no
Lorentz invariance in the medium.)
These groups are not symmetries
of the Dirac Lagrangian. This implies that the medium is not
a quark gluon plasma which is a system of weakly interacting
partons and where only chiral symmetries exist.

The chiral spin group  is a symmetry of the chromoelectric
part of the QCD Lagrangian. The approximate chiral spin symmetry
 can emerge  only
when the quark-electric interaction strongly dominates over
the quark-magnetic interaction and over the quark kinetic
term. This symmetry is characteristic of  quark-antiquark
systems with chirally symmetric quarks bound by the chromoelectric
field (presumably by a chromoelectric flux tube, that is why
this regime was dubbed a stringy fluid).
The emergent $SU(2)_{CS}$ and $SU(4)$ symmetries seen at $T_{ch} - 3 T_{ch}$
suggest that the physical degrees of freedom at these temperatures
are chirally symmetric quarks bound into color singlets by the
chromoelectric field.

The stringy fluid regime arises above $T_{ch}$ and extends to
approximately $3 T_{ch}$, as illustrated in Fig. \ref{fig:sketch}.
Above these temperatures the chiral spin symmetry smoothly
disappears because the confining electric field gets screened
and one observes a smooth transition to a quark gluon plasma.

\subsection{Is $U(1)_A$ restored simultaneously with $SU(2)_L \times SU(2)_R$ in hot QCD?}

The idea that above the chiral phase transition the $U(1)_A$ symmetry is still
broken is an old one \cite{GPY}. It was suggested that the instanton gas picture  could be 
valid in the chirally
restored and deconfined phase. It would still induce the $U(1)_A$ breaking.  
The quark condensate and the $U(1)_A$ breaking are differently sensitive to the
 near-zero modes of the Dirac operator:  
\begin{equation}
 \langle \bar{\psi}\psi\rangle=  -\lim_{m\rightarrow 0} \int_0^\infty d\lambda
 \rho(\lambda,m) \frac{2m}{m^2 + \lambda^2}\;.
\label{qcon}\\
\end{equation} 

 \begin{equation}
  \int d^4x <O_\pi(x)O_\pi^\dagger(0)> - \int d^4x <O_{a0}(x)O^\dagger_{a0}(0)> = \lim_{m\rightarrow 0}
     \int_0^\infty d\lambda
 \rho(\lambda,m) \frac{4m^2}{(m^2 + \lambda^2)^2}\;.
\label{qcon}\\
\end{equation}
 Here $\rho(\lambda,m)$ is a density of modes of the Dirac operator with the quark mass $m$ (see (\ref{banks})).
 It is possible to reconcile the vanishing quark condensate and a non vanishing difference
 of $\pi$ and $a_0$ correlators by assuming the  non analytical form of the Dirac spectral
 density,
 $\rho(\lambda,m)\sim m^2 \delta(\lambda)$. Here the $\delta(\lambda)$ reflects
 the near-zero modes arising from the well isolated instantons.
 Consequently, it is in principle possible that in the chirally symmetric regime
 with vanishing quark condensate the $U(1)_A$ susceptibility is not zero.
 
However, the existence of a confining electric field above $T_{ch}$, discussed in this review,
rules out the instanton gas picture at temperatures below few times $T_{ch}$,
because a dilute instanton gas cannot provide
a confining electric field. Still, a possibility of a presence
of rare topological fluctuations on top of a confining field is not excluded.

The modern topic and argument  was initiated by Cohen \cite{Cohen1} who insisted that in QCD in the chiral limit
restoration  of $SU(2)_L \times SU(2)_R$ at a critical temperature  requires actually
 an effective restoration of $U(2)_L \times U(2)_R$, i.e., an effective restoration
 of anomalously broken $U(1)_A$ symmetry. This $U(1)_A$ effective restoration means
 that at least two-point correlation functions of operators connected by the $U(1)_A$
 transformation  must be identical.\footnote{The degree of $U(1)_A$
 symmetry breaking at the chiral restoration point may have physical consequences. 
 Analysis of Ref. \cite{PW} suggests that
 if this
 breaking is large in QCD with two massless flavors, then the transition might be second
 order. In the case of simultaneous restoration of both $SU(2)_L \times SU(2)_R$ and
 $U(1)_A$ it should be of first order. The same analysis tells, however, that the phase transition
 with $N_F=3$ should be of first order, while recent lattice results with staggered
 fermions demonstrate second order phase transition at $N_F = 3,4,5,6$ \cite{CF}.}
 Cohen's argument was challenged in Ref. \cite{LH}:
 Exact zero modes arising from the topological configurations with nonzero topological
 charge would violate
 the identity of the correlators connected by $U(1)_A$. However, it is known that
 in the thermodynamic limit $V \rightarrow \infty$ the contribution of the exact zero
 modes vanishes, see, e.g.,  Ref. \cite{AFT}.
 Cohen further suggested that a finite gap in the Dirac spectrum
 might emerge above the chiral phase transition. 
 Such a gap  would automatically induce the effective
 restoration of $U(1)_A$ in meson and baryon two-point functions \cite{lang,cgl}.
 
 A behavior of the near-zero modes and the question whether a finite gap in the Dirac spectrum 
 arises or not is a delicate issue
 and can be answered only in nonperturbative lattice calculation. However,
 it is  a rather complicated task that  could not be completely accomplished
 so far, because it requires  a Dirac operator with perfect chiral properties,
  a very large lattice volume, and  a very small quark mass.

Existing lattice results related to this question could be grouped into
three categories: (i) hybrid calculations that use   staggered fermions for the vacuum 
configurations while the overlap Dirac operator for valence quarks \cite{Dick,Kac}, (ii) the same chirally
symmetric Dirac operator (either domain wall or overlap) is employed for both sea and valence quarks \cite{bhot,JLQCD1,JLQCD3,JLQCD4,JLQCD5} and (iii) staggered sea and valence fermions \cite{Ding}.
One should also always keep in mind that the staggered fermions rely
on the rooting procedure, which causes questions about its validity at small quark masses.

In works of categories (i)
and (iii) a big peak near $\lambda=0$ is seen in the Dirac eigenvalue spectrum at temperatures
significantly above $T_{ch}$, which implies a serious violation of $U(1)_A$\footnote{Such a violation, if large, should
be seen in spatial and temporal correlators. However, it is not observed, as discussed in the present chapter.}, such a peak is not observed in papers from the category (ii). 
A search of a possible gap in the Dirac spectrum above $T_{ch}$ 
  was performed by the JLQCD collaboration in Refs. \cite{JLQCD1,JLQCD3,JLQCD4,JLQCD5}.

In Ref. \cite{JLQCD1} the $N_F=2$ QCD with the overlap Dirac operator in the trivial topological
 sector $Q=0$ was studied.  The Dirac spectrum in the quenched
 case (overlap valence quark Dirac operator on pure glue Q=0 vacuum configurations) showed
 a sharp peak at the smallest eigenvalues $\lambda$,  in agreement with papers from the
 category (i). The full QCD calculation  demonstrates, however, absence of a peak and
 even  a gap opens at $T > 200$ MeV.  This suggests that the peak could be a quenching
 lattice artifact.
 
\begin{figure}
  \centering
  \includegraphics[scale=1]{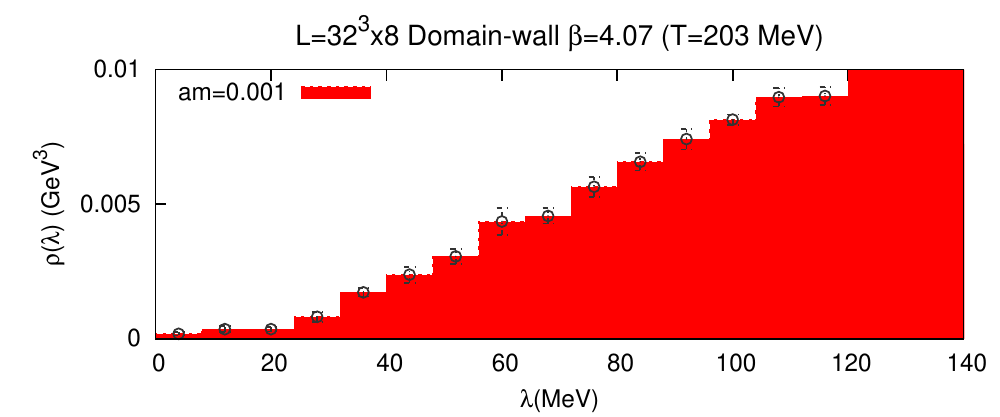} 
  \includegraphics[scale=1]{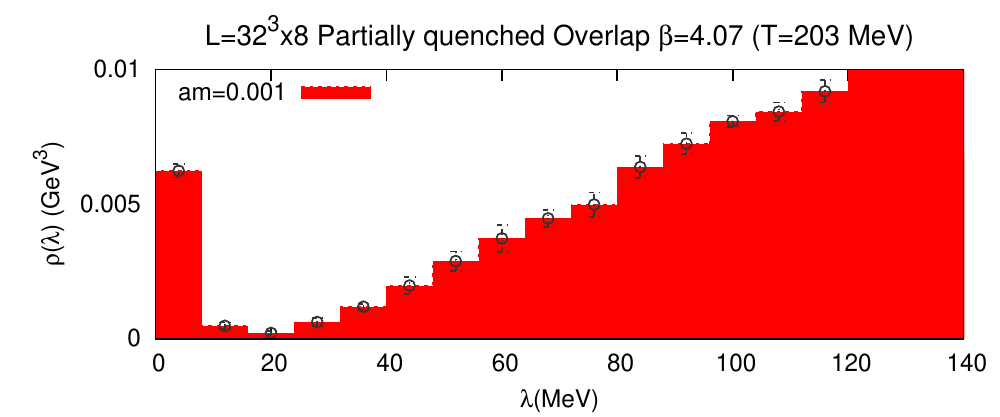} 
  \includegraphics[scale=1]{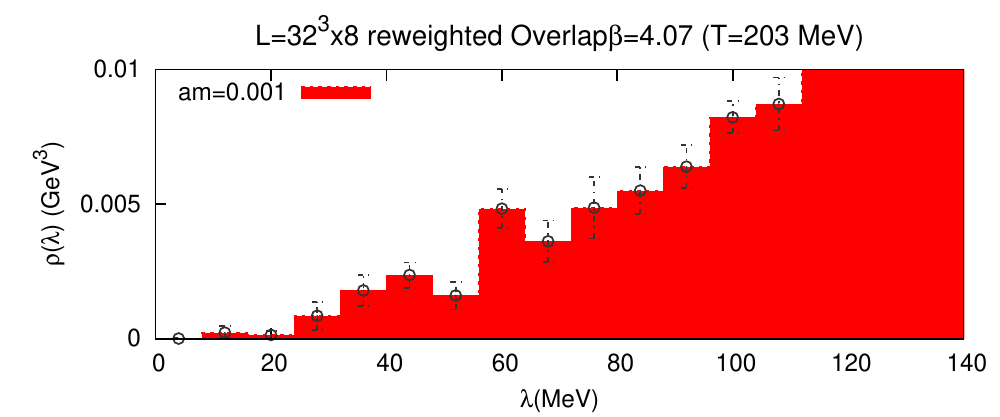}
\caption{ Eigenvalue spectrum of the domain wall (top panel),
partially quenched valence overlap with domain wall sea quarks (middle panel)
and reweighted overlap Dirac operator for both valence and sea quarks (bottom panel) with the near physical quark masses.
 From Ref. \cite{JLQCD3}.
}
\label{sp}
\end{figure}

 This issue was further investigated in Refs. \cite{JLQCD3,JLQCD4,JLQCD5} with $N_F=2$. In Fig. \ref{sp}
 we show a typical result of these studies at $T=203$ MeV with physical degenerate $u$ and $d$ quark masses.
 In the top panel a Dirac spectrum with the
 domain wall Dirac operator for both valence and sea
 quarks is shown. No peak at small values of $\lambda$ is visible. The domain wall operator
 still has small residual effects of violation of the Ginsparg-Wilson relation, i.e. of exact chiral symmetry. The authors suggest that a small non vanishing density of Dirac eigenvalues near zero in the lowest bin could
 be connected with these small residual chiral symmetry breaking effects. To control the latter
 issue they reweight the domain wall eigenmodes
 with the overlap eigenmodes (bottom panel).
  Here the density of eigenmodes vanishes even with nonzero quark masses and
  a gap near zero is seen, as suggested in Ref. \cite{AFT}.
In the middle panel of Fig. \ref{sp} a partially quenched result is shown,
where the valence overlap operator is combined with the domain wall sea quarks.
A sharp peak is found in the lowest bin, in agreement with studies from
the category (i). This result suggests that even a small partial
quenching could induce spurious effects in the near zero Dirac modes.

While these results on restoration of $U(1)_A$  above
the chiral crossover are interesting and convincing, they do not prove yet a simultaneous
restoration of both $U(1)_A$ and $SU(2)_L \times SU(2)_R$ symmetries
in the chiral limit. 
The precise temperature where $U(1)_A$ effectively restores, is not yet conclusively
determined. We cannot exclude that it may  be a bit larger than the $SU(2)_L \times SU(2)_R$
restoration temperature.

\section{Screening masses and the equation of state}

As we have discussed in Introduction, in QCD with light quarks
there is no obvious definition of "deconfinement" and of the
corresponding order parameter. The only sensible question is
about effective degrees of freedom that drive the hot QCD matter
at a given temperature. Emergence of approximate chiral spin
symmetry rules out weakly interacting (quasi)quarks and (quasi)gluons as the
only symmetry of perturbation theory is chiral symmetry, that is
the symmetry of the Dirac equation. The chiral spin symmetry points
out a true nonperturbative regime where dynamics is dominated by
the nonperturbative chromoelectric field. Since only the color-singlet
states can survive the gauge averaging (and hence propagate) this means that these color-singlet states are chirally symmetric quark-antiquark systems bound
by the chromoelectric field. 

Still another observables that would be consistent with the above
picture and that would discriminate degrees of freedom are highly welcome.
Screening masses of spatial correlators are among  such observables.
Results of the present Section are based on Ref. \cite{GPP}.

The screening masses are defined as asymptotic exponential slope
of spatial correlators (\ref{eq:c_z}) at $z \rightarrow \infty$ \cite{tar}:

\begin{equation}
C_\Gamma^s(z) \rightarrow const \cdot e^{-m_{scr} z}.
\label{mscr}
\end{equation}
It is very well seen from Fig. (\ref{spatial}) that indeed
the spatial correlators are driven at large $z$  by the
exponential asymptotic. This asymptotic determines the ground
state of a "Hamiltonian" $H_z$ that acts on a Hilbert space defined
over the $x,y,t$ Euclidean coordinates. $H_z$ generates
translations in  $z$-direction (\ref{Hz}). 
If this asymptotic pure exponential,
then this ground state corresponds to a bound state of $H_z$. 
A spectrum of $H_z$ is sensitive to the temperature as it is sensitive
to the compactified Euclidean time direction $t$. The boundary
conditions along the finite time direction, $ T^{-1} = a N_t$,
are fixed: periodic for gauge field
and anti-periodic for fermionic fields. The thermodynamic limit is defined
as $N_{x,y,z} \rightarrow \infty$ at the given temperature. Real lattice
calculations are done on a finite lattice, hence either periodic or
anti-periodic boundary condition should be imposed along the
spatial axes $x,y,z$. In the thermodynamic limit results will not depend
on a particular choice of spatial boundary conditions. In the limit $T=0$
the spectrum of $H_z$ is identical to that of the Hamiltonian $H$ which provides translations
in Euclidean time direction. In the opposite limit $T \rightarrow \infty$
a dimensional reduction to a 3d theory takes place  and the spectrum
of $H_z$ reduces to the spectrum  of 3d QCD. For either unstable or multiparticle states the exponential in (\ref{mscr}) gets multiplied
by the inverse power law factors. 

On a Euclidean lattice the thermal partition function can
be represented in two equivalent ways, either via the spectrum
of $H$, which is $T$-independent, or via the spectrum of $H_z$,
which is explicitly $T$-dependent

\begin{eqnarray}
e^{pV/T}=Z&=&{\rm Tr}(e^{-aHN_t})\nonumber\\
&=&{\rm Tr}(e^{-aH_zN_z})=\sum_{i}e^{-{E^{z}}_i a N_z}\;,
\label{eq:ham}
\end{eqnarray} 
where ${E^{z}}_i$ is a spectrum of $H_z$. The full spectrum
of $H_z$ defines the partition function and is directly related
to the equation of state. Consequently the
screening masses that represent the ground states of $H_z$
are also directly related to the equation of state.

Information about effective degrees of freedom at any temperature
is encoded in the thermal partition function (\ref{eq:ham}).
If the thermal partition function and the equation of state
are described by the parton dynamics one naturally speaks of the quark-gluon 
plasma. For a thermal equilibrium system,
screening masses are accessible by perturbative and non-perturbative
(lattice)  calculations. If the non-perturbative  lattice results
for screening masses
are well described with the perturbative parton language,  one then
concludes that effective degrees  of freedom in the system
are quarks and gluons. 

\begin{figure}
  \centering
  \includegraphics[scale=0.5]{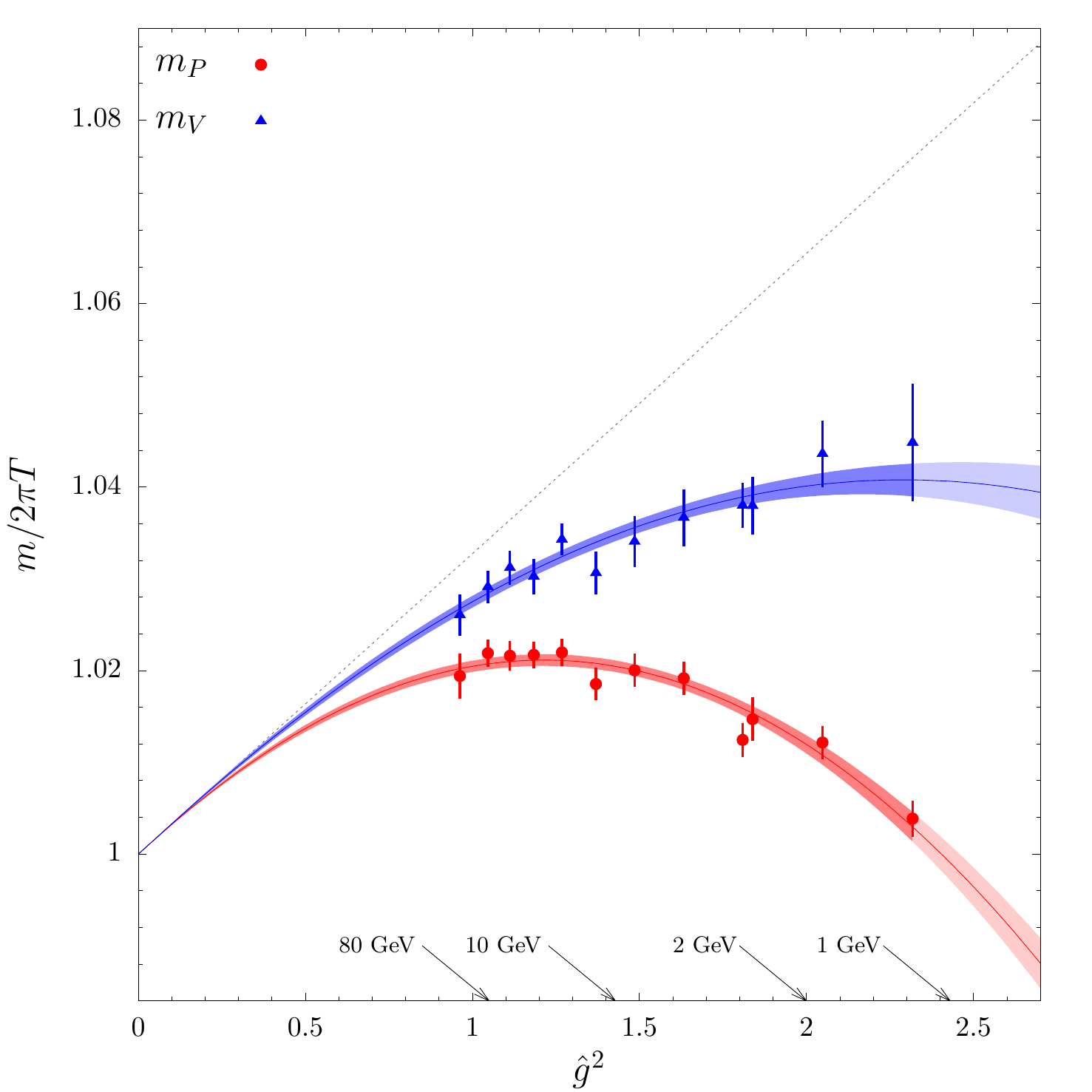}  
\caption{ Temperature dependence of the pseudoscalar and vector
screening masses at large temperatures. The shadow bands represent a fit according
to Eq. \ref{eq:psv}.  From Ref. \cite{brida}.
}
\label{tdep}
\end{figure}

Very recently the pseudo-scalar and vector screening masses
have been calculated on the lattice at high temperatures
$T= 1 - 160$ GeV  which are shown in Fig. \ref{tdep} \cite{brida}.
Over two orders of magnitude in temperature the lattice
data are well parameterized by

\begin{eqnarray}
\frac{m_{PS}}{2\pi T}&=&1+p_2 \,\hat{g}^2(T) + p_3 \, \hat{g}^3(T)+p_4 \,\hat{g}^4(T)\;,\nonumber \\
\frac{m_{V}}{2\pi T}&=&\frac{m_{PS}}{2\pi T} + s_4\, \hat{g}^4(T)\;,
\label{eq:psv}
\end{eqnarray}
where $\hat{g}^2(T)$ denotes the temperature-dependent
running coupling constant renormalized in the $\overline{\mathrm{MS}}$-scheme  at $\mu = 2\pi T$. The  value of $p_2$ is fixed by the EQCD calculation \cite{Laine},
while $p_3,p_4,s_4$ are not yet known analytically and consequently fitted to the lattice data. Note that
$p_2,p_3,p_4,s_4$ are numbers and the temperature dependence of
the screening masses resides in the coupling constant. The temperature dependence of the coupling constant is logarithmically slow which is the reason
why the screening masses vary very little in the large temperature interval.
A perturbative description of screening masses and of the equation of state
suggests  partonic degrees of freedom, which is a signal of the quark-gluon plasma.

\begin{figure}
\centering
\includegraphics[scale=1.]{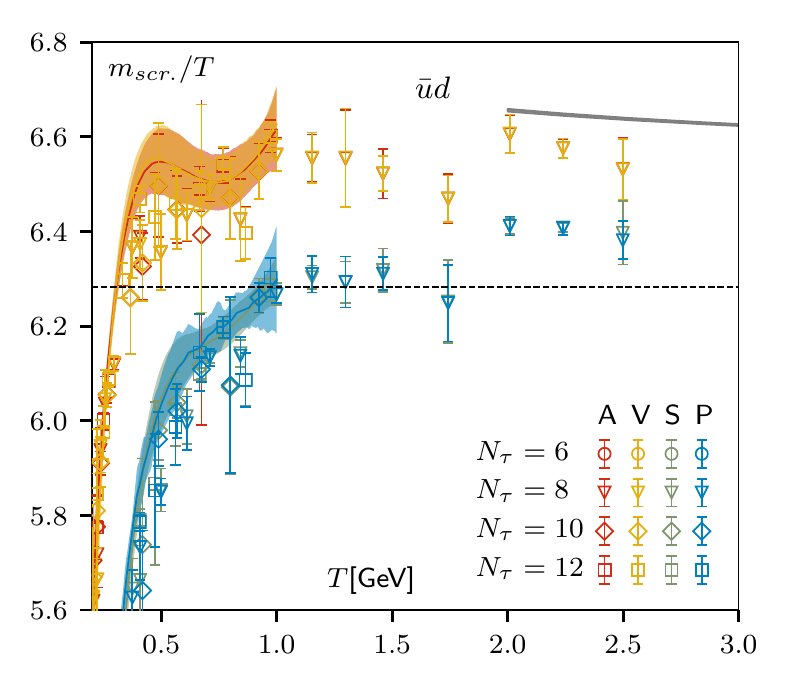}
\caption{ Screening masses of the lightest $\bar u d$ mesons. 
The solid line represents the screening masses at the order
$\sim g^2$ obtained within EQCD \cite{Laine}. From Ref.\cite{hotscr}.}
\label{scrlow}
\end{figure}

Screening masses at lower temperatures above the chiral restoration crossover
up to $ T  \sim 2.5$ GeV are shown in Fig. \ref{scrlow} \cite{hotscr}. One observes
that above $ T  \sim 1$ GeV the screening masses in Figs.  \ref{tdep}
and \ref{scrlow} match with each other and a temperature dependence of screening masses above $ T  \sim 1$  GeV is flat. What immediately attracts our
attention  is the rapid bending of curves within $ T  \sim 0.5-0.6$ GeV,
from a steep increase with temperature to a flat behavior.
Since the temperature dependence of the partonic description (\ref{eq:psv})
is only in  the coupling constant $\hat{g}^2(T)$, the partonic description
cannot explain the nearly vertical parts of the plots. This feature is
observed in all $J=0,1$ mesons with strangeness as well \cite{hotscr}.
The screening masses in the $u,d,s$ sector are the dominant contributions
to the partition function (\ref{eq:ham}). We observe an apparent change of dynamics at $ T  \sim 0.5-0.6$ GeV from the parton dynamics at higher
temperatures to another one at $T < 0.5$ GeV. The temperature at which
we see a change of dynamics in the partition function
coincides with the temperature where
chiral spin symmetry disappears and the chromoelectric confining
interaction gets screened. At $ T  \sim 0.5-0.6$ GeV and below the strong
coupling constant is large, as can be concluded  from Fig. \ref{tdep},
so it is not surprising that a non-perturbative confining dynamics
is operative.

One should raise the question whether the  description
within the EQCD \cite{Ginsparg:1980ef,Appelquist:1981vg,Nadkarni:1982kb} is consistent or not with the chiral spin symmetric 
regime below $T \sim 500 - 600$ MeV. The EQCD is an effective 
bosonic description of QCD at high temperatures obtained upon perturbative dimensional reduction
of the four dimensional QCD with both fermion and gluon degrees of freedom
to effective bosonic degrees of freedom in a three dimensional space. At the asymptotically 
high temperatures it should be an accurate representation of QCD. The perturbative dimensional
reduction at small coupling constants relies on the QCD Lagrangian where no approximate
chiral spin symmetry can exist because within the perturbative description the quark kinetic
term is of primary importance and which breaks the chiral spin symmetry. Consequently in
a validity range of EQCD one would not expect an approximate chiral spin symmetry. Observation
of the approximate chiral spin symmetry at $T < 500 - 600$ MeV restricts then application of EQCD
to higher temperatures where the CS symmetry disappears. This simple consideration is
consistent with results depicted in Fig. \ref{tdep} with the $p_2$ term to be consistent with
lattice results only at rather high temperatures, of the order  of 1 GeV and larger.

We conclude that the behavior of meson screening masses from 12
different quantum number channels in $N_F =2+1$ QCD provides an
independent demonstration of the existence of the temperature
window  below 500 - 600 MeV in which chiral symmetry is restored but
the dynamics is inconsistent with a partonic description.

\begin{figure}
\centering
\includegraphics[scale=0.6]{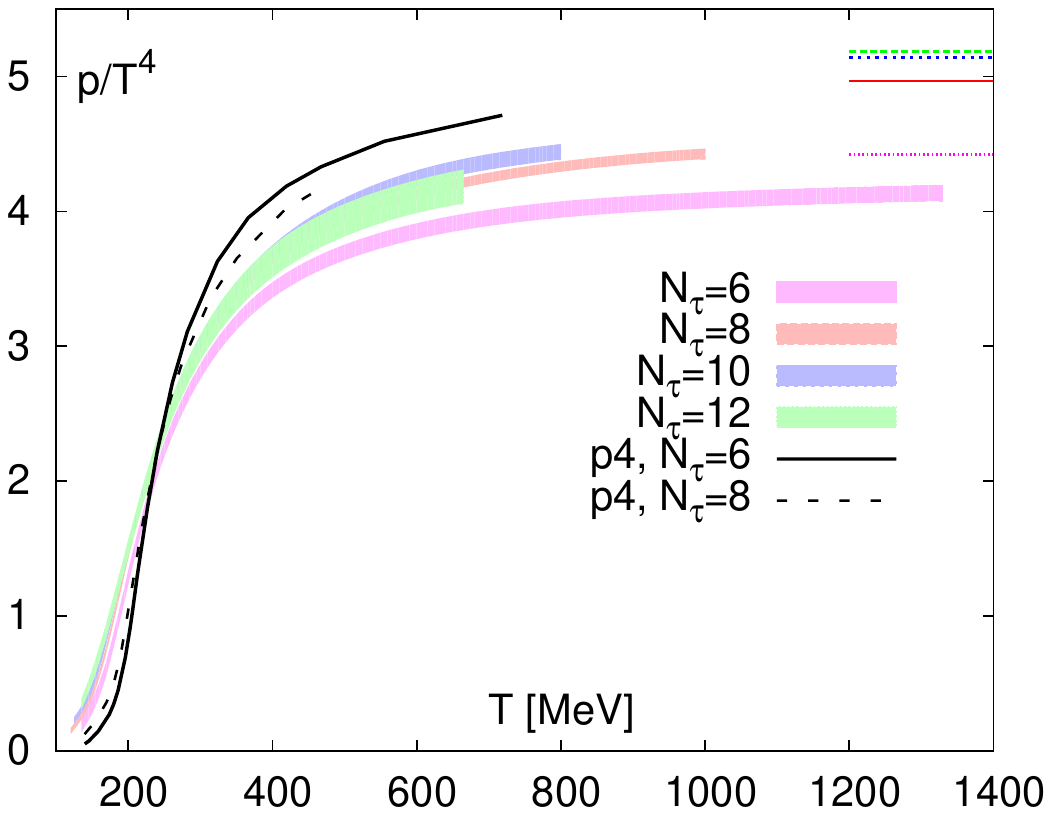}
\caption{ The pressure calculated with HISQ action for
$N_F =2+1$ QCD. From Ref.\cite{bpw}.}
\label{pre}
\end{figure}

The discussed behavior of screening masses below $ T  \sim 0.5-0.6$ GeV
must also be reflected in the equation of state. Indeed, a very steep 
increase  of $p/T^4$  with temperature in the same temperature interval
is observed 
\cite{bpw} which is shown in Fig. \ref{pre}.
Weakly interacting partons in the quark-gluon plasma require $p/T^4 \sim const$,
which is detected at higher temperatures.

\section{Pion states above the chiral crossover}

A direct evidence for the hadron-like degrees of freedom in the
stringy fluid should be observation of the corresponding states
in spectral functions. A break-through in this direction was
done in Ref. \cite{LP}. This section is devoted to the results obtained
in this paper.

Typically  attempts to reconstruct a spectral function at large temperatures
relied on  Euclidean temporal correlators that contain a small number
of points. It is an ill-posed problem.   It is not clear a-priori
to which extent these reconstructions can be credible as there is
no control of results. In Ref. \cite{LP} instead the pion spectral
function was extracted from the spatial PS correlators depicted in Fig.
\ref{spatial}. In this paper the approach was used which was developed
in Refs. \cite{bb1,bb2,bb3,bb4,bb5} and which is based on locality of QCD.
Given the pion spectral function extracted from the spatial correlators,
the temporal correlators can be directly predicted according to (\ref{eq:corr})
and compared with the lattice results.

It is well known that locality (causality) of QFT at T=0 requires
 existence of the K\"{a}llen-Lehmann spectral representation.
In Ref. \cite{bb1} this representation 
 for scalar spectral density was generalized to arbitrary
 temperature and is

\begin{align}
\rho(\omega,\vec{p}) = \int_{0}^{\infty} \! ds \int \! \frac{d^{3}\vec{u}}{(2\pi)^{2}} \ \epsilon(\omega) \, \delta\!\left(p^{2}_{0} - (\vec{p}-\vec{u})^{2} - s \right)\widetilde{D}_{\beta}(\vec{u},s),
\label{commutator_rep}
\end{align}
with $\widetilde{D}_{\beta}(\vec{u},s)$ being the thermal spectral density
which completely determines the properties of scalar particles in the medium.
For stable particle that has in vacuum a discrete pole at $\sqrt s = m$
and which is well separated from the continuum contributions, the
following Ansatz can be used for $\widetilde{D}_{\beta}(\vec{u},s)$ if one
looks for this particle in the medium at a temperature $T$:

\begin{align}
\widetilde{D}_{\beta}(\vec{u},s)= \widetilde{D}_{m,\beta}(\vec{u})\, \delta(s-m^{2}) + \widetilde{D}_{c, \beta}(\vec{u},s),
\label{decomp}
\end{align} 
where $\widetilde{D}_{c, \beta}(\vec{u},s)$ is continuous in $s$. Refs.~\cite{bb1,bb2,bb3,bb4,bb5}
discussed  several reasons for why the discrete component in Eq.~\eqref{decomp}  (the first term)
provides a natural description of a particle state in the medium.  The  \textit{damping factor} $\widetilde{D}_{m,\beta}(\vec{u})$  causes $\rho(p_{0},\vec{p})$ to have contributions outside of the mass shell $p^{2}=m^{2}$, and hence the $T=0$  peak of the spectral function gets broadened, which is a natural expectation.
The precise nature of this broadening is controlled by the underlying interactions between the particle state and the constituents of the thermal medium~\cite{bb5}.  The factorization of the $(\vec{u},T)$ and $s$ dependence ensures that this representation can distinguish between particle decays brought about by dissipative thermal effects, controlled by $(\vec{u},T)$, and those due to any intrinsic instability of the $T=0$ particle. The  damping factors in specific models were  explored in Ref.~\cite{bb5}, and  recently in Refs.~\cite{Lowdon:2021ehf,Lowdon:2022keu,Lowdon:2022ird}. 

\begin{figure}
\centering
\includegraphics[scale=0.4]{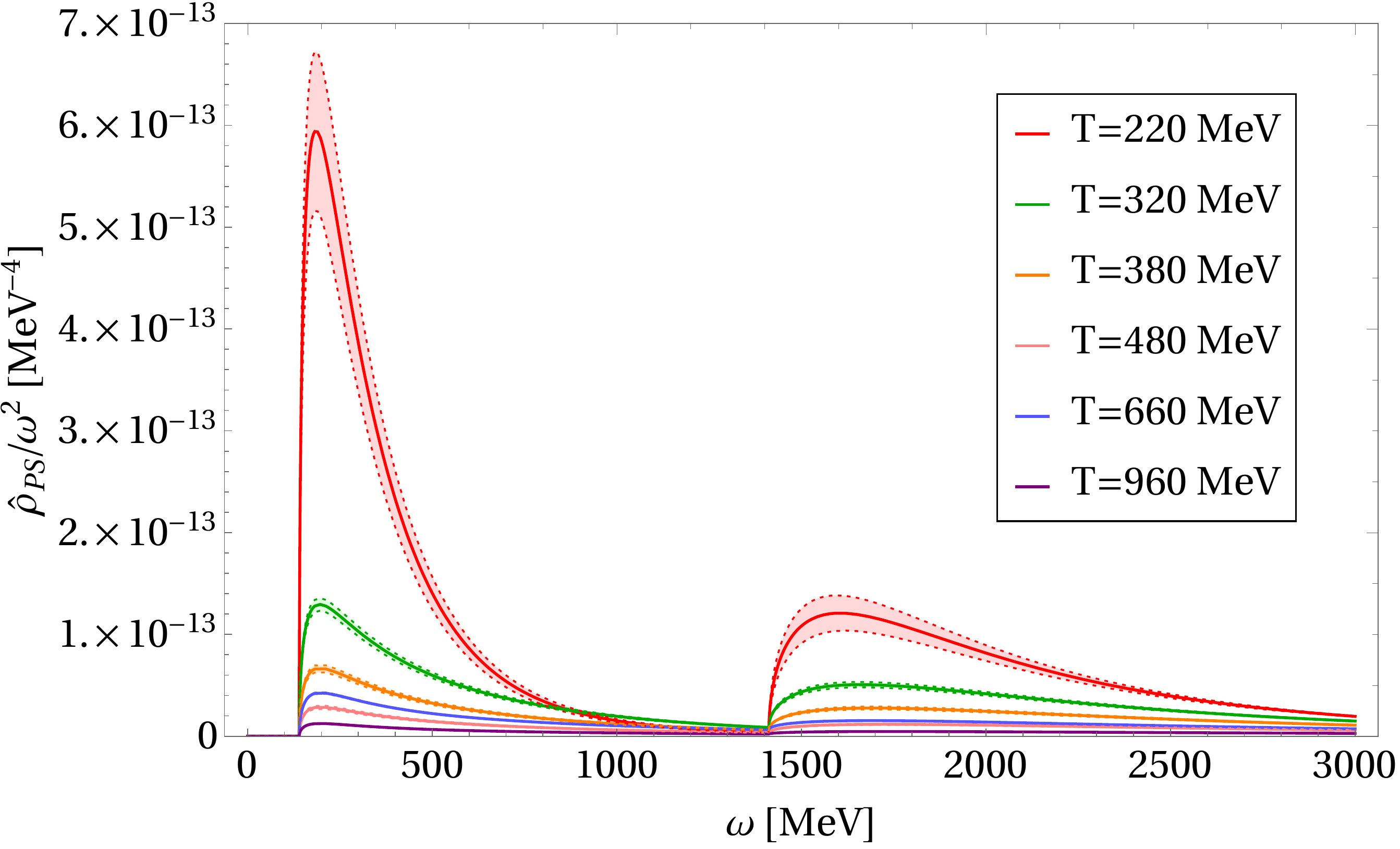}
\caption{ The pion spectral density at different temperatures
extracted from the spatial correlators of Fig. \ref{spatial}. From Ref. \cite{LP}.}
\label{lpsf}
\end{figure}

Using these ideas Ref. \cite{LP} established a bridge between the spatial pion
correlators (\ref{eq:c_z}) and the rest frame spectral density
$\rho(\omega,\vec{p} = 0)$. From a two-exponent fit of the spatial correlator with a very good quality the pion spectral function was reconstructed at different
temperatures. These two exponents are interpreted as contributions from two
(quasi)discrete levels $\pi,\pi'$ in the medium and the continuum part
in Eq. (\ref{decomp}) is neglected.
The results are shown in Fig. \ref{lpsf}.
The spectral function demonstrates two distinct peaks that correspond
to the pion and its first radial excitation in the medium. These peaks
get broader with temperature and melt above $T \sim 500 - 600$ MeV out.

\begin{figure}
\centering
\includegraphics[scale=0.4]{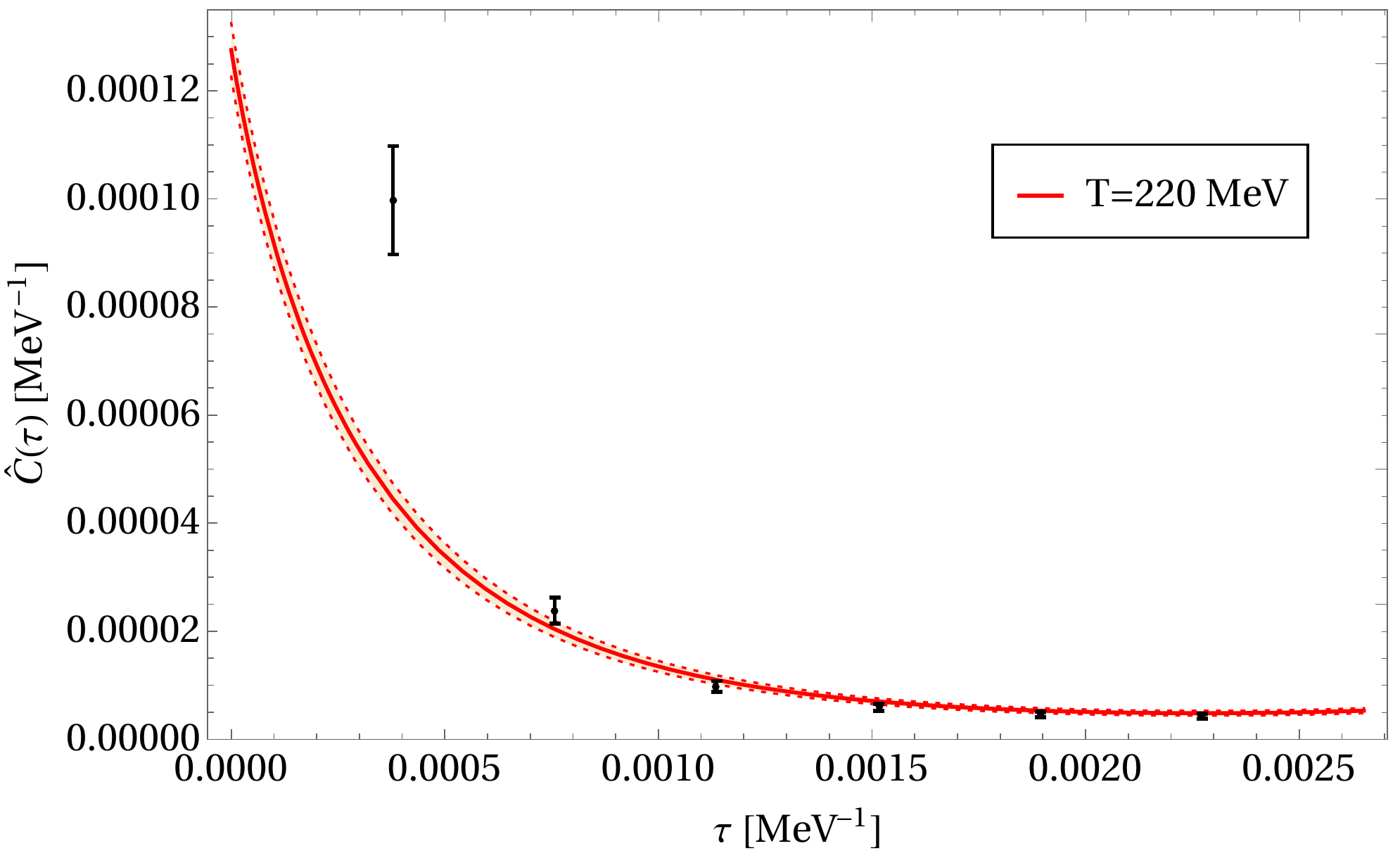}
\caption{ The pion temperoral correlator calculated
from the reconstructed spectral density of Fig. \ref{lpsf}
in comparison with the lattice temporal correlator of
Fig. \ref{tcorr}. From ref. \cite{LP}}
\label{comparison}
\end{figure}

This spectral function extracted from the spatial correlators
 can be controlled since a temporal correlator 
can be calculated
according to Eq. (\ref{eq:corr}) and compared with the lattice results
of Fig. \ref{tcorr}. The output is shown in Fig. \ref{comparison}.

It is a truly remarkable result. The large $t$ part of the correlator
is accurately reproduced. A deviation is seen only in the small $t$
part that is sensitive to a contribution of the higher excited states, $\pi'',...$ and to  contributions
of neglected continuum. The latter contributions cannot  be picked  with
the two exponential fit of the spatial correlators up.
This test suggests that the low energy part of the spectral function 
with $\pi,\pi'$
presented in
Fig. \ref{lpsf} is close to reality, though the omitted contributions from $\pi'',...$
and continuum should influence a small $t$ part of the correlator and the spectral function
beginning from $\omega \sim 1600-1700$ MeV. Their effect should increase with temperature.

It is instructive to compare the spectral function of Fig. \ref{lpsf}
with a typical result obtained earlier from the temporal
correlators using the maximum entropy method with an additional
constraint that at some critical temperature the spectral
function is described by perturbative QCD \cite{petr}.
While the latter spectral function also shows two distinct peaks
at high temperatures, their position is proportional
to the temperature
and the width of these excitations remains constant
with temperature, in contrast to the results presented in Fig. \ref{lpsf}.

To summarize, 
the results of this section  imply that degrees of freedom 
in the medium above
the chiral crossover and below $T \sim 500 - 600$ MeV are hadrons. 
This is entirely consistent with the conclusions obtained in previous
sections based on
symmetries of correlators and on screening masses and the equation of state.

\section{Bottomonium spectrum above $T_{ch}$}

\begin{figure}
\centering
\includegraphics[scale=0.6]{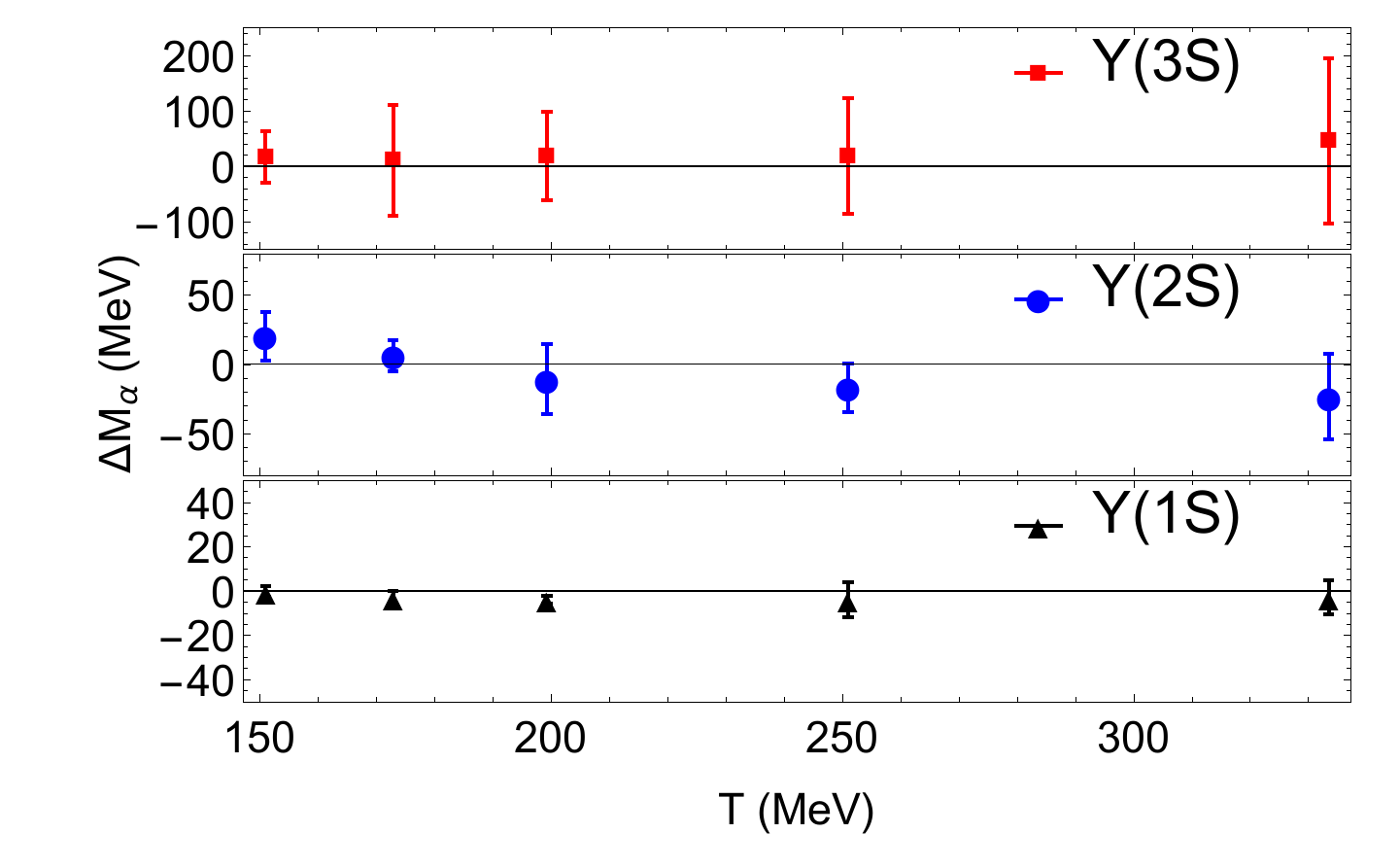}
\caption{Shifts of the bottomonium levels in medium at temperatures above
$T_{ch}$ with respect to the masses in vacuum. From Ref. \cite{Lar}.}
\label{massshifts}
\end{figure}

Another evidence that  at temperatures significantly 
above $T_{ch}$ there is no "deconfinement",
is the observation on the lattice of the 1S,2S,3S and 1P,2P radial and orbital excitations
of bottomonium \cite{Lar}. The states
were obtained with the standard variational analysis of the corresponding
correlators using non-relativistic QCD lattice framework. It is important
to stress that no potential picture is assumed here, it is an output of QCD.

The results for the mass shifts of the 1S,2S,3S levels with respect
to the vacuum masses of the corresponding states are shown in Fig. 
\ref{massshifts}.
We see that masses of the bottomonium states in the medium 
remain stable and agree with those in vacuum. The same feature
was observed in previous Section for the pion spectral function.
\begin{figure}
\centering
\includegraphics[scale=0.4]{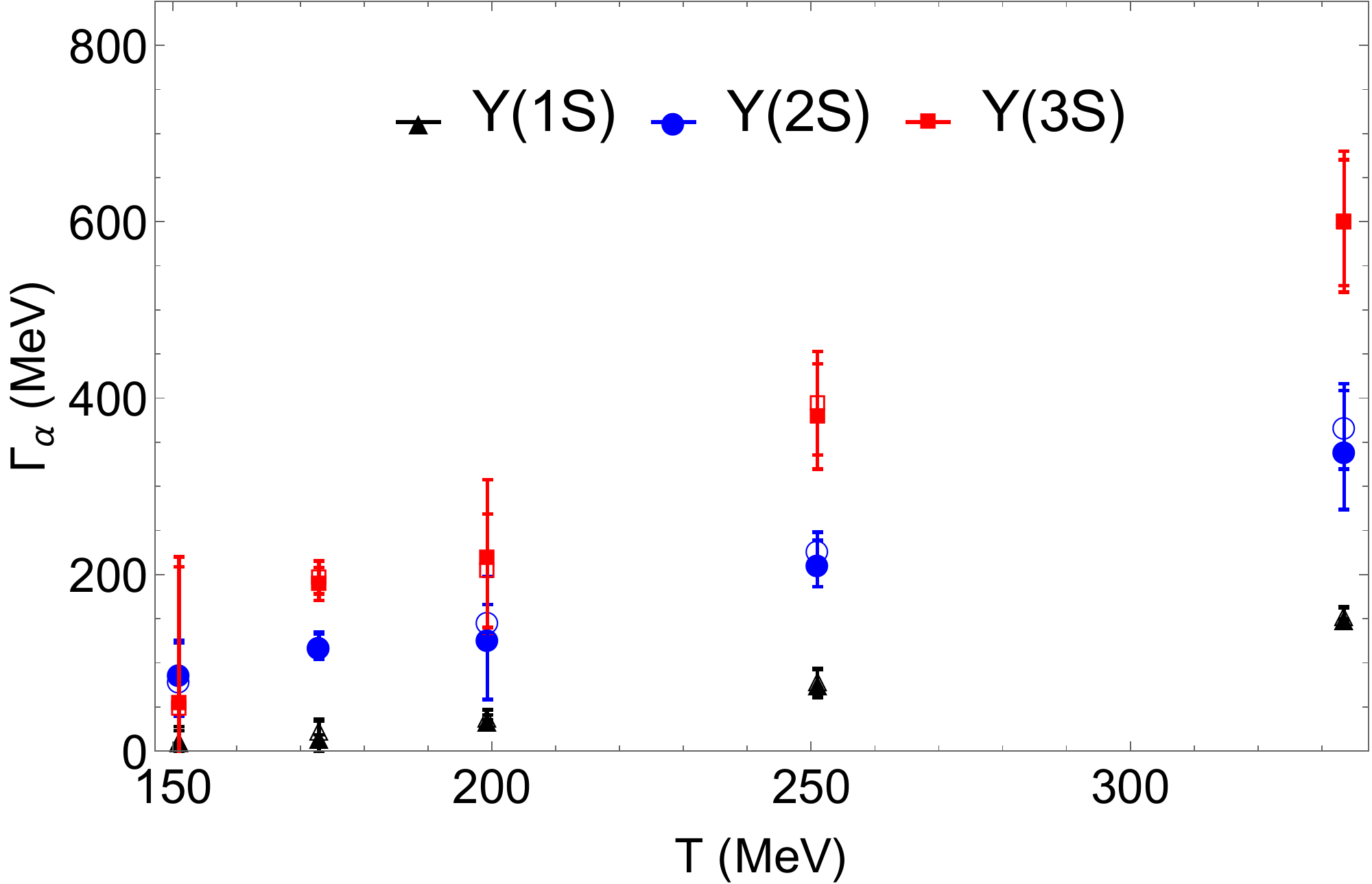}
\caption{
The temperature dependence of widths of the bottomonium levels in medium  above
$T_{ch}$. From Ref. \cite{Lar}.}
\label{bottomoniumwidth}
\end{figure}
The widths of the corresponding states are shown in Fig. \ref{bottomoniumwidth}. The widths increase with temperature.
Again the same feature is seen for pions.

The observation of the radial and orbital excitations in a heavy 
quark-antiquark system is an unambiguous evidence
for confinement. According to the Matsui-Satz prediction \cite{MS},
deconfinement at a critical temperature would mean that the confining
Coulomb plus linear potential becomes Debye screened and gets weaker
than the Coulomb potential,

\begin{equation}
\sim -1/r \exp( - m_D r).
\label{deb}
\end{equation}
Such potential does not support any bound state in
a system of heavy quarks and would evidence a deconfinement.
The Coulomb potential supports only the 1S state (positronium)
and no P-levels. Above there would be a quark - antiquark continuum.

\begin{figure}
\begin{center}
\includegraphics[width=6cm]{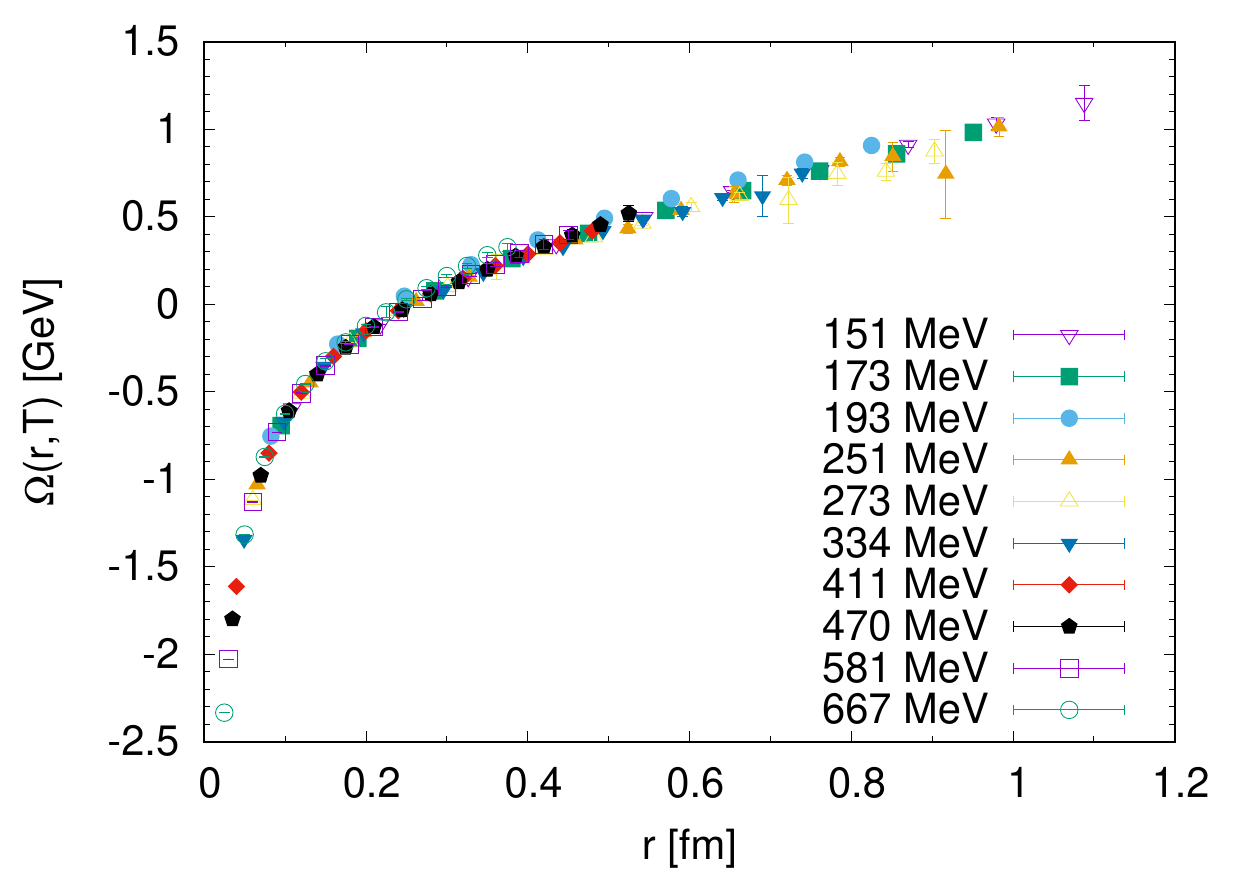}
\includegraphics[width=6cm]{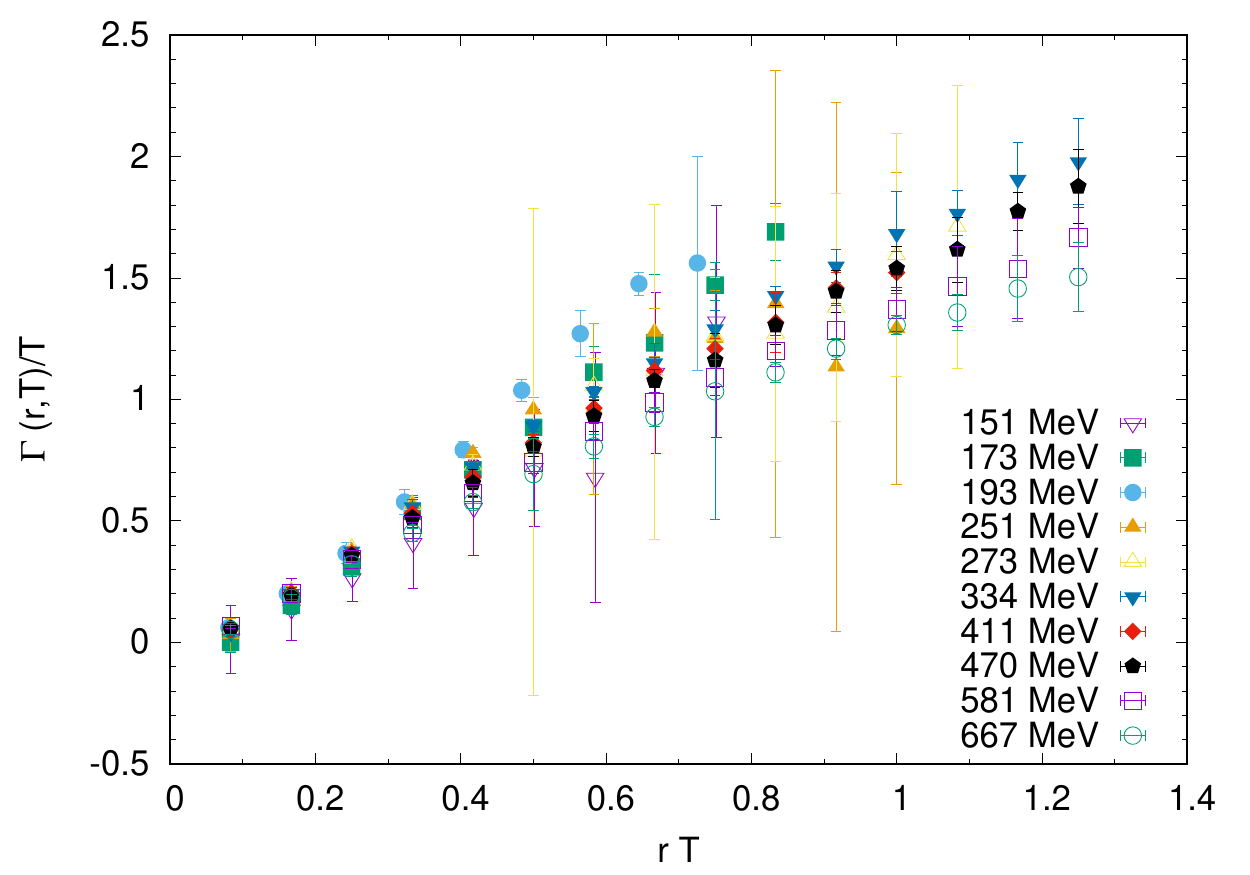}
\caption{Real (left panel) and imaginary (right panel) parts of the optical potential. From Ref. \cite{optical}}
\label{opt}
\end{center}
\end{figure}

The survival of the radial and orbital excitations above the chiral
restoration temperature points to a confining interaction
at these temperatures. A phenomenological model picture consistent
with these results is an optical potential that consists of  
a real linear confining potential, 
that is not modified with temperature, and an imaginary part,
that increases with $T$ \cite{optical}, see Fig. \ref{opt}. The T-independent
real part is responsible for the stability of the energy levels
in the medium, and the rising with  temperature  imaginary part provides
increasing widths.

As a conclusion, 
these results stress that degrees of freedom in the medium  above $T_{ch}$ are
color-singlet hadrons.

\section{Why is the stringy fluid  stringy?}

It is not accidental that the name "stringy fluid" was
given to a chirally symmetric and approximately chiral
spin symmetric QCD matter above the chiral restoration crossover \cite{G4}.
The reason is that a simple stringy picture of confined hadrons,
provided that the chiral symmetry is restored \cite{G6}, very
naturally accommodates the chiral spin symmetry.

The celebrated  approximately linear Regge trajectories

\begin{equation}
M^2 (n,L) = c_n n + c_L L + corrections,
\label{rtr}
\end{equation}

\noindent
where $n$ and $L$ are the radial quantum number and angular momentum
of the string, respectively, represent the most important achievement of the string description of hadrons. 
The slope of the angular  trajectories, $c_L$, is fixed by the
string tension $\sigma$
which is a fundamental parameter of the Nambu-Goto action.

What is  missing in this description is a degeneracy of states
with opposite parity, i.e., a presence of the chiral multiplets in the
spectrum. This is because the
spin degree of freedom of  quarks at the ends of the string
 is missing in the standard open bosonic string description.

\begin{figure}
\begin{center}
\includegraphics[width=10cm]{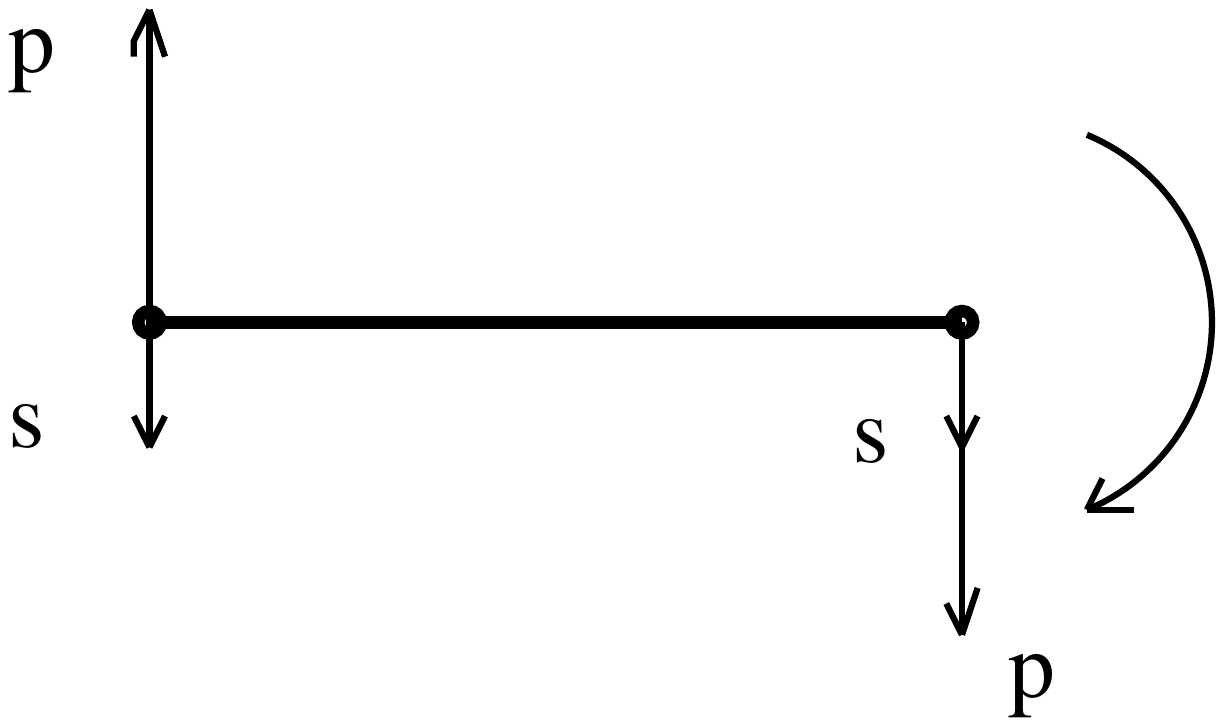}
\caption{The rotating string with quarks with definite chirality
at the ends.}
\label{str}
\end{center}
\end{figure}

If chiral symmetry is restored, as it does above $T_{ch}$, then one
naturally views a string
with massless quarks at the ends that have definite chiralities \cite{G6},
see Fig. \ref{str}. 

Explicitly all eigenvectors of chiral symmetry for mesons
were constructed in Ref. \cite{G1}.
The chirally symmetric $\bar q q$ states are specified with: ${r;IJ^{PC}}$, where $r$ denotes a representation
of the parity-chiral group and all  other quantum numbers are
isospin, spin, spatial and charge parities.
The states   fill out the following 
irreducible representations of the parity-chiral group $SU(2)_L \times SU(2)_R \times {\cal C}_i$,
where  ${\cal C}_i$ consists of the space inversion and identity.
A product with the latter group is required to construct states of definite parity:

{\bf (i)~~~ (0,0):}

\begin{equation}
|(0,0); \pm; J \rangle = \frac{1}{\sqrt 2} |\bar R R \pm \bar L L\rangle_J.
\label{00}
\end{equation} 

\noindent
Here $I=0$, $R$ and $L$ denote the right-handed  $SU(2)_R$ 
($R^T=(u_R,d_R)$) and the left-handed
 $SU(2)_L$   ($L^T=(u_L,d_L)$)   vectors.
The subscript $J$ means that a definite spin and its projection ($J$ and $M$)
are ascribed
to the given quark-antiquark system according to the 
relativistic spherical helicity formalism \cite{LANDAU}:

\begin{equation}
|\lambda_q\lambda_{\bar q}\rangle_J=D^{(J)}_{\lambda_q - \lambda_{\bar q}, M}(\vec n)\sqrt {\frac{2J+1}{4\pi}}
|\lambda_q\rangle  |-\lambda_{\bar q}\rangle,
\label{hel}
\end{equation}
with $D^{(J)}_{MM'}(\vec n)$ being the Wigner $D$--function describing rotation from the quantization axis to
the quark momentum direction $\vec n=\vec{p}/p$ and 
$\lambda_{q}$ ($\lambda_{\bar q}$) are the quark (antiquark) helicities. Note
 that the quark chirality and helicity coincide, while for the antiquark
 they are just opposite. 
 The parity of the quark-antiquark state is then 

\begin{equation}
\hat P |(0,0); \pm; J \rangle = \pm (-1)^J  |(0,0); \pm; J \rangle.
\label{P00}
\end{equation}

{\bf (ii)~~~ $(1/2,1/2)_a$ and $(1/2,1/2)_b$:} 

\begin{equation}
|(1/2,1/2)_a; +;I=0; J \rangle = \frac{1}{\sqrt 2} |\bar R L + \bar L R\rangle_J,
\label{I0+1}
\end{equation} 

\begin{equation}
|(1/2,1/2)_a; -;I=1; J \rangle = \frac{1}{\sqrt 2} |\bar R \vec \tau L -
\bar L \vec \tau R\rangle_J,
\label{I0+2}
\end{equation} 

\noindent
and

\begin{equation}
|(1/2,1/2)_b; -;I=0; J \rangle = \frac{1}{\sqrt 2} |\bar R L - \bar L R\rangle_J,
\label{I0-1}
\end{equation}

\begin{equation}
|(1/2,1/2)_b; +;I=1; J \rangle = \frac{1}{\sqrt 2} |\bar R \vec \tau L +
\bar L \vec \tau R\rangle_J.
\label{I0-2}
\end{equation}

\noindent
Here $\vec \tau$ are isospin Pauli matrices.
The parity of  all states in these representations is determined as

\begin{equation}
\hat P |(1/2,1/2); \pm; I; J \rangle = \pm (-1)^J |(1/2,1/2); 
\pm; I; J \rangle.
\label{P12}
\end{equation}

Note that a sum of the two independent $(1/2,1/2)_a$ and $(1/2,1/2)_b$
 irreducible representations
of $SU(2)_L \times SU(2)_R$ forms an irreducible representation
of the $U(2)_L \times U(2)_R$ or $SU(2)_L \times SU(2)_R \times U(1)_A$ groups.

{\bf (iii)~~~ (0,1)$\oplus$(1,0):} 

\begin{equation}
|(0,1)+(1,0); \pm; J \rangle = \frac{1}{\sqrt 2} |\bar R \vec \tau R 
\pm \bar L  \vec \tau L \rangle_J,
\label{10}
\end{equation} 

\noindent
with $I=1$ and   parities

\begin{equation}
\hat P |(0,1)+(1,0); \pm; J \rangle = \pm (-1)^J 
 |(0,1)+(1,0); \pm; J \rangle.
\label{P10}
\end{equation} 

The $J=0$ states
are connected by the $SU(2)_L \times SU(2)_R $ and $U(1)_A$ transformations
 and cannot be constructed from the $(0,0)$ and $(0,1)+(1,0)$
representations, because the total spin projection onto the momentum
direction of the quark  is $\pm 1$ for the latter representations.

Now comes a key point.
The states  (\ref{00}),(\ref{I0+1}),(\ref{I0+2}),(\ref{I0-1}),(\ref{I0-2})
and (\ref{10}) with
$J > 0$ transform into each other
upon the $SU(2)_{CS}$ and $SU(4)$ transformations of the vector
(\ref{fourcomp}), for different irreducible
representation of the latter groups see Fig. \ref{algebra}.
We automatically incorporate the $SU(2)_L \times SU(2)_R \times U(1)_A$ 
chiral symmetry as well as the  $SU(2)_{CS}$ and $SU(4)$ symmetries \cite{G1}.
Namely,  all hadrons with different chiral configurations of 
quarks at the ends of the string  that belong to the same intrinsic quantum
state of the string must be degenerate.

There are important implications. The spin-orbit  interactions of 
quarks should vanish at the classical level. 
 Indeed, if
the quark has a definite chirality, then its spin is necessarily parallel
(or anti-parallel) with its momentum. Hence the
spin-orbit force, $\sim \vec L \cdot \vec S$, is necessarily zero. 
This is also true for the spin-orbit force due to the Thomas precession.

For a rotating $\bar q q$ string the tensor force also vanishes. 
Indeed, the tensor force consists of the scalar products 
$\vec S_i \cdot \vec R_j$, where $\vec R_j$ is the radius-vector
of the given quark in the center-of-mass frame.

There is more than that.
Both the spin-orbit and tensor forces are effects of the
magnetic field. However, emergent chiral spin and $SU(4)$ symmetries
indicate that  the magnetic field in the medium is highly suppressed
with respect to the confining electric field, as was discussed in chapter 2.
Then the absence of the spin-orbit and tensor interactions between
quarks is consistent  with the latter emerged symmetries.

To summarize: Approximate chiral spin and $SU(4)$
symmetries seen above $T_{ch}$ are entirely consistent
with a simple and intuitive picture that here hadrons are electric
strings with chiral quarks at the ends.

\section{Is the stringy fluid a gas or a liquid?}

An ideal gas of  quarks and gluons is characterized by 
the Stefan-Boltzmann behavior 

\begin{equation}
P \sim T^4.
\label{stb}
\end{equation} 
From Fig. \ref{pre} we can conclude that the system
is close to this limit at temperatures above $T \sim 1$
GeV. However at lower temperatures, below 500-600 MeV (but above the chiral restoration
temperature $T_{ch}$) the pressure rises with $T$ much faster. This 
indicates that  interaction
between the constituents is very important.  
The same feature is also seen in Fig. \ref{scrlow}. A flat temperature
dependence
of the screening masses  is observed above $T \sim 600$ MeV.
This  is  consistent with a perturbative HTL re-summation
and is characteristic of a quark-gluon plasma. However,
 between $T_{ch}$ and 500 - 600 MeV  a steep increase is seen, that
 is inconsistent with the parton description.
 
 From the chiral spin symmetry of an effective QCD action (and
 of the thermal QCD partition function)  at $T_{ch} - 3 T_{ch}$ 
 as well as from the pion and
  bottomonium spectral properties we have concluded that  in this temperature window degrees of freedom
 in the medium  should be chirally symmetric
 and approximately chiral spin symmetric hadron-like systems. Then the
 question arises, 
 whether these hadrons
 interact strongly and what evidence exists for this?
  Here we will give a qualitative answer
 to this question and explain that the observed spectral properties
 of pions do imply a strong interaction between the hadrons in the medium.
 We stress that we do not attempt to construct an effective
 theory of pions in the stringy fluid. It is a task for future.

The key point is that the discrete pion level at $\sqrt s =m$
in vacuum at zero temperature becomes a resonance
with a finite width in the QCD medium at a temperature $T > T_{ch}$, see
Fig. \ref{lpsf}. 
Let us simplify the system and assume  that a
hadron resonance gas at  temperatures below $T_{ch}$ consists only of
noninteracting pions. Such a system is described by the Klein-Gordon
Lagrangian: 

\begin{equation}
{\cal {L}} = \frac{1}{2}\partial^\mu \phi \partial_\mu \phi -\frac{1}{2} m^2 \phi^2.
\label{phi}
\end{equation}

Naively the finite width of the pion seen in Fig. \ref{lpsf} can be
connected to a decay of the pion into a quark-antiquark pair since the pion is the
lightest hadron. If it were so this would point to absence of a confining
interaction above $T_{ch}$. However, it is not so and 
the presence of a finite width of the pion in Fig. \ref{lpsf}  implies actually
that  above $T_{ch}$ the medium cannot be  a gas of noninteracting pions and instead the pions
should strongly interact. The simplest possible interaction term in the effective Lagrangian 
above $T_{ch}$ should be the $\phi^4$ term (or, some factors of $\phi$ in the interaction term should
be substituted by  derivative of the pionic field according to the power counting). For example, the well known $\phi^4$
theory of interacting scalars is given by the following Lagrangian

\begin{equation}
{\cal {L}} = \frac{1}{2}\partial^\mu \phi \partial_\mu \phi -\frac{1}{2} m^2\phi^2 - \frac{\lambda}{4!}\phi^4. 
\label{phi4}
\end{equation}

The $\phi^4$ term in the  Lagrangian
describes a collision of four pions above $T_{ch}$, respectively,
and graphically corresponds to an interaction vertex with  four pion
legs. Such vertices imply that a strong decay of the pion into three
pions is possible. Consequently this theory would require that the
pion is not  a stable particle and should be a resonance. 
We conclude that a finite decay width of the pion above $T_{ch}$
points to a strong interaction of pions and to  $\phi^4$-like
terms in an effective Lagrangian. We repeat, that it is not yet an
attempt of an effective theory of interacting pions above chiral restoration
temperature. Such a theory should
rely on the power counting and be constrained by the emerged symmetry. 

A presence of collisions of  four particles in a system
implies that
the system is more a liquid rather than a gas, if the effective coupling constant $\lambda$ is large
enough\footnote{
 Recall that
a  condition for a condensation of  vapor (gas) into water (liquid)
at some temperature and pressure is a presence of collisions of at least
three $H_2O$ molecules. This can happen only if the system is rather dense and a typical
distance between molecules is small.}.

\section{Baryonic parity doublets and chiral spin symmetry}

Could a chiral spin symmetric regime exist in the baryon rich
region at large chemical potentials and low temperatures?
It turns out that the manifestly chirally symmetric free parity doublet Lagrangian  \cite{LEE}
has precisely a $SU(4)$ symmetry with a $SU(2)$ subgroup that performs a
rotation in the space of right-handed and left-handed fields 
\cite{cat}.

Consider a Dirac Lagrangian for a massless fermion field

\begin{equation}
{\cal L} = i \bar \psi \gamma_\mu \partial^\mu \psi =
i \bar \psi_L \gamma_\mu \partial^\mu \psi_L +
i \bar \psi_R \gamma_\mu \partial^\mu \psi_R,
\label{L0}
\end{equation}
where 
\begin{equation}
\psi_R = \frac{1}{2}\left( 1+\gamma_5 \right ) \psi,~~
\psi_L = \frac{1}{2}\left( 1-\gamma_5 \right ) \psi.
\label{RL}
\end{equation}
This Lagrangian is obviously invariant upon independent
$U(1)$ rotations of the $\psi_R$ and $\psi_L$ components,
which represent the $U(1)_V \times U(1)_A$ symmetry.

If the fermion field $\psi$ is an isodoublet, this Lagrangian
is also  $SU(2)_L \times SU(2)_R$ invariant under two independent
isospin rotations of the right-handed and left-handed components 
(\ref{RL}):
\begin{equation}
\psi_R \rightarrow 
\exp \left( \imath \frac{\theta^a_R \tau^a}{2}\right)\psi_R; ~~
\psi_L \rightarrow 
\exp \left( \imath \frac{\theta^a_L\tau^a}{2}\right)\psi_L,
\label{ROT}
\end{equation}

\noindent
with $\tau^a$ being the isospin Pauli matrices
  and  $\theta^a_R$ and $\theta^a_L$ parameterize rotations
of the right- and left-handed components. The transformation
(\ref{ROT}) defines the  $(0,1/2) \oplus (1/2,0)$ representation 
of the chiral group, where $0$ and $1/2$ represent isospins of the
left- and right-handed components. A direct sum of two independent
irreducible representations of the $SU(2)_L \times SU(2)_R$ group is
required to get a field of a fixed spatial parity because under the
spatial reflection one has $L \leftrightarrow R$.

It is known for a long time that it is possible to construct
a chirally symmetric Lagrangian for a massive fermion field if
there are two independent mass-degenerate fermions of opposite parity - parity doublets \cite{LEE}:

\begin{equation}
\Psi = \left(\begin{array}{c}
\Psi_+\\
\Psi_-
\end{array} \right),
\label{doub}
\end{equation}

\noindent
where {\it independent} Dirac bispinors
$\Psi_+$ and $\Psi_-$ have positive and negative parity, respectively.
The parity doublet  is a spinor constructed from
two independent Dirac bispinors and contains eight components. There
is in addition an isospin index which is suppressed.

The right- and left-handed fields are
directly connected with the opposite parity fields

\begin{equation}
\Psi_R = \frac{1}{\sqrt{2}}\left( \Psi_+ + \Psi_-\right);~~
 \Psi_L = \frac{1}{\sqrt{2}}\left( \Psi_+ - \Psi_-\right).
\label{RLL}
\end{equation}
\noindent
Notice a difference with the definition of the right- and
left-handed components (\ref{RL}) of a single massless Dirac
field.
The vectorial and axial parts of the $SU(2)_L \times SU(2)_R$ transformation 
 under the $(0,1/2) \oplus (1/2,0)$ representation  is 
\begin{equation}
\Psi \rightarrow 
\exp \left( \imath \frac{\theta^a_V \tau^a}{2}
\otimes \mathds{1}\right)\Psi; ~~
\Psi \rightarrow 
\exp \left(  \imath \frac{\theta^a_A\tau^a}{2}\otimes \sigma_1
\right)\Psi,
\label{VAD}
\end{equation}

\noindent
where $\sigma_i$ is a Pauli matrix that acts in the  
space of the parity doublet. The  axial part of the chiral transformation
law (\ref{ROT}) mixes the massless Dirac spinor $\psi$
with $\gamma_5 \psi$ , while the chiral rotation of the parity doublet  
 provides a mixing of two independent fields $\Psi_+$ and $\Psi_-$. 
 
The chiral-invariant Lagrangian of the free parity doublet 
can be written in two equivalent forms as

\begin{equation}
\begin{split}
\mathcal{L} & = i \bar{\Psi} \gamma^\mu \partial_\mu \Psi - m \bar{\Psi}
\Psi  \\
& =  i \bar{\Psi}_+ \gamma^\mu \partial_\mu \Psi_+ + 
i \bar{\Psi}_- \gamma^\mu \partial_\mu \Psi_-
- m \bar{\Psi}_+ \Psi_+ - m \bar{\Psi}_- \Psi_-  \\
\end{split}
\label{lag}
\end{equation}
or
\begin{equation}
\begin{split}
\mathcal{L} 
& =  i \bar{\Psi}_L \gamma^\mu \partial_\mu \Psi_L + 
i \bar{\Psi}_R \gamma^\mu \partial_\mu \Psi_R
- m \bar{\Psi}_L \Psi_L - m \bar{\Psi}_R \Psi_R.
\end{split}
\label{llag}
\end{equation}

\noindent
The latter form demonstrates  that the right-
and left-handed degrees of freedom are completely
decoupled and the Lagrangian is manifestly chiral-invariant.
It is also manifestly Lorentz-invariant.

This Lagrangian can also be written in another forms
\cite{DETAR,TIT}. Now we will demonstrate  \cite{cat} its equivalence
to the "mirror" assignment of Ref. \cite{TIT}.
The Lagrangian (2.36) of  Ref. \cite{TIT} with two Dirac fermions

\begin{equation}
\mathcal{L} = i\bar{\psi}_1 \gamma_{\mu}\partial^{\mu}\psi_1 + 
i\bar{\psi}_2 \gamma_{\mu}\partial^{\mu}\psi_2 
-m(\bar{\psi}_1\psi_2 + \bar{\psi}_2 \psi_1),
\label{eq:mirror_lagr}
\end{equation}

\noindent
is invariant under chiral transformation with the ``mirror assignment":

\begin{equation}
\psi_1 \rightarrow \exp\left(i \frac{\alpha^a \tau^a}{2}\gamma^5\right)\psi_1 ,\qquad \psi_2\rightarrow \exp\left(-i \frac{\alpha^a \tau^a}{2}\gamma^5 \right) \psi_2.
\label{eq:mirror_trans}
\end{equation}

\noindent
This is exactly equivalent to the chiral transformation law
(\ref{VAD})
of the doublet (\ref{doub}) where 

\begin{equation}
\begin{split}
&\Psi_+ = \frac{1}{\sqrt{2}}(\psi_1 + \psi_2),\\
&\Psi_- = \frac{1}{\sqrt{2}}(\gamma_5\psi_1 -\gamma_5 \psi_2).
\end{split}
\end{equation}

\noindent
Then upon the ``mirror" transformation (\ref{eq:mirror_trans}) the parity doublet transforms as:

\begin{equation}
\left(\begin{matrix}
\Psi_+\\
\Psi_-
\end{matrix}\right)
\rightarrow \exp\left(i\frac{\alpha^a \tau^a}{2}\otimes\sigma^1\right) 
\left(\begin{matrix}
\Psi_+\\
\Psi_-
\end{matrix}\right).
\end{equation}

It turned out, however, that the free parity doublet Lagrangian 
 (\ref{lag}-\ref{llag}) has a larger symmetry than the 
$SU(2)_L \times SU(2)_R$ symmetry. It is 
manifestly  $SU(4)$ symmetric \cite{cat}.
Indeed, given  Eq.
(\ref{RLL}), the parity doublet
(\ref{doub}) can be unitarily transformed into
a doublet
\begin{equation}
\tilde{\Psi} = \left(\begin{array}{c}
\Psi_R\\
\Psi_L
\end{array} \right)\;.
\label{doubtr}
\end{equation}
It is a two-component spinor composed of Dirac bispinors
$\Psi_R$ and $\Psi_L$ (i.e., altogether there are  eight components).

The Lagrangian (\ref{lag}-\ref{llag}) is obviously invariant under
the $SU(2)$ rotations that mix $\Psi_R$ and $\Psi_L$ ,
\begin{equation}
\left(\begin{array}{c}
\Psi_R\\
\Psi_L
\end{array}\right)\; \rightarrow
\exp \left(i  \frac{\varepsilon^n \sigma^n}{2}\right) \left(\begin{array}{c}
\Psi_R\\
\Psi_L
\end{array}\right)\; .
\label{eq:su2cstra}
\end{equation}
Then  the parity doublet Lagrangian is not only
chirally invariant under the transformation (\ref{VAD}), but also
$SU(4)$-invariant with the generators of $SU(4)$
being 
\begin{align}
\{
(\tau^a \otimes \mathds{1}),
(\mathds{1} \otimes \sigma^n),
(\tau^a \otimes \sigma^n)
\}.
\end{align}
Since the rotation (\ref{eq:RL} ), that mixes the right- and
left-handed Weyl spinors, defines the $SU(2)_{CS}$  chiral spin
group, the transformation (\ref{eq:su2cstra}) is also
a chiral spin transformation, defined, however, on the space
of eight-dimensional parity doublets.

We conclude that a system of (quasi)free parity doublets,
perhaps with a phenomenologically introduced  short
range repulsion (which would still preserve the $SU(4)$ symmetry), is a good
candidate for a chiral spin symmetric regime in a baryon rich
medium at large chemical potentials and low temperatures.

This Lagrangian can be supplemented by  
the pion and
sigma-fields of the linear sigma model \cite{DETAR,TIT}.
A coupling of parity doublets to the $\pi,\sigma$ field
lifts the $SU(4)$ symmetry and only chiral symmetry is
left in the Lagrangian. This is because the $\pi,\sigma$
Lagrangian is chirally invariant but not a $SU(4)$-singlet.

The chiral symmetry breaking order parameter, $\langle 0|\sigma|0\rangle \neq 0$, generates a mass
splitting of the positive and negative parity baryons. I.e. the chiral symmetry
of the Lagrangian (\ref{lag}-\ref{llag}) is lifted. 
This regime
is reminiscent of  nuclear matter, where physics at large distances
is guided
by a coupling of nucleons of positive parity with $\pi,\sigma$
fields. 
However, a short range repulsion between nucleons is still missing
in this model, which is important for properties of nuclear matter.

The parity doublets coupled to the $\pi,\sigma$ fields 
have  been used 
in baryon spectroscopy \cite{Gallas:2009qp,Olbrich:2015gln} and  for
study of  chiral symmetry 
restoration at high temperature or density,  where  baryons
with non-zero mass do not vanish upon a chiral restoration, 
see e.g.~\cite{Zschiesche:2006zj,Steinheimer:2011ea,Sasaki:2017glk,Larionov:2021ycq} 
and references therein.
The chiral restoration transition  
can be either of first or second oder \cite{Zschiesche:2006zj}. 
This Lagrangian is only chiral invariant since a coupling of the parity doublets to pion and sigma fields destroys the $SU(4)$ symmetry. The latter
symmetry would approximately persist only if the coupling to the
$\pi,\sigma$ fields were suppressed, i.e. there would be no
baryon - baryon-hole excitations with pion quantum numbers.
It is a very interesting question whether the approximately
chiral spin symmetric matter at low temperatures and large baryon
chemical potential realized in nature or not.

\section{Chiral spin symmetric band of the QCD phase diagram}

From the  temperature dependence of the spatial correlators of Fig. \ref{spatial}, from the T-behavior
of screening masses in Fig. \ref{scrlow} and pressure in Fig. \ref{pre},
as well as from the   pion spectral density in Fig. \ref{lpsf} it is naturally
to assume that there is no critical line between the stringy fluid and QGP
regimes and both regimes are connected by a smooth analytic crossover.
This is precisely the reason why we call it regimes, but not phases.
However, to rule out a non-analytic phase transition a finite size
scaling study would be necessary to demonstrate that no discontinuity develops
in the thermodynamic limit. At the moment our knowledge of the $T$-dependence
of observables  above and below $3T_{ch}$ is not sufficiently detailed
to claim a crossover or a phase transition.
In the case of crossovers, there are no sharp phase boundaries and
a position of the crossover line, that "separates" two regimes, 
necessary varies with its definition.

\begin{figure}
\centering
\includegraphics[scale=0.3]{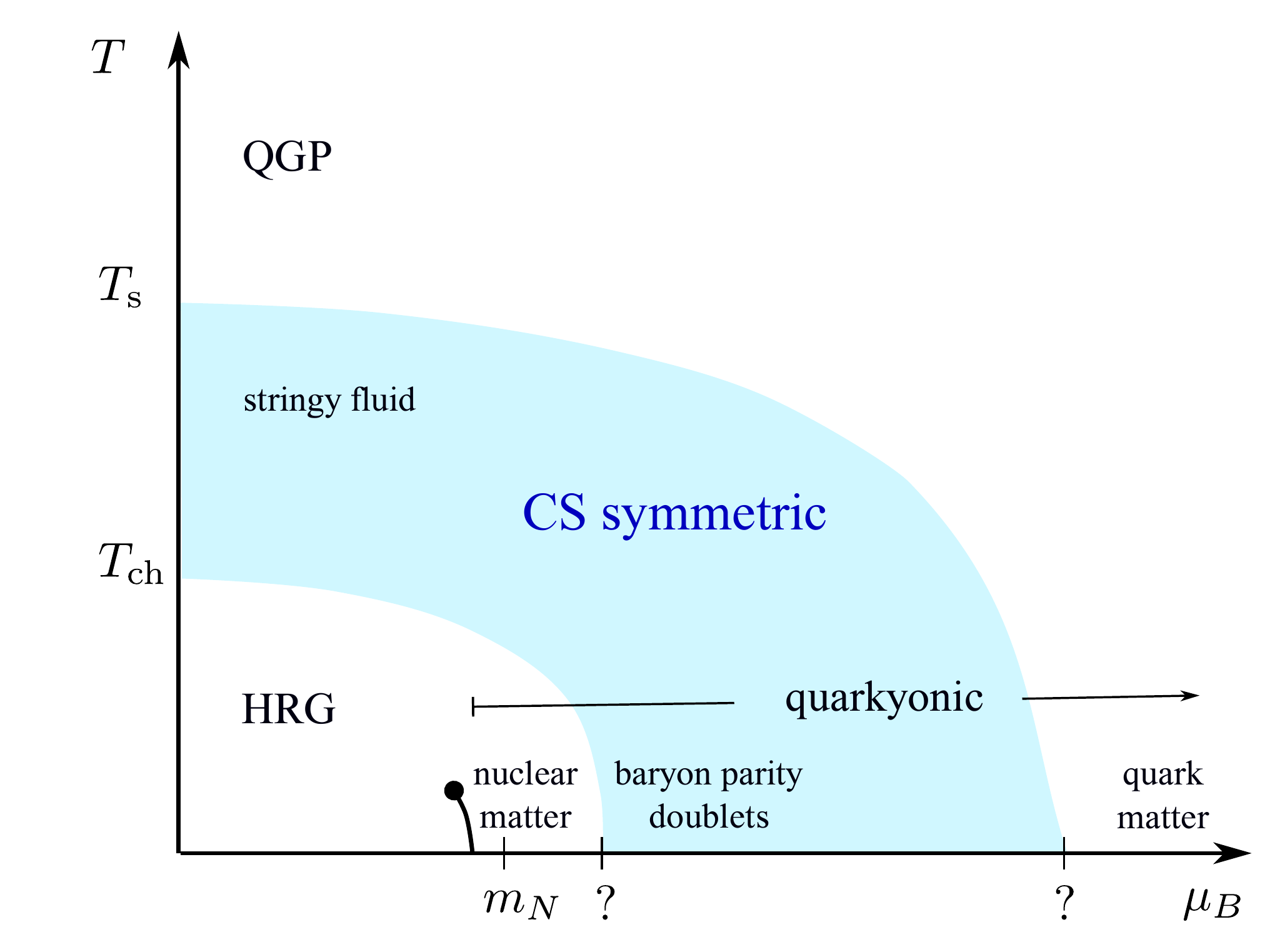}
\caption{Qualitative sketch of a possible QCD phase
diagram with a band of approximate chiral spin symmetry.
At zero density, both transitions between the regimes are smooth
crossovers. The lower boundary corresponds to the chiral symmetry
restoration crossover, which could be a true phase transition at larger
density. From. Ref. \cite{GPP}.}
\label{bandgpp}
\end{figure}

A possible definition could be a position of the bend   
of the vector 
screening masses at $T_s  \sim 500$ MeV
seen in Fig. \ref{scrlow} \cite{GPP}. Then there are three different regimes
in QCD at vanishing chemical potential, as illustrated in Fig. \ref{fig:sketch}. At $T < T_{ch}$ we have a hadron resonance gas
with broken chiral symmetry. In the window $ T_{ch} < T < T_s$
the QCD medium is a stringy fluid with restored chiral symmetries and approximate chiral spin symmetry and still with hadron-like degrees of freedom. Above $T_s$ the chiral spin symmetry disappears,
the hadron degrees of freedom  melt down and one can speak of
a quark gluon plasma with parton degrees of freedom.

The next question is a fate of the chiral spin symmetric regime
at non vanishing baryon chemical potential. The quark chemical
potential term in the quark-gluon part of the QCD action
\begin{equation}
S = \int_{0}^{\beta} d\tau \int d^3x
\overline{\Psi}  [ \gamma_{\mu} D_{\mu} + \mu \gamma_4 + m] \Psi,
\label{chpot}
\end{equation}
is manifestly chiral spin and $SU(4)$ symmetric \cite{G4}. This suggests that
these symmetries, observed at $\mu=0$, should also persist at finite $\mu$.

We know from lattice simulations how the chiral crossover temperature, which constitutes
a lower bound for  the chiral spin symmetric regime,
behaves for small $\mu_B\lsi 3T$: 
\begin{equation}
\frac{T_\mathrm{ch}(\mu_B)}{T_\mathrm{ch}(0)}
= 1-0.016(5)\left(\frac{\mu_B}{T_\mathrm{pc}(0)}\right)^2+\ldots\;,
\end{equation} 
with the sub-leading term not yet statistically significant \cite{Bellwied:2015rza,Cea:2015cya,Bonati:2018wdn,HotQCD:2018pds}.
The qualitative behavior of the upper boundary 
of the chiral spin symmetric band can be inferred from
the value of a chosen vector meson screening 
mass at the temperature $T_\mathrm{s}$,
\beq
\frac{m_V(T_\mathrm{s})}{T_\mathrm{s}}=C_0\;.
\eeq
Then,  $CP$-symmetry requires that mesonic screening
masses are even functions of $\mu_B/T$, and therefore
\beq
\frac{m_V(\mu_B)}{T}=C_0+C_2\left(\frac{\mu_B}{T}\right)^2+\ldots\;.
\label{eq:scr_mu}
\eeq
 Keeping this value constant as
chemical potential is varied, $dm_V\stackrel{!}{=}0$,
one finds
\beq
\frac{dT_\mathrm{s}}{d\mu_B}=-\frac{2C_2}{C_0}\frac{\mu_B}{T}-\frac{2C_2^2}{C_0^2}\left(\frac{\mu_B}{T}\right)^3+\ldots\;.
\eeq 
We know from analytic calculations \cite{Vepsalainen:2007ke} as well as 
lattice simulations \cite{Hart:2000ef,Pushkina:2004wa} that $C_2>0$. Then
the upper boundary of the chiral spin symmetric regime leaves the temperature axis with zero slope and negative
curvature. This implies  that a chiral spin symmetric band  bends downwards with chemical
potential, as sketched in Fig. \ref{bandgpp} \cite{GPP}. 

\begin{figure}
\centering
\includegraphics[scale=0.3]{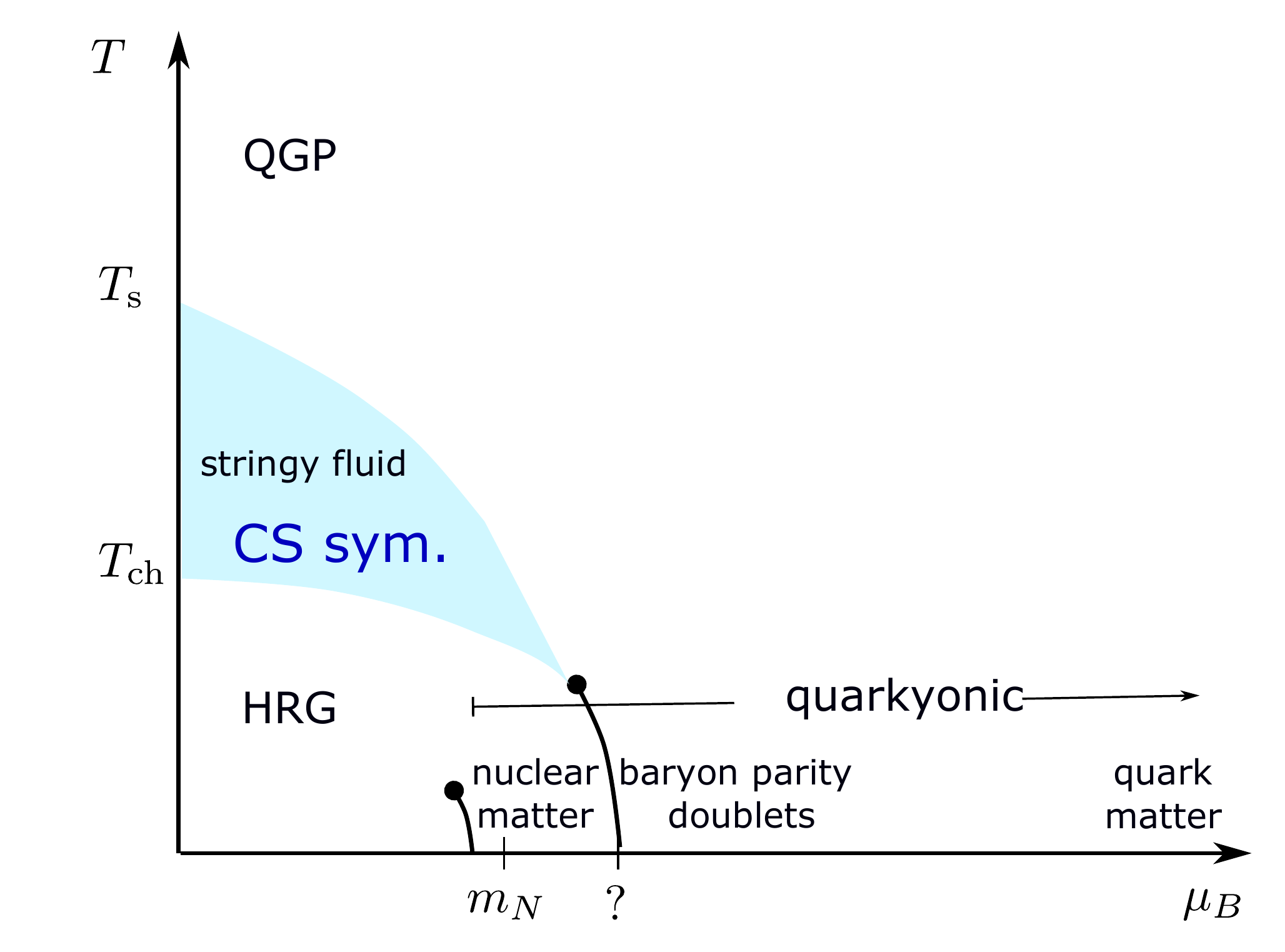}
\caption{Qualitative sketch of a possible QCD phase
diagram with a band of approximate chiral spin symmetry
terminating at the critical end point of a non-analytic
chiral phase transition. From Ref. \cite{GPP}}
\label{bandpr}
\end{figure}

However, our expectations for the upper boundary of
the CS symmetric band are based on sufficiently small
$\frac{\mu_B}{T}$. Consequently we cannot exclude that
at larger chemical potentials the upper and lower boundaries
merge at some point. It could be expected, for example,
at a possible critical end point of the first-order
chiral phase transition at reasonably large chemical potential
\cite{fis,paw},
which is not yet excluded both by the lattice data and experiments.
Then a CS symmetric band could be modified as sketched  in Fig. \ref{bandpr}
\cite{GPP}.
The trend for the upper boundary of the CS-symmetric band
could be studied on the lattice similar to what was done
for $T_{ch}(\mu)$ using  the imaginary chemical potential,
Taylor expansion, etc. 

\section{Dileptons and the chiral spin symmetric band}
In vacuum the electron - positron annihilation into hadrons
shows a powerful resonance peak from $\rho$- and $\omega$-mesons, 
then a sharp  peak from $\phi$-meson. Above these peaks there are
 oscillations about  perturbative $  e^+ + e^- \rightarrow \bar q + q$ curve. These oscillations arise from  broad higher-lying resonances
$\rho', \rho'', ...$. The existence of the resonance peaks and oscillations
around perturbative curve reflects the
confining and chiral symmetry breaking properties of
the QCD vacuum \cite{Shifman:2000jv}. These properties should also persist in a dilute
hadron resonance gas. The 30-years long experimental study in heavy ion collisions
at different
temperatures and chemical potentials 
employ 
the inverse process, with the final state being
the electron-positron pair. This study intends to shed light on the question
to what extent  a hot or dense medium differs from the
vacuum.

The  dilepton production rate is determined by the spectral function of 
the electromagnetic current in the medium which is proportional
to the imaginary part, $\mathrm{Im} [\Pi_\mathrm{em}(M,q;T,\mu_B)]$, 
of the two-point correlator of the electromagnetic current
\cite{Kapusta:2006pm}:
\begin{equation}
\frac{dN}{d^4q d^4x}= - \frac{\alpha^2}{\pi^3 M^2} f^{B}(q_0,T)
\mathrm{Im} [\Pi_\mathrm{em}(M,q;T,\mu_B)].
\end{equation} 
Here $M$ is the invariant mass of the $e^+e^-$ pair with the four-momentum
$ q= (q_0, \vec q)$, $f^{B}(q_0,T)$ is the Bose-Einstein distribution
in the thermalized medium and $\alpha$ the fine structure constant.

The absence of sharp resonance peaks within the fireball
was usually taken as a signal of chiral symmetry
restoration and deconfinement\footnote{Actually  sharp $\rho$,$\omega$
peaks are seen in heavy ion collisions
\cite{STAR:2015zal,PHENIX:2015vek,ALICE:2018ael}. They
are interpreted as arising from the $\rho$,$\omega$-decay into dileptons
beyond the fireball, at the final stage of the heavy ion collision. 
The vacuum cross sections for dilepton production in different elementary
reactions (the so-called cocktail) are subtracted from the full rate, and 
the sharp 
$\rho$,$\omega$ peaks  become much smoother. It is not yet
clear what is left in reality after such subtraction.}. 
For correct interpretation  some care is in order.

The finite temperature $\rho$- 
spectral function is  encoded in the temporal $\rho_{(1,0)+(0,1)}$
correlators of Fig. \ref{tcorr} as well as in the spatial correlators
$V_x$ of Fig. \ref{fig:e2_withfreedata}.
If a Euclidean correlator evaluated in full QCD is essentially
different from that calculated with non-interacting
quarks, one can safely state that the spectral
density  will not be dual to a perturbative description, but should
contain some remnant resonance structure. 
Comparing results for full QCD with those for free quark gas in
Figs. \ref{tcorr} and \ref{fig:e2_withfreedata}  one notices such a 
difference, very clearly seen especially in spatial correlators, up to
temperatures $T \sim 500$ MeV. One then expects some broad structure
in the $\rho$- spectral function. Obviously, it should be essentially
broader than in vacuum.
This could be caused by a fast decay of the $J=1$ excitation into $J=0$ excitations, $\rho\rightarrow \pi+\pi$. 
This is consistent with the less pronounced  $\rho$-peak in the spectral function representing the fireball
above the chiral restoration temperature, as possibly observed at RHIC \cite{STAR:2015zal,PHENIX:2015vek}, 
SPS \cite{NA60:2006ymb,CERES:2006wcq} and LHC \cite{ALICE:2018ael}.
We thus conclude that 
the absence of sharp $\rho$ and $\omega$ peaks in high temperature dilepton spectra coming from the fireball
is entirely consistent with the hadronic description above the chiral restoration. It is an important task for future to establish
quantitative experimental signatures for violation of the
quark-hadron duality in dileptons, i.e. persistence of a broad $\rho$-lke
state above the chiral restoration crossover.
 This violation would imply a measurable
difference between the perturbative $ \bar q + q \rightarrow e^+ + e^-$
contribution and total experimental result within the fireball.

Approximate $SU(4)$ symmetry requires the isoscalar $\omega_{(0,0)}$ correlator to be 
close to the isovector correlator $\rho_{(1,0)+(0,1)}$. Hence, what was said about the $\rho$ peak above chiral
restoration line, should also be true with respect to the
$\omega$ peak. 

\begin{figure}[t]
    \centering
    \includegraphics[width=0.5\columnwidth, clip, trim=3mm 2mm 0 0]{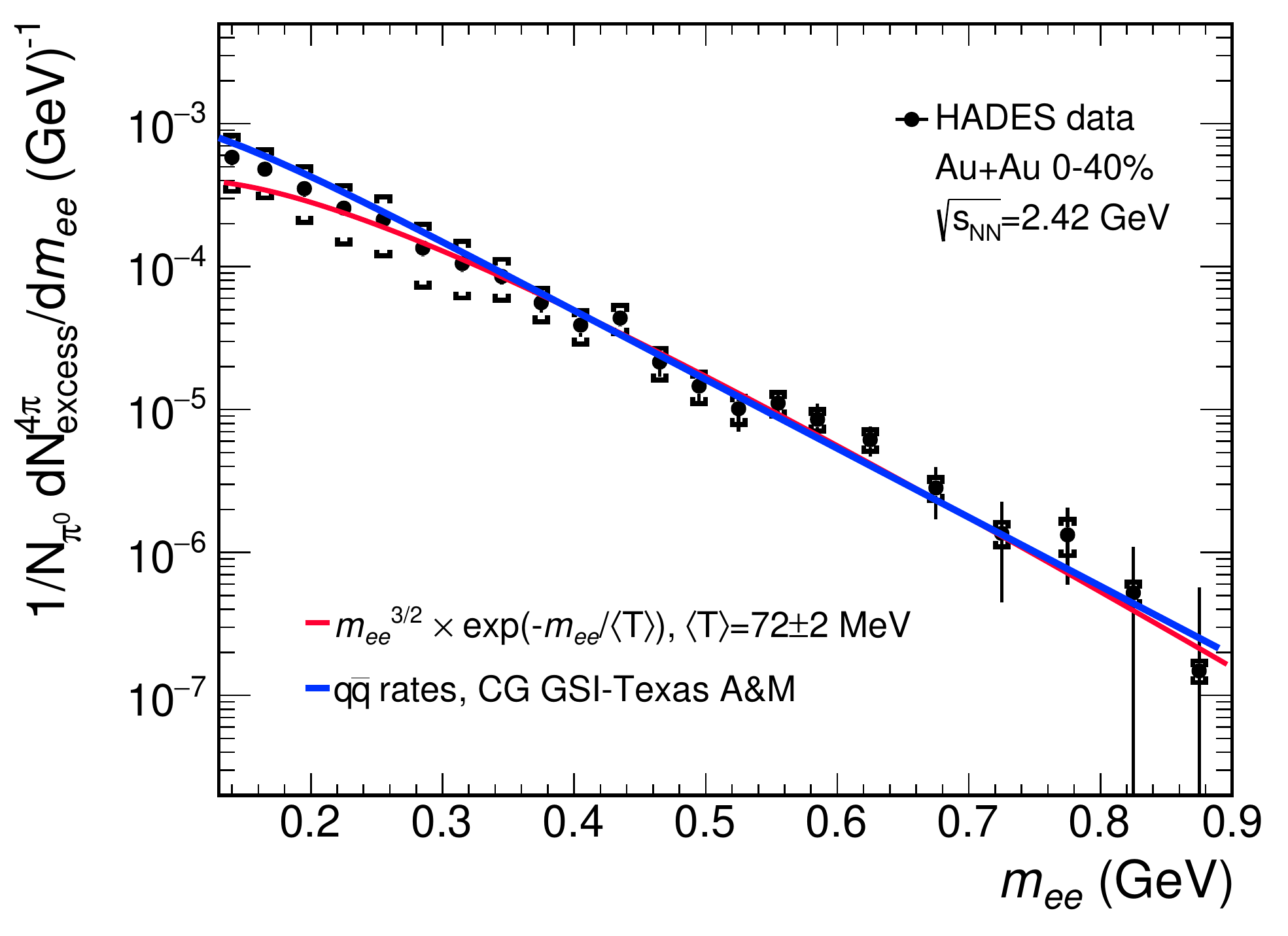}
    \caption{Acceptance-corrected dilepton excess yield obtained in Au-Au collisions at $\sqrt{s_{NN}}=2.42$ GeV \cite{HADES:2019auv}. From Ref. \cite{GPP}.}
    \label{fig:hades}
\end{figure}

The dilepton production at essentially lower
temperature $T \sim 72$ MeV and reasonably large baryon chemical potential 
 $\mu_B \sim 900$ MeV has been studied by HADES
collaboration in Au-Au collisions
at $\sqrt{s_{NN}}=2.42$~GeV
\cite{HADES:2019auv}.
The excess yield extracted by subtracting
the $\eta, \omega$ contributions, which are produced beyond the fireball, 
is shown
in  Fig. \ref{fig:hades}. It exhibits a nearly exponential fall-off
that can be well described by the black-body spectral distribution
\begin{equation}
\frac{dN}{dM} \sim M^{3/2} e^{-M/T}.
\end{equation} 
The latter fit allows HADES to extract the temperature $T \sim 72$ MeV.
 
 No pronounced $\rho$-structure is visible.
The data are well described by the leading order $ \bar q + q \rightarrow e^+ + e^-$
diagram (blue curve).
A slight oscillation
about the perturbative curve might also be visible in Fig. \ref{fig:hades}, that
would hint  at the quark-hadron duality violations. Notice that
such violations can appear provided that there are still contributions
from a very broad $\rho$-state. A broad $\rho$-state in the medium can exist  in the chirally broken regime \cite{Rapp:1999ej}. It  can also exist
in the chirally symmetric and chiral spin symmetric regime like
at zero chemical potential. 

The experimental results
are also consistent with perturbative $ \bar q + q \rightarrow e^+ + e^-$
curve without any oscillations. This might be explained by a perturbative
quark matter. However, there is an alternative explanation.
 A possibility for a broad $\rho$-peak
in a dense chirally symmetric baryonic medium or its absence
is provided by the chiral spin and $SU(4)$ 
symmetric parity doublet matter. 
Within such matter the baryons of positive and
negative parity to leading order decouple from pions and sigmas
as discussed in Chapter 12. However, they can be coupled with the
15-plet of vector mesons from Fig. \ref{algebra}, what would keep the
$SU(4)$ symmetry.
Hence a baryonic medium becomes
a  Fermi gas, except, perhaps, possible corrections from a short range repulsion
between baryons. 
Electromagnetic   baryon - baryon hole excitations guarantee
 an equilibrium between the baryonic Fermi gas and the photonic  Bose
gas. 
This is consistent with the black-body radiation description
of the excess shown in Fig. \ref{fig:hades}. The dilepton production
of the chiral spin symmetric baryonic parity doublet matter is hence very similar to that of  thermalized
quark matter.
This suggests that the HADES point at $T \sim 72$ MeV, $\mu_B \sim 900$ MeV
might be just above the
chiral restoration line, $T_\mathrm{ch}(\mu_B)$, and could possibly be within the
chiral spin and $SU(4)$ symmetric band. 

\section{\label{sec:conclusions}Conclusions}

In this review we have presented  lattice evidences that above
the chiral symmetry restoration crossover around $T_{ch} \sim 155$ MeV
the QCD medium at vanishing baryon chemical potential is populated with the hadron-like (mostly mesons with $J=0,1$) degrees of freedom. This regime is dubbed a stringy fluid because these  states
are chirally symmetric quarks connected into color-singlet hadrons
by a confining chromo-electric field. These hadrons are chirally symmetric and approximately chiral spin symmetric. The chiral spin and $SU(4)$ symmetries  
are symmetries
of the electric part of the QCD Lagrangian which are larger than the chiral
symmetries of the QCD Lagrangian as a whole. The chiral spin and $SU(4)$
symmetries can approximately emerge only if both $U(1)_A$ and $SU(2)_L \times SU(2)_R$
chiral symmetries are restored (at least approximately)  and at the same time
the chromoelectric contributions into energy strongly dominate over the
chromomagnetic contributions and the quark kinetic terms. A direct evidence of
approximate $SU(2)_{CS}$ and $SU(4)$ symmetries of the thermal QCD partition
function and of an effective QCD action above $T_{ch} \sim 155$ MeV is a multiplet
structure observed in spatial and temporal meson correlators calculated on the lattice with
$N_F=2$ QCD with a chirally symmetric Dirac operator at physical quark masses.

These $SU(2)_{CS}$ and $SU(4)$ symmetries smoothly disappear above $T \sim 500$ MeV and correlators of full QCD  approach correlators calculated with a free quark gas. This can happen only if the contributions from
the quark kinetic terms become dominant and the confining chromoelectric
field gets screened. Consequently at zero baryon chemical potential
we can distinguish three different regimes according to symmetries
of the thermal partition function and degrees of freedom. Below
$T_{ch} \sim 155$ MeV the QCD matter is a dilute meson gas with spontaneously
broken chiral symmetry. Within the
window $T_{ch} - 3 T_{ch}$ the hot QCD is represented by the stringy
fluid with restored chiral and approximate chiral spin symmetries.
Above $ \sim 3 T_{ch}$ the chiral spin symmetry disappears and one observes a smooth transition to partonic degrees of freedom, i.e. to a quark-gluon plasma.\footnote{After 
completion of the revised version of this
review  a new study of the chiral
spin symmetry with domain wall fermions in
2+1+1 QCD has appeared with qualitatively similar results
\cite{chiu}.}

The symmetry arguments have been supported by the behavior of screening masses
and of the equation of state. While the screening masses and the equation
of state are compatible with the partonic description at temperatures
of $\sim 1$ GeV and above, at temperatures below $\sim 500$ MeV they demonstrate
a radically different behavior, not consistent with the perturbative
description, indicating a truly non-perturbative regime. 

A direct evidence of hadron degrees freedom in the stringy fluid is
a pion spectral function extracted from the spatial lattice correlators using a generalized K\"{a}llen-Lehmann representation. This spectral function
demonstrates a distinct pion state and its first radial excitation
at temperatures significantly above $T_{ch}$. They become broader
with temperature and melt above $\sim 500$ MeV down. It is important that
this spectral function allows one to predict temporal Euclidean correlators.
The latter correlators can be compared with the lattice data and this comparison
shows a satisfactory agreement. This test implies that the extracted
spectral function is close to reality. Another direct evidence of hadron degrees of freedom in the stringy fluid is existence of 1S,2S,3S and 1P,2P radial and orbital
bottomonium states seen on the lattice at temperatures above $T_{ch}$, that become broader
with temperature. This excitation spectrum is consistent with an optical
potential that consists of its real part with a Coulomb
plus linear confining potential, which is temperature independent,
and an imaginary part that increases with temperature.

The very fact that the pion state, that is discrete in vacuum, becomes
a broad state in  medium above $T_{ch}$, points to a strong interactions
between hadrons in  the stringy fluid. This strong interaction is induced
by a small separation distance between the hadrons. The stringy fluid
is a system of densely packed mesons mainly with $J=0,1$ and is closer
to a liquid rather than to a gas.

The quark chemical potential in the QCD action is manifestly $SU(2)_{CS}$ and
$SU(4)$ symmetric. This suggests that the chiral spin symmetric regime
seen on the lattice at zero chemical potential
between $T_{ch}$ and $3T_{ch}$ extends into the QCD phase diagram as a band
that bends downwards with the chemical potential. In the cold and dense region
a $SU(4)$ symmetric parity doublet matter could be a good candidate
for a chiral spin symmetric matter.

Finally we discussed  available experimental data on dilepton production
at temperatures above $T_{ch}$.  The lattice correlators hint at the existence
of a very broad $\rho$ state that fastly decays into two pions. We also 
discussed
recent results of HADES at smaller temperature and reasonably large baryonic
chemical potential.

 \newpage
	\section*{Acknowledgements}

Some results presented in this review have been obtained together
with Y. Aoki, M. Catillo, G. Cossu, M. Denissenya, H. Fukaya, C. Gattringer, 
S. Hashimoto, C.B. Lang, M. Pak, O. Philipsen, R. Pisarski, S. Prelovsek and C. Rohrhofer. The author is grateful to all of them for a fruitful
collaboration.

\end{document}